\documentclass[prd,aps,superscriptaddress,nofootinbib,letterpaper,twocolumn,10pt]{revtex4-1}
\usepackage{color}
\usepackage{graphicx}
\usepackage{subfigure}
\usepackage{acronym}
\usepackage{amsmath}
\usepackage{SIunits}
\addunit{\invyear}{yr^{-1}}

\def\thercsid{\relax}
\def\rcsid#1{\def\next##1#1{\def\thercsid{##1}}\next}
\rcsid$Id: pipeline.tex,v 1.100 2012/08/13 16:45:16 vallis Exp $

\def\ihope{\texttt{ihope}}

\renewcommand{\today}{\number\day\space\ifcase\month\or
  January\or February\or March\or April\or May\or June\or
  July\or August\or September\or October\or November\or December\fi
  \space\number\year}

\newcommand{\UWM}{University of Wisconsin--Milwaukee, Milwaukee, WI  53201, USA}
\newcommand{\SU}{Syracuse University, Syracuse, NY  13244, USA}
\newcommand{\CU}{Cardiff University, Cardiff, CF24 3AA, United Kingdom}
\newcommand{\UofM}{The University of Mississippi, University, MS 38677, USA}
\newcommand{\LSU}{Louisiana State University, Baton Rouge, LA  70803, USA}
\newcommand{\PI}{Perimeter Institute for Theoretical Physics, Ontario, Canada, N2L 2Y5}
\newcommand{\AEI}{Albert-Einstein-Institut, Max-Planck-Institut f\"ur
Gravitationsphysik, D-30167 Hannover, Germany}
\newcommand{\Leibniz}{Leibniz Universit\"at Hannover, D-30167 Hannover, Germany}
\newcommand{\LIGOCIT}{LIGO Laboratory, California Institute of Technology, Pasadena, CA  91125, USA}
\newcommand{\JPL}{Jet Propulsion Laboratory, California Institute of Technology, Pasadena, CA  91109, USA}
\newcommand{\LIGOMIT}{LIGO Laboratory, Massachusetts Institute of Technology, Cambridge, MA 02139, USA}
\newcommand{\CITA}{Canadian Institute for Theoretical Astrophysics, University of Toronto, Toronto, Ontario, M5S 3H8, Canada}
\newcommand{\UMD}{University of Maryland, College Park, MD 20742 USA}
\newcommand{\GaTech}{Center for Relativistic Astrophysics and School of Physics,
Georgia Institute of Technology, Atlanta, GA 30332, USA}
\newcommand{\iit}{IIT Gandhinagar, VGEC Complex, Chandkheda Ahmedabad 382424, Gujarat, India}

\newcommand{\golm}{Max-Planck-Institut f\"ur Gravitationsphysik -
Albert-Einstein-Institut Am M\"uhlenberg 1, 14476 Potsdam-Golm, Germany}
\newcommand{\UTB}{The University of Texas at Brownsville and Texas Southmost College, Brownsville, TX  78520, USA}

\begin{document}

\title{Searching for gravitational waves from binary coalescence}

\author{S.~Babak}\affiliation{\CU}\affiliation{\golm}
\author{R.~Biswas}\affiliation{\UWM}\affiliation{\UTB}
\author{P.~R.~Brady}\affiliation{\UWM}
\author{D.~A.~Brown}\affiliation{\SU}\affiliation{\LIGOCIT}\affiliation{\UWM}
\author{K.~Cannon}\affiliation{\CITA}\affiliation{\LIGOCIT}\affiliation{\UWM}
\author{C.~D.~Capano}\affiliation{\SU}\affiliation{\UMD}
\author{J.~H.~Clayton}\affiliation{\UWM}
\author{T.~Cokelaer}\affiliation{\CU}
\author{J.~D.~E.~Creighton}\affiliation{\UWM}
\author{T.~Dent}\affiliation{\CU}\affiliation{\AEI}
\author{A.~Dietz}\affiliation{\UofM}\affiliation{\CU}\affiliation{\LSU}
\author{S.~Fairhurst}\affiliation{\CU}\affiliation{\LIGOCIT}\affiliation{\UWM}
\author{N.~Fotopoulos}\affiliation{\LIGOCIT}\affiliation{\UWM}
\author{G.~Gonz\'alez}\affiliation{\LSU}
\author{C.~Hanna}\affiliation{\PI}\affiliation{\LIGOCIT}\affiliation{\LSU}
\author{I.~W.~Harry}\affiliation{\CU}\affiliation{\SU}
\author{G.~Jones}\affiliation{\CU}
\author{D.~Keppel}\affiliation{\AEI}\affiliation{\Leibniz}\affiliation{\LIGOCIT}
\author{D.~J.~A.~McKechan}\affiliation{\CU}
\author{L.~Pekowsky}\affiliation{\SU}\affiliation{\GaTech}
\author{S.~Privitera}\affiliation{\LIGOCIT}
\author{C.~Robinson}\affiliation{\CU}\affiliation{\UMD}
\author{A.~C.~Rodriguez}\affiliation{\LSU}
\author{B.~S.~Sathyaprakash}\affiliation{\CU}
\author{A.~S.~Sengupta}\affiliation{\iit}\affiliation{\LIGOCIT}\affiliation{\CU}
\author{M.~Vallisneri}\affiliation{\JPL}
\author{R.~Vaulin}\affiliation{\LIGOMIT}\affiliation{\UWM}
\author{A.~J.~Weinstein}\affiliation{\LIGOCIT}

\begin{abstract}
We describe the implementation of a search for gravitational waves from
compact binary coalescences in LIGO and Virgo data. This all-sky,
all-time, multi-detector search for binary coalescence has been used to
search data taken in recent LIGO and Virgo runs.  The search is built
around a matched filter analysis of the data, augmented by numerous
signal consistency tests designed to distinguish artifacts of
non-Gaussian detector noise from potential detections.   We demonstrate
the search performance using Gaussian noise and data from the fifth LIGO
science run and demonstrate that the signal consistency tests are
capable of mitigating the effect of non-Gaussian noise and providing a
sensitivity comparable to that achieved in Gaussian noise.
\end{abstract}

\maketitle
\acrodef{LIGO}{Laser Interferometer Gravitational-wave Observatory}
\acrodef{CBC}{Compact binary coalescence}
\acrodef{LVC}{LIGO Scienctific Collaboration and the Virgo Collaboration}
\acrodef{S5}{LIGO's fifth science run}
\acrodef{S6}{LIGO's sixth science run}
\acrodef{VSR1}{Virgo's first science run}
\acrodef{VSR2}{Virgo's second science run}
\acrodef{VSR3}{Virgo's third science run}
\acrodef{VSR23}{Virgo's second and third science runs}
\acrodef{DAG}{directed acyclic graph}
\acrodef{IFO}{interferometer}
\acrodef{HIPE}{Hierarchical Inspiral Pipeline Executable}
\acrodef{PN}{post-Newtonian}
\acrodef{SNR}{signal-to-noise ratio}
\acrodef{GW}{gravitational-wave}
\acrodef{HIPE}{Hierarchical Inspiral Pipeline Executable}
\acrodef{FAR}{false alarm rate}
\acrodef{FAP}{false alarm probability}
\acrodef{PSD}{power spectral density}
\acrodef{SPA}{stationary phase approximation}
\acrodef{ISCO}{inner-most stable circular orbit}
\acrodef{GEO}{GEO}
\acrodef{FFT}{FFT}
\acrodef{DQ}{Data Quality}
\acrodef{BH}{black hole}
\acrodef{BBH}{binary black hole}
\acrodef{BNS}{binary neutron star}
\acrodef{NSBH}{neutron star--black hole}
\acrodef{LAL}{LAL}
\acrodef{NS}{neutron star}
\acrodef{GRB}{GRB}
\acrodef{IMR}{IMR}

\section{Introduction}
\label{sec:intro}

Coalescing binaries of compact objects such as \acp{NS} and stellar-mass
\acp{BH} are promising \ac{GW} sources for ground-based, kilometer-scale
interferometric detectors such as LIGO \cite{Abbott:2007kv}, Virgo
\cite{Accadia:2012zz}, and GEO600 \cite{Grote:2008}, which are sensitive to
waves of frequencies between tens and thousands of Hertz. Numerous searches for
these signals were performed on data from the six LIGO and GEO science runs
(S1--S6) and from the four Virgo science runs (VSR1--4)
\cite{Abbott:2003pj,LIGOS2iul,LIGOS2bbh,LIGOS2macho,ligotama,
S3_BCVSpin,LIGOS3S4all,Collaboration:2009tt,Abbott:2009qj,S5LowMassLV,Collaboration:S6CBClowmass}.

Over time, the software developed to run these searches and evaluate the
significance of results evolved into a sophisticated pipeline, known as \ihope.
An early version of the pipeline was described in \cite{brown-2005-22}. In this
paper, we describe the \ihope\ pipeline in detail and we characterize its detection
performance by comparing the analysis of a month of real data with the analysis
of an equivalent length of simulated data with Gaussian stationary noise.

\acfp{CBC} consist of three dynamical phases: a gradual
\emph{inspiral}, which is described accurately by the post-Newtonian
approximation to the Einstein equations \cite{Blanchet:2002av}; a nonlinear
\emph{merger}, which can be modeled with numerical simulations (see
\cite{Centrella:2010,Hannam:2009rd,2011arXiv1107.2819S} for recent reviews); and the final ringdown of the
merged object to a quiescent state \cite{Berti:2007gv}. For the lighter NS--NS
systems, only the inspiral lies within the band of detector sensitivity. Since
\ac{CBC} waveforms are well modeled, it is natural to search for them by
matched-filtering the data with banks of theoretical \emph{template} waveforms
\cite{wainstein:1962}.

The most general \ac{CBC} waveform is described by seventeen parameters,
which include the masses and intrinsic spins of the binary components, as well
as the location, orientation, and orbital elements of the binary.
It is not feasible to perform a search by placing templates across such a
high-dimensional parameter space.
However, it is astrophysically reasonable to neglect orbital eccentricity
\cite{Cokelaer:2009hj,Brown:2009ng}; furthermore, \ac{CBC} waveforms that omit the effects of spins have
been shown to have acceptable phase overlaps with spinning-binary waveforms, and
are therefore suitable for the purpose of detecting \acp{CBC}, if not to estimate
their parameters accurately \cite{VanDenBroeck:2009gd}.

Thus, \ac{CBC} searches so far have relied on nonspinning waveforms that are
parameterized only by the component masses, by the location and orientation of
the binary, by the initial orbital phase, and by the time of coalescence. Among
these parameters, the masses determine the intrinsic phasing of the waveforms,
while the others affect only the relative amplitudes, phases, and timing
observed at multiple detector sites \cite{Allen:2005fk}.
It follows that templates need to be placed only across the two-dimensional
parameter space spanned by the masses \cite{Allen:2005fk}. Even so, past \ac{CBC}
searches have required many thousands of templates to cover their target ranges
of masses.
(We note that \ihope\ could be extended easily to nonprecessing binaries with
aligned spins. However, more general precessing waveforms
would prove more difficult, as discussed in
\cite{PhysRevD.49.6274,Apostolatos:1995,BuonannoChenVallisneri:2003b,Pan:2003qt}.)

In the context of stationary Gaussian noise, matched-filtering would directly yield the most
statistically significant detection candidates. In practice, environmental and
instrumental disturbances cause non-Gaussian noise transients (\emph{glitches})
in the data.
Searches must distinguish between the candidates, or \textit{triggers}, resulting from glitches and
those resulting from true \acp{GW}.
The techniques developed for this challenging task include \emph{coincidence}
(signals must be observed in two or more detectors with consistent mass
parameters and times of arrival), \emph{signal-consistency} tests (which
quantify how much a signal's amplitude and frequency evolution is consistent
with theoretical waveforms \cite{Allen:2004}),
and \emph{data quality vetoes} (which identify time periods when the detector 
glitch rate is elevated). We describe these in detail later.

The \emph{statistical significance} after the consistency tests have been
applied is then quantified by computing the \ac{FAP} or \ac{FAR} of each
candidate; we define both below. For this, the background of noise-induced
candidates is estimated by performing \emph{time shifts}, whereby the
coincidence and consistency tests are run after imposing relative time offsets
on the data from different detectors. Any consistent candidate found in this way
must be due to noise; furthermore, if the noise of different detectors is
uncorrelated, the resulting background rate is representative of the rate at
zero shift.

The \emph{sensitivity} of the search to \ac{CBC} waves is estimated by adding
simulated signals (\emph{injections}) to the detector data, and verifying which
are detected by the pipeline. With this diagnostic we can tune the search to a
specific class of signals (e.g., a region in the mass plane), and we can give an
astrophysical interpretation, such as an upper limit on \ac{CBC} rates
\cite{Fairhurst:2007qj}, to completed searches.

As discussed below, commissioning a \ac{GW} search with the \ihope\ pipeline requires a number of
parameter tunings, which include the handling of coincidences, the
signal-consistency tests, and the final ranking of triggers.  To avoid biasing
the results, \ihope\ permits a blind analysis:  the results of the
non-time-shifted analysis can be sequestered, and tuning performed using only
the injections and time-shifted results.  Later, with the parameter tunings
frozen, the non-time-shifted results can be unblinded to reveal the candidate GW
events.

This paper is organized as follows. In Sec.\ \ref{sec:coinc_search} we provide a
brief overview of the \ihope\ pipeline, and describe its first few stages (data conditioning,
template placement, filtering, coincidence), which would be sufficient to
implement a search in Gaussian noise but not, as we show, in real detector data.
In Sec.\ \ref{sec:nongauss} we describe the various techniques that have been
developed to eliminate the majority of background triggers
due to non-Gaussian noise.
In Sec.\ \ref{sec:interpretation} we describe how the \ihope\ results are used
to make astrophysical statements about the presence or absence of signals in the
data, and to put constraints on \ac{CBC} event rates.
Last, in Sec.\ \ref{sec:discussion} we discuss ways in which the analysis can be
enhanced to improve sensitivity, reduce latency, and find use in the
advanced-detector era. 

Throughout this paper we show representative \ihope\ output, taken from a search
of one month of LIGO data from the S5 run (the third month in
\cite{Abbott:2009qj}), when all three LIGO detectors (but not Virgo) were
operational. The search focused on low-mass \ac{CBC} signals with component masses $>
1 \, M_{\odot}$ and total mass $< 25 \, M_{\odot}$. For comparison, we also run
the same search on Gaussian noise generated at the design sensitivity of the
\ac{LIGO} detectors (using the same data times as the real data). Where we
perform \ac{GW}-signal injections (see Sec.\ \ref{ssec:injections}), we adopt a
population of binary-neutron-star inspirals, uniformly distributed in distance,
coalescence time, sky position and orientation angles.

\section{IHOPE, part 1: setting up a matched-filtering search with
multiple-detector coincidence}
\label{sec:coinc_search}

\begin{figure}[tp]
\includegraphics[width=\linewidth]{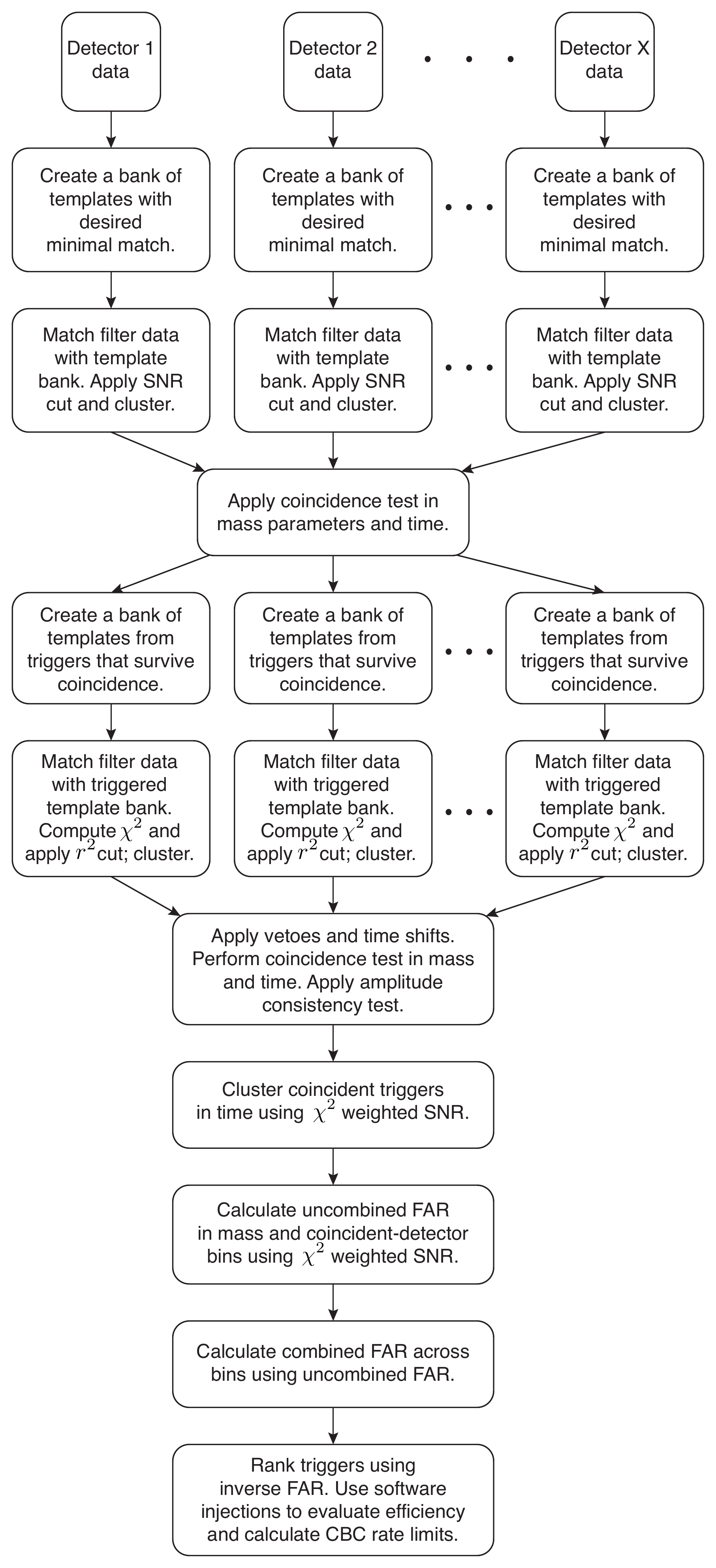}
\caption{\label{fig:cbc_ihopepipe}
Structure of the \ihope\ pipeline.
}
\end{figure}

The stages of the \ihope\ pipeline are presented schematically in Fig.\
\ref{fig:cbc_ihopepipe}, and are described in detail in Secs.\
\ref{sec:coinc_search}--\ref{sec:interpretation} of this paper.
First, the \emph{science data} to be analyzed is identified and split into
\(\unit{2048}{\second}\) \emph{blocks}, and the power spectral density is estimated for each
block (see Sec.\ \ref{ssec:segment_psd}).
Next, a template bank is constructed independently for each detector and each
block (Sec.\ \ref{ssec:tmpltbank}).
The data blocks are matched-filtered against each bank template, and the times
when the \ac{SNR} rises above a set threshold are recorded as \emph{triggers} (Sec.\
\ref{ssec:inspiral}).
The triggers from each detector are then compared to identify
\emph{coincidences}---that is, triggers that occur in two or more detectors with
similar masses and compatible times (Sec.\ \ref{ssec:thinca}).

If detector noise was Gaussian and stationary, we could proceed directly to the
statistical interpretation of the triggers. Unfortunately, non-Gaussian noise
glitches generate both an increase in the number of low-SNR triggers as well as
high-SNR triggers that form long tails in the distribution of SNRs.  The
increase in low-SNR triggers will cause an small, but inevitable, reduction in
the sensitivity of the search.   It is, however, vital to distinguish the
high-SNR background triggers from those caused by real \ac{GW} signals.  To
achieve this, the coincident triggers are used to generate a reduced template
bank for a second round of matched-filtering in each detector (see the beginning
of Sec.\ \ref{sec:nongauss}).
This time, signal-consistency tests are performed on each trigger to help
differentiate background from true signals (Secs.\ \ref{ssec:chisq},
\ref{ssec:rsq}). These tests are computationally expensive, so we reserve them
for this second pass.
Single-detector triggers are again compared for coincidence, and the final list
is clustered 
and ranked (Sec.\ \ref{ssec:effective_snr}), taking into account
signal consistency, amplitude consistency among detectors (Sec.\
\ref{ssec:amplitude_consistency}), as well as the times in which the detectors
were not operating optimally (Sec.\ \ref{ssec:vetotimes}). These steps leave
coincident triggers that have a quasi-Gaussian distribution; they can now be
evaluated for statistical significance, and used to derive event-rate upper
limits in the absence of a detection.

To do this, the steps of the search that involve coincidence are repeated many
times, artificially shifting the time stamps of triggers in different detectors,
such that no true \ac{GW}
signal would actually be found in coincidence (Sec.\ \ref{ssec:background}).
The resulting \emph{time-shift} triggers are used to calculate the \ac{FAR}
of the \emph{in-time} (zero-shift) triggers.
Those with FAR lower than some threshold are the \ac{GW}-signal
candidates (Sec.\ \ref{ssec:far}). 
Simulated \ac{GW} signals are then injected into the data,
and by observing which injections are recovered as triggers
with FAR lower than some threshold, we
can characterize detection efficiency as a function of distance 
and other parameters (Sec.\
\ref{ssec:injections}), providing an astrophysical interpretation for the
search.
Together with the \acp{FAR} of the loudest triggers, the efficiency yields the upper
limits (Sec.\ \ref{ssec:ul}).

\subsection{Data segmentation and conditioning, power-spectral-density
generation}
\label{ssec:segment_psd}

As a first step in the pipeline, \ihope\ identifies the stretches of detector
data that should be analyzed: for each detector, such \emph{science segments}
are those for which the detector was locked (i.e., interferometer laser light was
resonant in Fabry--Perot cavities \cite{Abbott:2007kv}), no other experimental work was being
performed, and the detector's ``science mode'' was confirmed by a human
``science monitor.'' \ihope\ builds a list of science-segment times by querying
a network-accessible database that contains this information for all detectors.

The \ac{LIGO} and Virgo \ac{GW}-strain data are sampled at \(\unit{16,384}{\hertz}\) and
\(\unit{20,000}{\hertz}\), respectively, but both are down-sampled to
\(\unit{4096}{\hertz}\) prior to analysis \cite{brown-2005-22}, since at
frequencies above \(\unit{1}{\kilo\hertz}\) to  \(\unit{2}{\kilo\hertz}\) detector noise overwhelms any likely \ac{CBC}
signal. This sampling rate sets the Nyquist frequency at \(\unit{2048}{\hertz}\);
to prevent aliasing, the data are preconditioned with a time-domain digital
filter with low-pass cutoff at the Nyquist frequency \cite{brown-2005-22}.
While \ac{CBC} signals extend to arbitrarily low frequencies, detector
sensitivity degrades rapidly, so very little \ac{GW} power could be observed below
\(\unit{40}{\hertz}\). Therefore, we usually suppress signals below \(\unit{30}{\hertz}\) with two rounds of
8th-order Butterworth high-pass filters, and analyze data only above \(\unit{40}{\hertz}\).

Both the low- and high-pass filters corrupt the data at the start and end of a
science segment, so the first and last few seconds of data (typically \(\unit{8}{\second}\)) are
discarded after applying the filters. Furthermore, SNRs are computed by
correlating templates with the (noise-weighted) data stream, which is only
possible if a stretch of data of at least the same length as the template is
available. 
Altogether, the data are split into \(\unit{256}{\second}\) segments, and the first
and last \(\unit{64}{\second}\)
of each segment are not used in the search. Neighboring segments are
overlapped by \(\unit{128}{\second}\) to ensure that all available data are analyzed.

The strain \ac{PSD} is computed separately for every \(\unit{2048}{\second}\)
\textit{block} of data (consisting of 15 overlapping \(\unit{256}{\second}\)
segments). The blocks themselves are overlapped by \(\unit{128}{\second}\). The
block \ac{PSD} is estimated by taking the median \cite{findchirp} (in each
frequency bin) of the segment PSDs, ensuring robustness against noise
transients and \ac{GW} signals (whether real or simulated). The \ac{PSD} is
used in the computation of SNRs, and to set the spacing of templates in the
banks. Science segments shorter than \(\unit{2064}{\second}\)
(\(\unit{2048}{\second}\) block length and \(\unit{16}{\second}\) to account
for the padding on either side) are not used in the analysis, since they cannot
provide an accurate \ac{PSD} estimate.

\subsection{Template-bank generation}
\label{ssec:tmpltbank}

Template banks must be sufficiently dense in parameter space to ensure a minimal
loss of matched-filtering \ac{SNR} for any \ac{CBC} signal within the mass range
of interest; however, the computational cost of a search is proportional to the
number of templates in a bank. The method used to place templates must balance
these considerations. This problem is well explored for nonspinning \ac{CBC}
signals
\cite{hexabank,BBCCS:2006,Owen:1998dk,Owen:1995tm,Balasubramanian:1995bm,
Dhurandhar:1992mw,SathyaDhurandhar:1991}, for which templates need only be
placed across the two-dimensional \emph{intrinsic}-parameter space spanned by
the two component masses. The other \emph{extrinsic} parameters enter only as
amplitude scalings or phase offsets, and the SNR can be maximized analytically
over these parameters after filtering by each template.

Templates are placed in parameter space so that the \emph{match} between any \ac{GW}
signal and the best-fitting template is better than a \emph{minimum match} MM
(typically 97\%). The match between signals $h$ with parameter vectors
$\boldsymbol{\xi}_1$ and $\boldsymbol{\xi}_2$ is defined as
\begin{equation}
\max_{t_{c}, \phi_{c}} 
\frac{\bigl(h(\boldsymbol{\xi}_1)\big|h(\boldsymbol{\xi}_2)\bigr)
}{
\sqrt{\bigl(h(\boldsymbol{\xi}_1)\big|h(\boldsymbol{\xi}_1)\bigr)}
\sqrt{\bigl(h(\boldsymbol{\xi}_2)\big|h(\boldsymbol{\xi}_2)\bigr)}}
\end{equation}
where $t_{c}$ and $\phi_{c}$ are the time and phase of coalescence of the
signal, $(\cdot|\cdot)$ is the standard noise-weighted inner product
\begin{equation}
\label{eq:inner_product}
(a|b) = 4 \, \mathrm{Re} \int_{f_{\mathrm{low}}}^{f_{\mathrm{high}}}
\frac{\tilde{a}^{*}(f) \, \tilde{b}(f)}{S_n(f)} df,
\end{equation}
with $S_n(f)$ the one-sided detector-noise \ac{PSD}. The MM represents the
worst-case reduction in matched-filtering SNR, and correspondingly the
worst-case reduction in the maximum detection distance of a search. Thus, under
the assumption of sources uniformly distributed in volume, the loss in  
sensitivity due to template-bank discreteness is bounded by $\mathrm{MM}^3$, or
$\simeq 10\%$ for $\mathrm{MM} = 97\%$.

It is computationally expensive to obtain template mismatches for pairs of
templates using Eq.\ \eqref{eq:inner_product}, so an approximation based on a
parameter-space \emph{metric} is used instead:
\begin{equation} 
1 - (h(\boldsymbol{\xi}) | h(\boldsymbol{\xi} + \delta\boldsymbol{\xi}))
\simeq \sum_{ij} g_{ij}(\boldsymbol{\xi}) \,\delta\xi^i \,\delta\xi^j,
\end{equation}
where
\begin{equation} 
\label{eq:cbc_metric}
g_{ij}(\boldsymbol{\xi}) = - \frac{1}{2} \frac{\partial^2
\left(h(\boldsymbol{\xi})|h(\boldsymbol{\xi})\right)}{\partial \xi^i \partial
\xi^j}.
\end{equation}
The approximation holds as long as the metric is roughly constant between bank
templates, and is helped by choosing parameters (i.e., coordinates $\boldsymbol{\xi}$) that make
the metric almost flat, such as the ``chirp times'' $\tau_{0}$, $\tau_{3}$ given
by \cite{Sathyaprakash:1994nj}
\begin{eqnarray}
 \label{eq:tau0tau3def}
 \tau_0 &=& \frac{5}{256\pi f_{\mathrm{low}} \eta} 
   \left(\frac{\pi G M f_{\mathrm{low}}}{c^3}\right)^{-5/3}, \\
 \tau_3 &=& \frac{5}{8f_{\mathrm{low}} \eta} \left(\frac{\pi G M f_{\mathrm{low}}}{c^3}\right)^{-2/3}.
\end{eqnarray}
Here $M$ is the total mass, $\eta$ is the symmetric mass ratio $\eta = m_1 m_2/M^2$
and $f_\mathrm{low}$ is the lower frequency cutoff used in the template generation.

\begin{figure}
    \includegraphics[width=\linewidth]{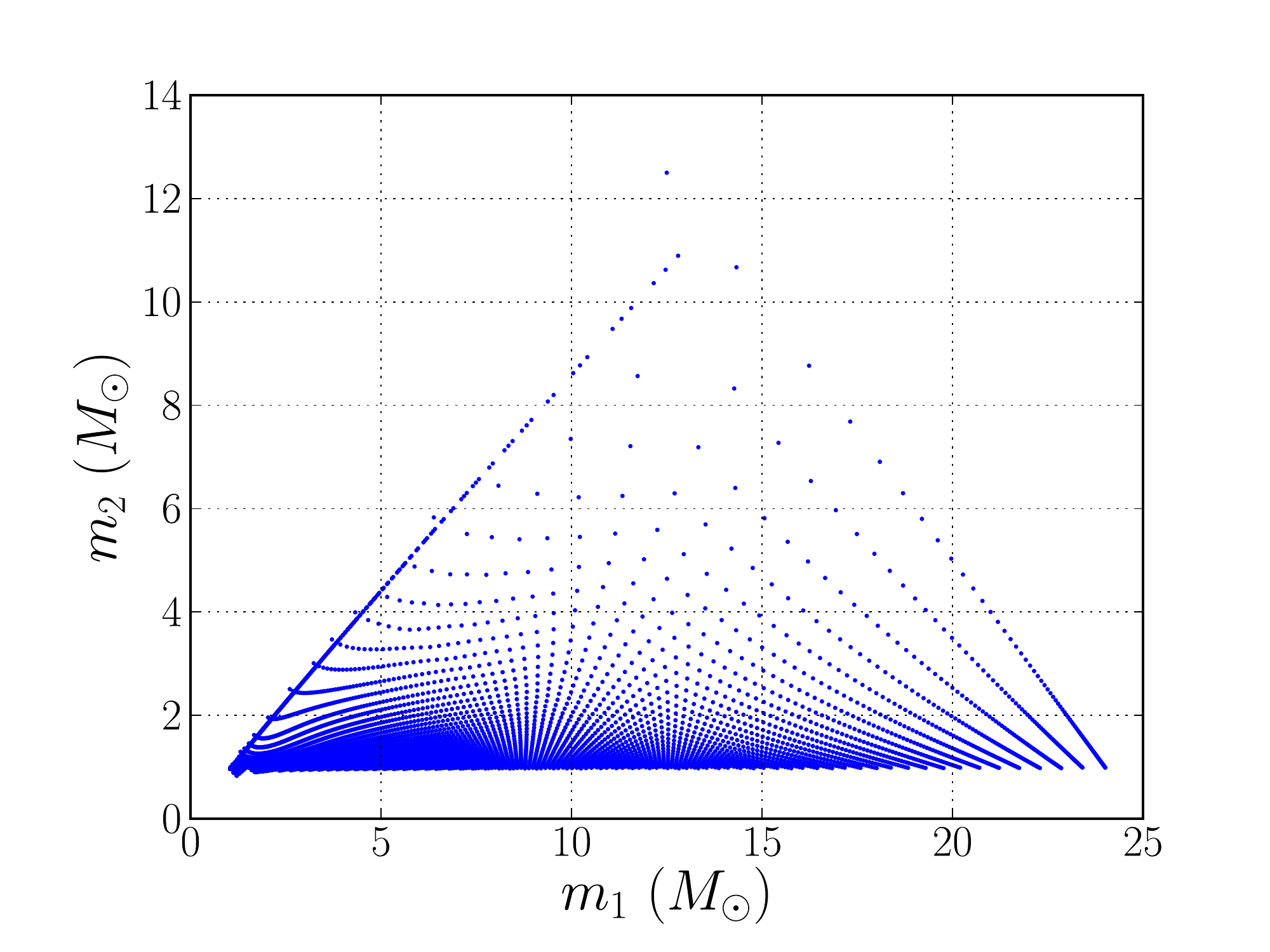}
    \includegraphics[width=\linewidth]{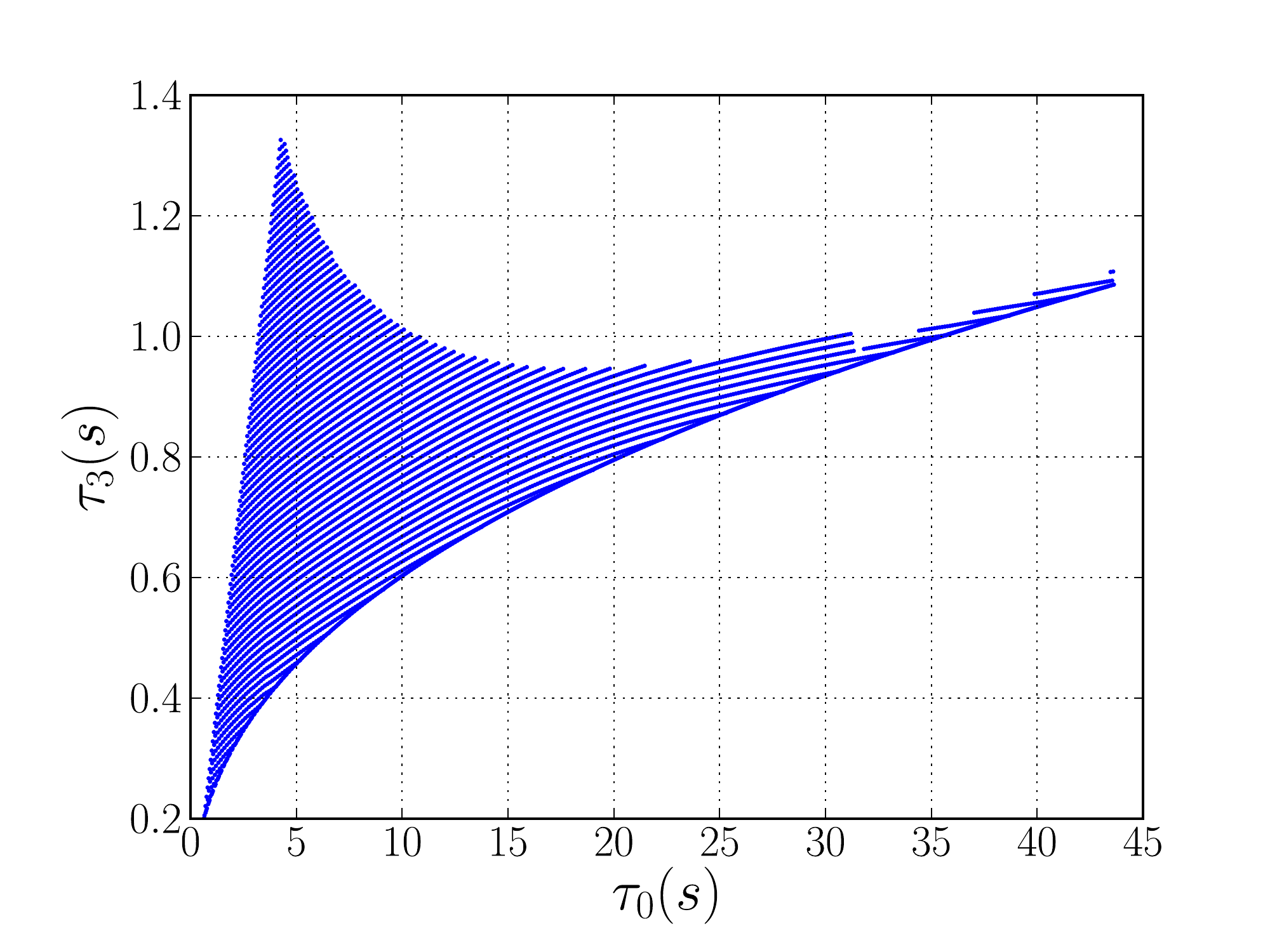}
\caption{
\label{fig:template_bank}
A typical template bank for a low-mass \ac{CBC} inspiral search, as plotted in
$m_1$--$m_2$ space (top panel) and $\tau_0$--$\tau_3$ space (bottom panel).
Templates are distributed more evenly over $\tau_0$ and $\tau_3$, since the
parameter-space metric is approximately flat in those coordinates.}
\end{figure}

For the S5--S6 and VSR1--3 \ac{CBC} searches, templates were placed on a regular
hexagonal lattice in $\tau_0$--$\tau_3$ space \cite{hexabank}, sized so that MM
would be 97\%
\cite{Collaboration:2009tt,Abbott:2009qj,Collaboration:S6CBClowmass}. The metric
was computed using inspiral waveforms at the second post-Newtonian
(2PN) order in phase. Higher-order templates are now used in searches (some including
merger and ringdown), but not for template placement; work is ongoing to
implement that. Figure \ref{fig:template_bank} shows a typical template bank in
both $m_1$--$m_2$ and $\tau_0$--$\tau_3$ space for the low-mass \ac{CBC} search.
 For a typical data block, the bank contains around 6000 templates (Virgo,
which has a a flatter noise PSD, requires more).

As Eqs.\ \eqref{eq:cbc_metric} and \eqref{eq:inner_product} imply, the metric
depends on both the detector-noise \ac{PSD} and the frequency limits
$f_\mathrm{low}$ and $f_\mathrm{high}$. We set $f_\mathrm{low}$ to
\(\unit{40}{\hertz}\), while
$f_\mathrm{high}$ is chosen naturally as the frequency at which waveforms end
(\(\unit{200}{\hertz}\) and \(\unit{2}{\kilo\hertz}\) for the highest- and lowest-mass signals, respectively). The
PSD changes between data blocks, but usually only slightly, so template banks
stay roughly constant over time in a data set.

\subsection{Matched filtering}
\label{ssec:inspiral}

The central stage of the pipeline is the matched filtering of detector data with
bank templates, resulting in a list of \emph{triggers} that are further analyzed
downstream. This stage was described in detail in Ref.\ \cite{findchirp}; here
we sketch its key features.

The waveform from a non-spinning \ac{CBC}, as observed by a ground-based detector and
neglecting higher-order amplitude corrections, can be written as
\begin{equation}
 h(\tau) = h_0(\tau) \cos \Phi_0 + h_{\pi/2}(\tau) \sin \Phi_0,
\end{equation}
with
\begin{equation}
\left(\begin{array}{c} h_0(\tau) \\ h_{\pi/2}(\tau) \end{array}\right) =
A f(\tau)^{2/3}
\left(\begin{array}{c} \cos\bigl(\Phi(\tau)\bigr) \\
-\sin\bigl(\Phi(\tau)\bigr) \end{array}\right).
\end{equation}
Here, $\tau$ is a time variable relative to the coalescence time, $t_c$.  The
constant amplitude $A$ and phase $\Phi_{0}$, between them, depend on all the
binary parameters: masses, sky location and distance, orientation, and
(nominal) orbital phase at coalescence. By contrast, the time-dependent
frequency $f(\tau)$ and phase $\Phi(\tau)$ depend only on the component masses
\footnote{Strictly, the waveforms depend upon the red-shifted component masses
$(1+z)m_{1,2}$.  Note, however, that this does not affect the search as one can
simply replace the masses by their redshifted values.}
and on the absolute time of coalescence.

The squared \ac{SNR} $\rho^2$ for the data $s$ and template $h$, analytically
maximized over $A$ and $\Phi_{0}$, is given by
\begin{equation}
\label{eq:cbc_snr}
\rho^2 = \frac{ (s|h_0)^2 + (s | h_{\pi/2})^2 }{(h_0|h_0)};
\end{equation}
here we assume that $\tilde{h}_{\pi/2}(f) = i \tilde{h}_0(f)$, which is
identically true for waveforms defined in the frequency domain with the
stationary-phase approximation \cite{Droz:1999qx}, and approximately true for
all slowly evolving \ac{CBC} waveforms.  

The maximized statistic $\rho^2$ of Eq.\ \eqref{eq:cbc_snr} is a function only
of the component masses and the time of coalescence $t_c$. Now, a time shift
can be folded in the computation of inner products by noting that $g(\tau) =
h(\tau - \Delta t_c)$ transforms to  $\tilde{g}(f) = e^{i2\pi f \Delta
t_c}\tilde{h}(f)$; therefore, the SNR can be computed as a function of $t_c$ by
the inverse Fourier transform (a complex quantity)
\begin{equation} 
\label{eq:cbc_fouriermagic}
(s|h)(\Delta t_c) = 4 \int_{f_{\mathrm{low}}}^{f_{\mathrm{high}}}\frac{\tilde{s}(f)
\tilde{h}^{*}(f)}{S_n (f)}
e^{2\pi i f \Delta t_c} df.
\end{equation}
Furthermore, if $\tilde{h}_{\pi/2}(f) = i \tilde{h}_0(f)$ then Eq.\
\eqref{eq:cbc_fouriermagic}, computed for $h = h_0$, yields $(s|h_0)(\Delta t_c)
+ i\,(s|h_{\pi/2})(\Delta t_c)$.

The \ihope\ matched-filtering engine implements the discrete analogs of 
Eqs.\ \eqref{eq:cbc_snr} and \eqref{eq:cbc_fouriermagic} \cite{Allen:2005fk}
using the efficient FFTW library \cite{FFTW}. The resulting SNRs are not stored
for every template and every possible $t_c$; instead, we only retain triggers
that exceed an empirically determined threshold (typically 5.5), and that
corresponds to maxima of the SNR time series---that is, a trigger above the
threshold is kept only if there are no triggers with higher SNR within a
predefined time window, typically set to the length of the template (this is
referred to as \emph{time clustering}).

\begin{figure}
\includegraphics[width=\linewidth]{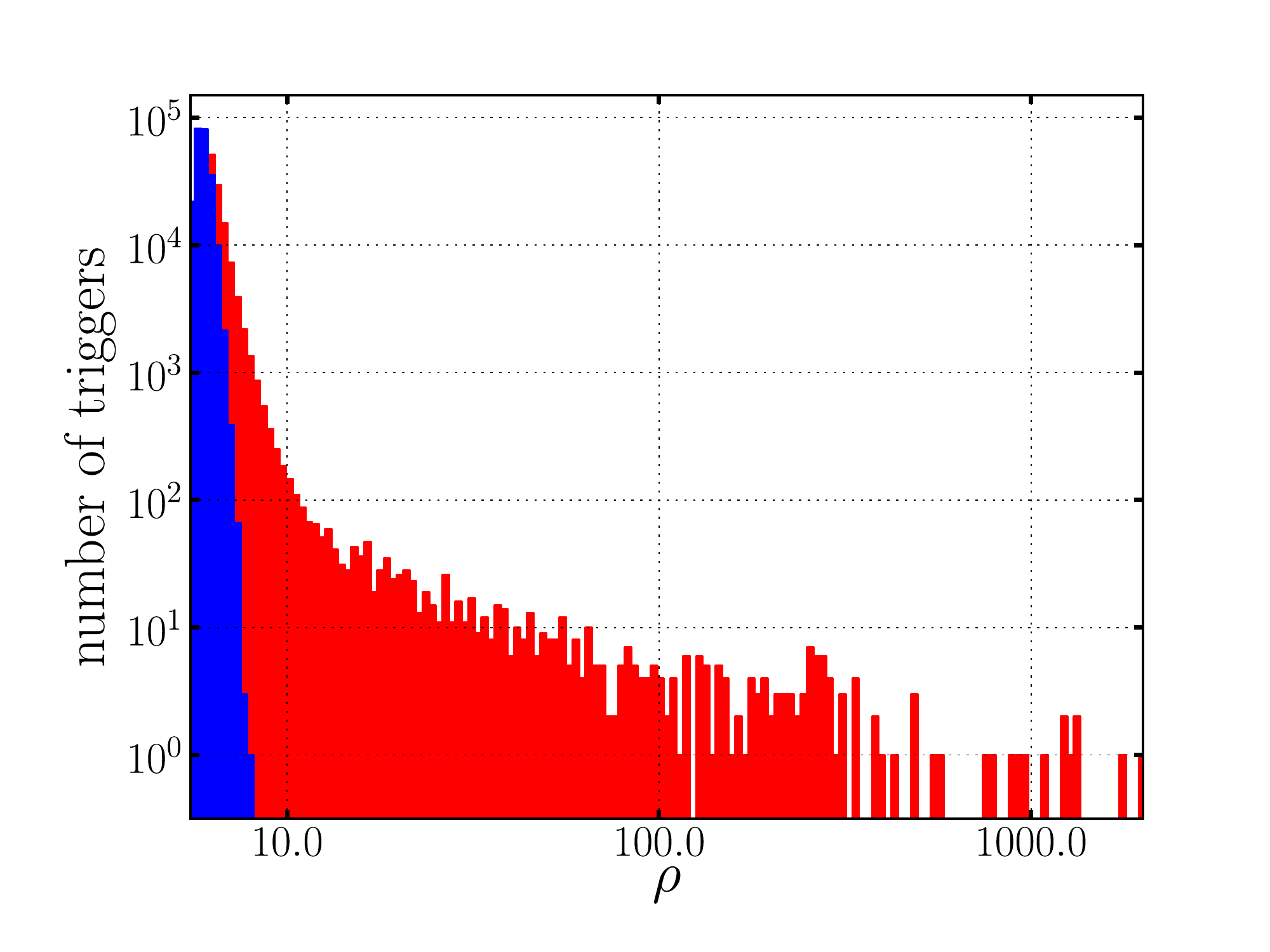}
\caption{\label{fig:firsttrigs}
Distribution of single detector trigger \acp{SNR} in a month of simulated Gaussian
noise (blue) and real \ac{S5} LIGO data (red) from the Hanford interferometer
H1.}
\end{figure}

For a single template and time and for detector data consisting of Gaussian
noise, $\rho^2$ follows a $\chi^2$ distribution with two degrees of freedom, which
makes a threshold of 5.5 seem rather large: $p(\rho > 5.5) = 2.7 \times
10^{-7}$. However, we must account for the fact that we consider a full template bank
and maximize over time of coalescence: the bank makes for, conservatively, a
thousand independent trials at any point in time, while trials separated by 0.1
seconds in time are essentially independent. Therefore, we expect to see a few
triggers above this threshold already in a few hundred seconds of Gaussian
noise, and a large number in a year of observing time. Furthermore, since the
data contain many non-Gaussian noise transients, the trigger rate will be even
higher.
In Fig.\ \ref{fig:firsttrigs} we show the distribution of triggers as a function
of SNR in a month of simulated Gaussian noise (blue) and real data (red) from LIGO's fifth science run (S5).
The difference between the two is clearly noticeable, with a tail of high
\ac{SNR} triggers extending to \acp{SNR} well over 1000 in real data.

It is useful to not just cluster in time, but also across the template bank.
When the \ac{SNR} for a template is above threshold, it is probable that it
will be above threshold also for many neighboring templates, which encode very
similar waveforms. The \ihope\ pipeline selects only one (or a few) triggers
for each event (be it a \ac{GW} or a noise transient), using one of two
algorithms.  In \emph{time-window} clustering, the time series of triggers from
all templates is split into windows of fixed duration; within each window, only
the trigger with the largest \ac{SNR} is kept. This method has the advantage of
simplicity, and it guarantees an upper limit on the trigger rate. However, a
glitch that creates triggers in one region of parameter space can mask a true
signal that creates triggers elsewhere.  This problem is remedied in
\emph{TrigScan} clustering \cite{SenguptaTrigScan:2008}, whereby triggers are
grouped by both time and recovered (template) masses, using the parameter-space
metric to define their proximity (for a detailed description see
\cite{Capano:2012}). However, when the data are particularly glitchy
\emph{TrigScan} can output a number of triggers that can overwhelm subsequent
data processing such as coincident trigger finding.

\subsection{Multi-detector coincidence}
\label{ssec:thinca}

The next stage of the pipeline compares the triggers generated for each of the
detectors, and retains only those that are seen in \emph{coincidence}. Loosely
speaking, triggers are considered coincident if they occurred at roughly the
same time, with similar masses; see Ref.\ \cite{Robinson:2008} for an exact
definition of coincidence as used in recent \ac{CBC} searches.  To wit, the
``distance'' between triggers is measured with the parameter-space metric of
Eq.\ \eqref{eq:cbc_metric}, maximized over the signal phase $\Phi_{0}$. Since
different detectors at different times have different noise PSDs and therefore
metrics, we construct a constant-metric-radius ellipsoid in
$\tau_{0}$--$\tau_{3}$--$t_{c}$ space, using the appropriate metric for every
trigger in every detector, and we deem pairs of triggers to be coincident if
their ellipsoids intersect. The radius of the ellipsoids is a tunable
parameter.  Computationally, the operation of finding all coincidences is
vastly sped up by noticing that only triggers that are close in time could
possibly have intersecting ellipsoids; therefore the triggers are first sorted
by time, and only those that share a small time window are compared.

When the detectors
are not co-located, the coincidence test must also take into account the light
travel time between detectors. This is done by computing the metric distance
while iteratively adding a small value, $\delta t_c$ to the end time of \textit{one} of the
detectors. $\delta t_c$ varies over the possible range of time delays due to light
travel time between the two detectors. The lowest value of the metric distance is
then used to determine if the triggers are coincident or not.

\begin{figure}
\includegraphics[width=\linewidth]{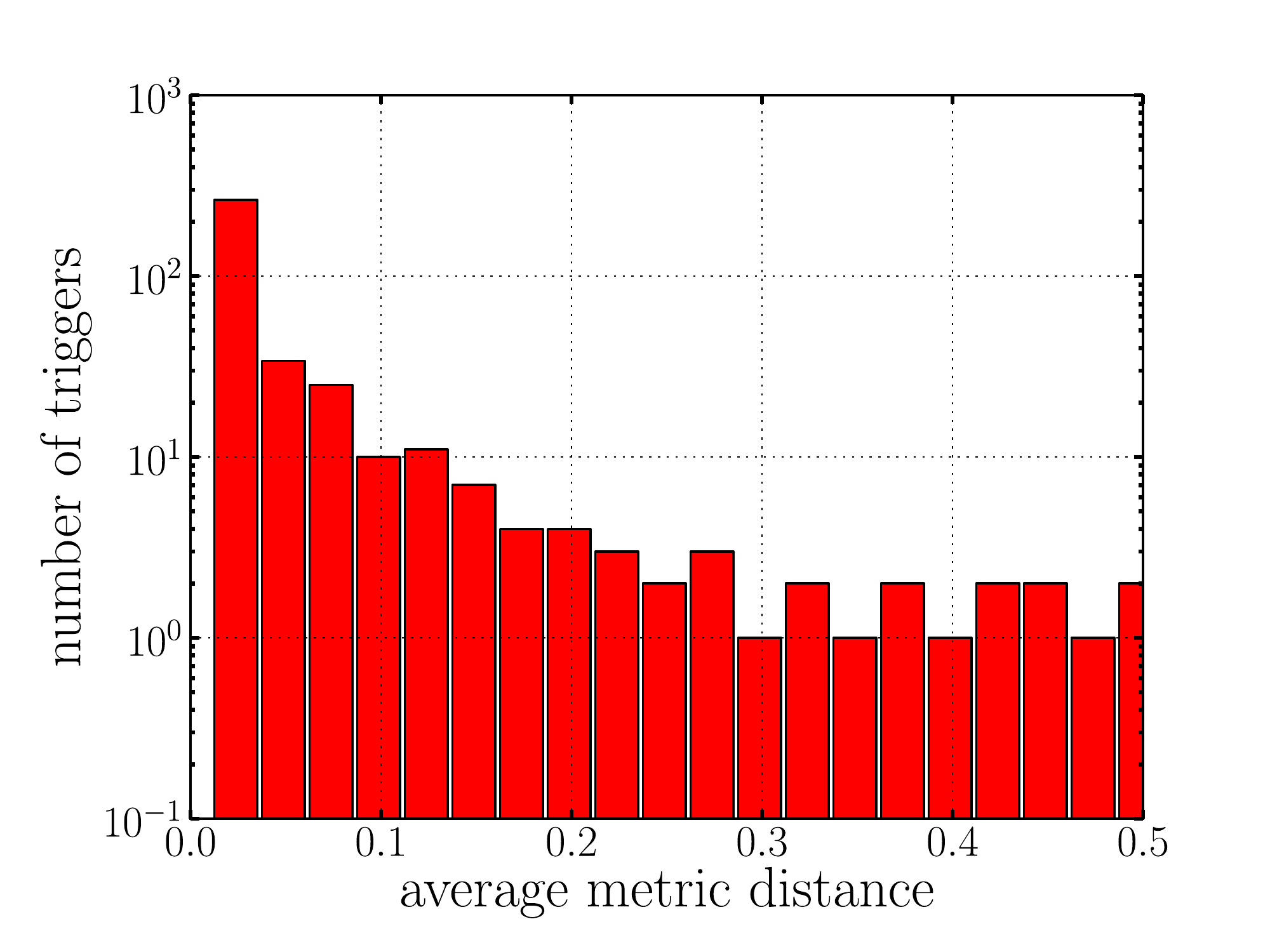}
\caption{\label{fig:ihope_ethinca}
Distribution of average parameter-space distance between coincident triggers
associated with simulated \ac{GW} signals in a month of representative S5 data,
as recovered by the LIGO H1 and L1 detectors.}
\end{figure}

In Fig.\ \ref{fig:ihope_ethinca} we show the distribution of metric distances
(the minimum value for which the ellipsoids centred on the triggers overlap)
for coincident triggers associated with simulated \ac{GW} signals (see
Sec \ref{ssec:injections}). The
number of coincidences falls off rapidly with increasing metric distances,
whereas it would remain approximately constant for \emph{background} coincident triggers
generated by noise.
However, it is the quieter triggers from farther \ac{GW} sources (which are
statistically more likely) that are recovered with the largest metric distances.
Therefore larger coincidence ellipsoids can improve the overall sensitivity of a
search.

The result of the coincidence process is a list of all triggers that have
\ac{SNR} above threshold in two or more detectors and consistent parameters
(masses and coalescence times) across detectors. When more than two detectors
are operational, different combinations and higher-multiplicity coincidences
are possible (e.g., three detectors yield triple coincidences and three types
of double coincidences).

\begin{figure}
\includegraphics[width=\linewidth]{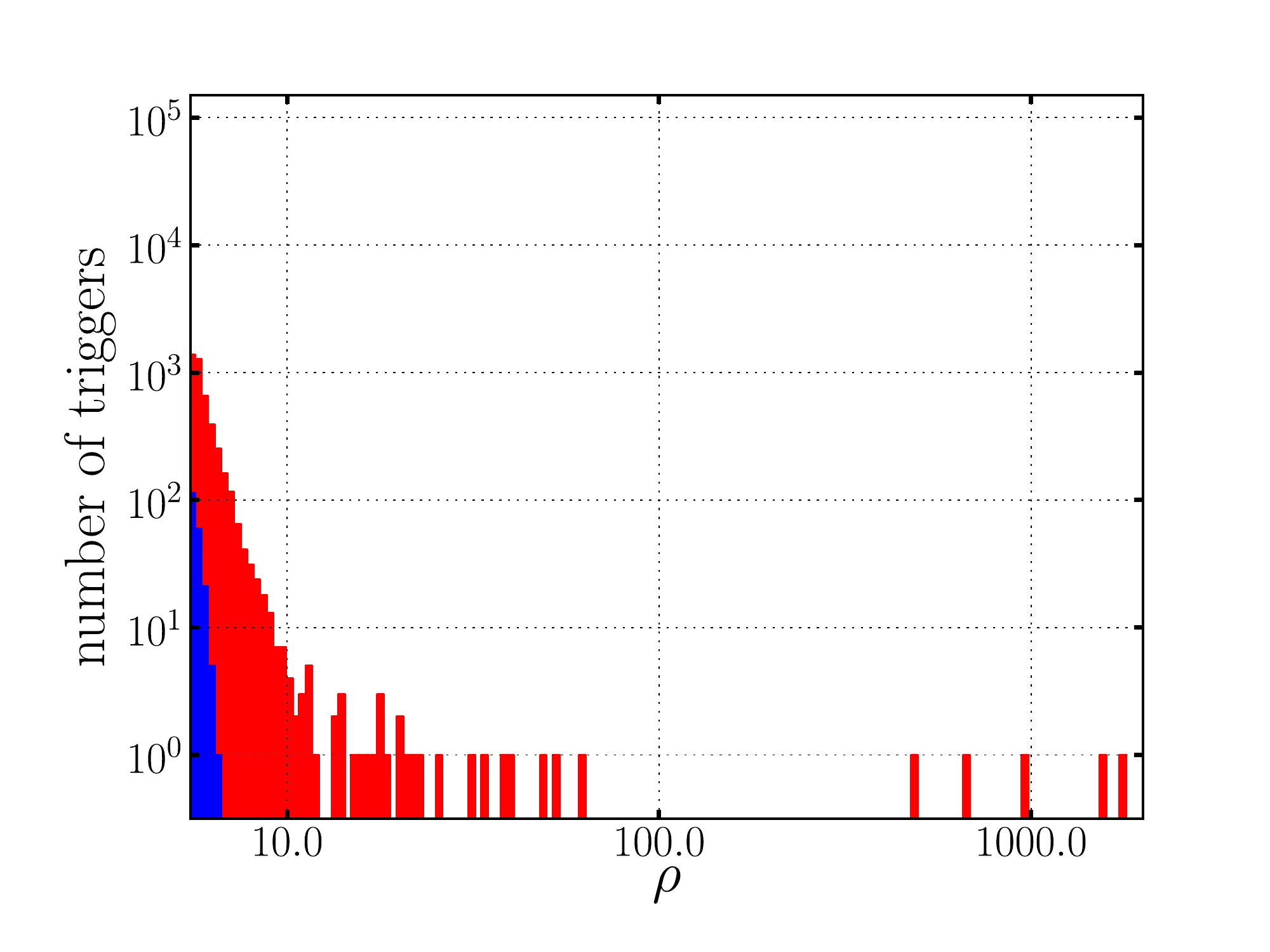}
\caption{\label{fig:coinctrigs}Distribution of single detector \acp{SNR} for H1
coincident triggers in a month of simulated Gaussian noise (blue) and
representative S5 data (red).
Coincidence was evaluated after time-shifting the SNR time series, so that only
background coincidences caused by noise would be included.
Comparison with Fig.\ \ref{fig:firsttrigs} shows that the coincidence
requirement reduces the high-\ac{SNR} tail, but by no means eliminates it.
} 
\end{figure}

In Fig.\ \ref{fig:coinctrigs} we show the distribution of coincident H1
triggers as a function of \ac{SNR} in a month of simulated Gaussian noise
(blue) and real \ac{S5} LIGO data (red).  The largest single-detector SNRs for
Gaussian noise are $\sim 7\mbox{--}8$, comparable (although somewhat larger)
with early theoretical expectations \cite{Schutz:1989cu, Cutler:1992tc}.
However, the distribution in real data is significantly worse, with SNRs of
hundreds and even thousands. If we were to end our analysis here, a \ac{GW}
search in real data would be a hundred times less sensitive (in distance) than
a search in Gaussian, stationary noise with the same PSD.

\section{IHOPE, part 2: mitigating the effects of non-Gaussian noise with
signal-consistency tests, vetoes, and ranking statistics}
\label{sec:nongauss}

To further reduce the tail of high-SNR triggers caused by the non-Gaussianity
and nonstationarity of noise, the \ihope\ pipeline includes a number of
\emph{signal-consistency} tests, which compare the properties of the data around
the time of a trigger with those expected for a real \ac{GW} signal. After removing
duplicates, the coincident triggers in each \(\unit{2048}{\second}\) block are used to create a
\emph{triggered template bank}. Any template in a given detector that forms
at least one coincident trigger in each \(\unit{2048}{\second}\) block will enter the triggered
template bank for that detector and chunk.
The new bank is again used to filter the data as described in Sec.\
\ref{ssec:inspiral}, but this time signal-consistency tests are also performed.
These include the $\chi^2$ (Sec.\ \ref{ssec:chisq}) and $r^2$ (Sec.\
\ref{ssec:rsq}) tests. Coincident triggers are selected as described in Sec.\
\ref{ssec:thinca}, and they are also tested for the consistency of relative
signal amplitudes (Sec.\ \ref{ssec:amplitude_consistency}); at this stage,
\emph{data-quality vetoes} are applied (Sec.\ \ref{ssec:vetotimes}) to sort
triggers into categories according to the 
quality of data at their times.

The computational cost of the entire pipeline is reduced greatly by applying the
expensive signal-consistency checks only in this second stage; the triggered
template bank is, on average, a factor of $\sim10$ smaller than the original template
bank in the analysis described in \cite{Abbott:2009qj}.
However, the drawback is
greater complexity of the analysis, and the fact that the coincident triggers
found at the end of the two stages may not be identical.

\subsection{The $\chi^{2}$ signal-consistency test}
\label{ssec:chisq}

The basis of the $\chi^{2}$ test \cite{Allen:2004} is the consideration that
although a detector glitch may generate triggers with the same \ac{SNR} as a \ac{GW}
signal, the manner in which the \ac{SNR} is accumulated over time and frequency
is likely to be different.
For example, a glitch that resembles a delta function corresponds to a burst of
signal power concentrated in a small time-domain window, but smeared out across
all frequencies. A \ac{CBC} waveform, on the other hand, will accumulate SNR across
the duration of the template, consistently with the \emph{chirp}-like morphology
of the waveform.

To test whether this is the case, the template is broken into $p$ orthogonal
subtemplates with support in adjacent frequency intervals, in such a way that
each subtemplate would generate the same SNR on average over Gaussian noise
realizations.
The actual SNR achieved by each subtemplate filtered against the data is compared to its
expected value, and the squared residuals are summed. Thus, the $\chi^2$ test
requires $p$ inverse Fourier transforms per template.
For the low-mass \ac{CBC} search, we found that setting $p =
16$ provides a powerful discriminator without incurring an excessive
computational cost \cite{Babak:2005iz}.

For a \ac{GW} signal that matches the template waveform exactly, the sum of
squared residuals follows the $\chi^{2}$ distribution with $2p - 2$ degrees of
freedom.  For a glitch, or a signal that does not match the template, the
expected value of the $\chi^{2}$-test is increased by a factor proportional to
the total $\mathrm{SNR}^2$, with a proportionality constant that depends on the
mismatch between the signal and the template. For signals, we may write the
expected $\chi^{2}$ value as
\begin{equation}
\langle \chi^{2} \rangle = (2p - 2) + \epsilon^{2} \rho^{2} \,,
\end{equation}
where $\epsilon$ is a measure of signal--template mismatch. Even if \ac{CBC} signals
do not match template waveforms perfectly, due to template-bank discreteness,
theoretical waveform inaccuracies \cite{BuonannoIyerOchsnerYiSathya2009}, 
spin effects \cite{VanDenBroeck:2009gd}, calibration uncertainties
\cite{S5calibration}, and so on, they will still yield significantly smaller
$\chi^2$ than most glitches. It was found empirically that a good fraction of glitches
are removed (with minimal effect on simulated signals) by imposing a
\ac{SNR}-dependent $\chi^{2}$ threshold of the form 
\begin{equation}\label{eq:chisq_thresh}
\chi^{2} \le \xi^{2} ( p + \delta \rho^{2}),
\end{equation}
with $\xi^2 = 10$ and $\delta = 0.2$.

\begin{figure}
\includegraphics[width=\linewidth]{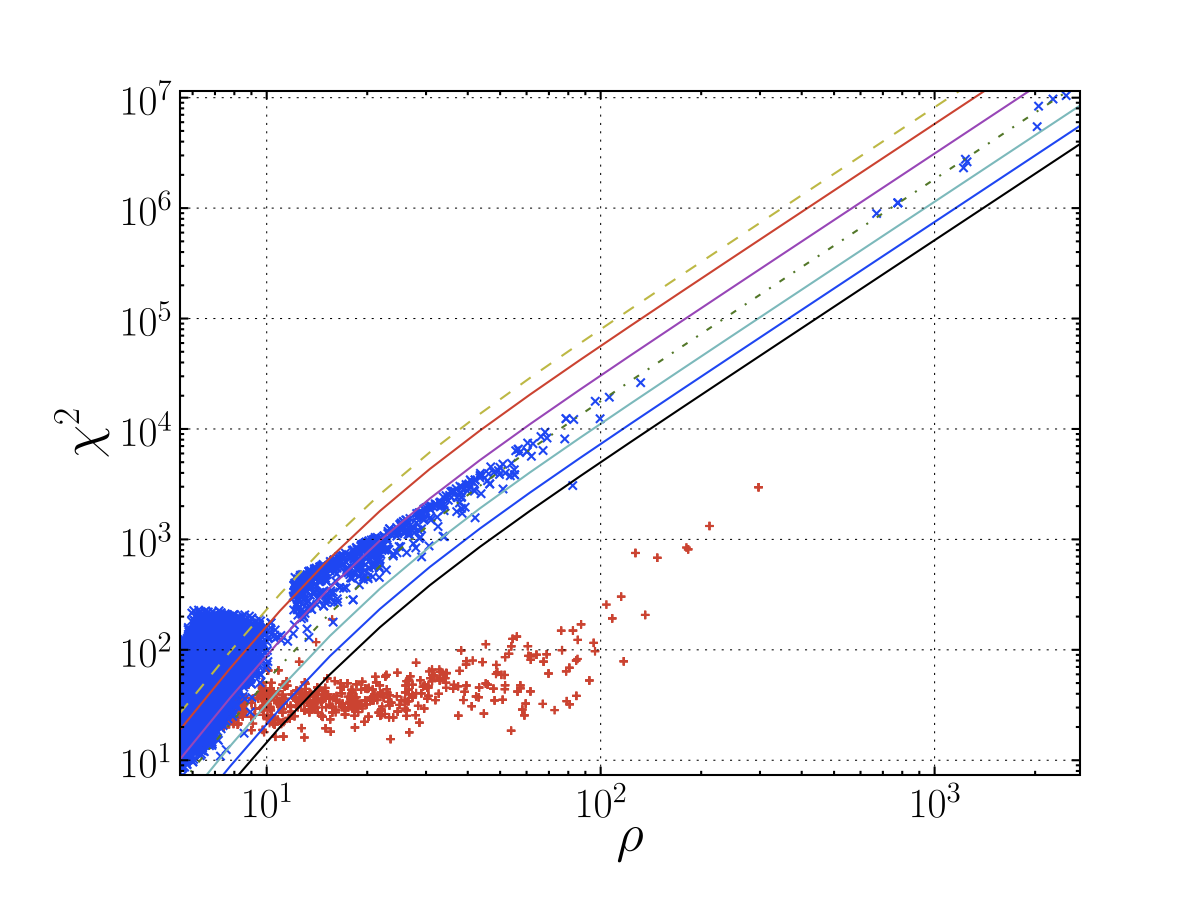}
\includegraphics[width=\linewidth]{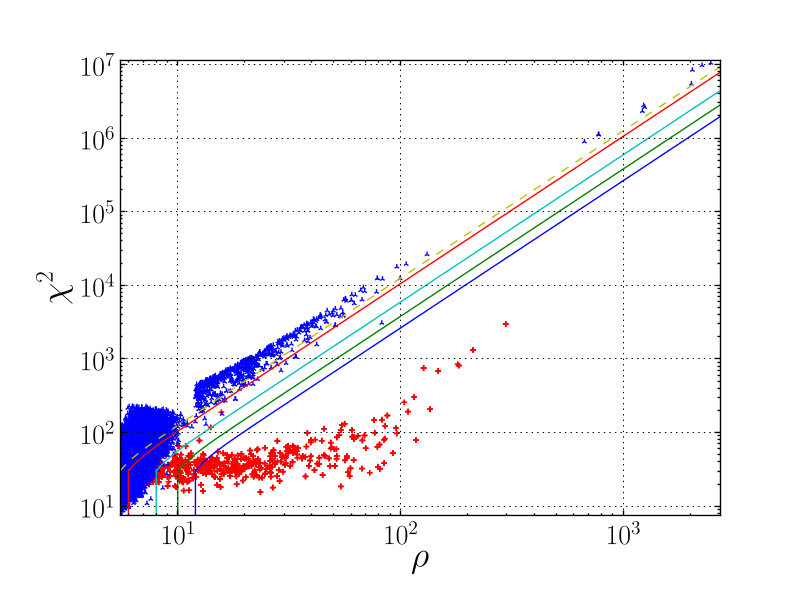}
\caption[The $\chi^2$ test plotted against
SNR.]{\label{fig:ihope_snr_chi} The $\chi^2$ test plotted against
\ac{SNR} for triggers in a month of representative S5 data
after the $\chi^2$ test has been applied, and the
$r^2$ cut has been applied for triggers with $\rho < 12$.
The blue crosses mark time shifted background triggers, the red pluses mark simulated-GW 
triggers.
The solid, colored lines on the plots indicate
lines of constant effective \ac{SNR} (top panel) and
new SNR (bottom panel), which are described in section \ref{ssec:effective_snr}.
Larger values of effective/new \ac{SNR} are at
the bottom and right end of the plots. The clearly visible notch in the
H1 and L1 plots is caused by the discontinuity in the $r^2$ cut at an
SNR of 12 (Section~\ref{ssec:rsq}). Here background triggers are
represented by blue crosses and injections by red pluses.
}
\end{figure}

In Fig.\ \ref{fig:ihope_snr_chi} we show the distribution of $\chi^2$ as a
function of \ac{SNR}.
A large number of triggers would have appeared in the upper left corner of the
plot (large $\chi^{2}$ value relative to the measured \ac{SNR}), but these have
been removed by the cut.  Even following the cut, a clear separation between
noise background and simulated signals can easily be observed.  This will be
used later in formulating a detection statistic that combines the values of
both $\rho$ and $\chi^{2}$.

\subsection{The $r^2$ signal-consistency test}
\label{ssec:rsq}

\begin{figure}
\includegraphics[width=\linewidth]{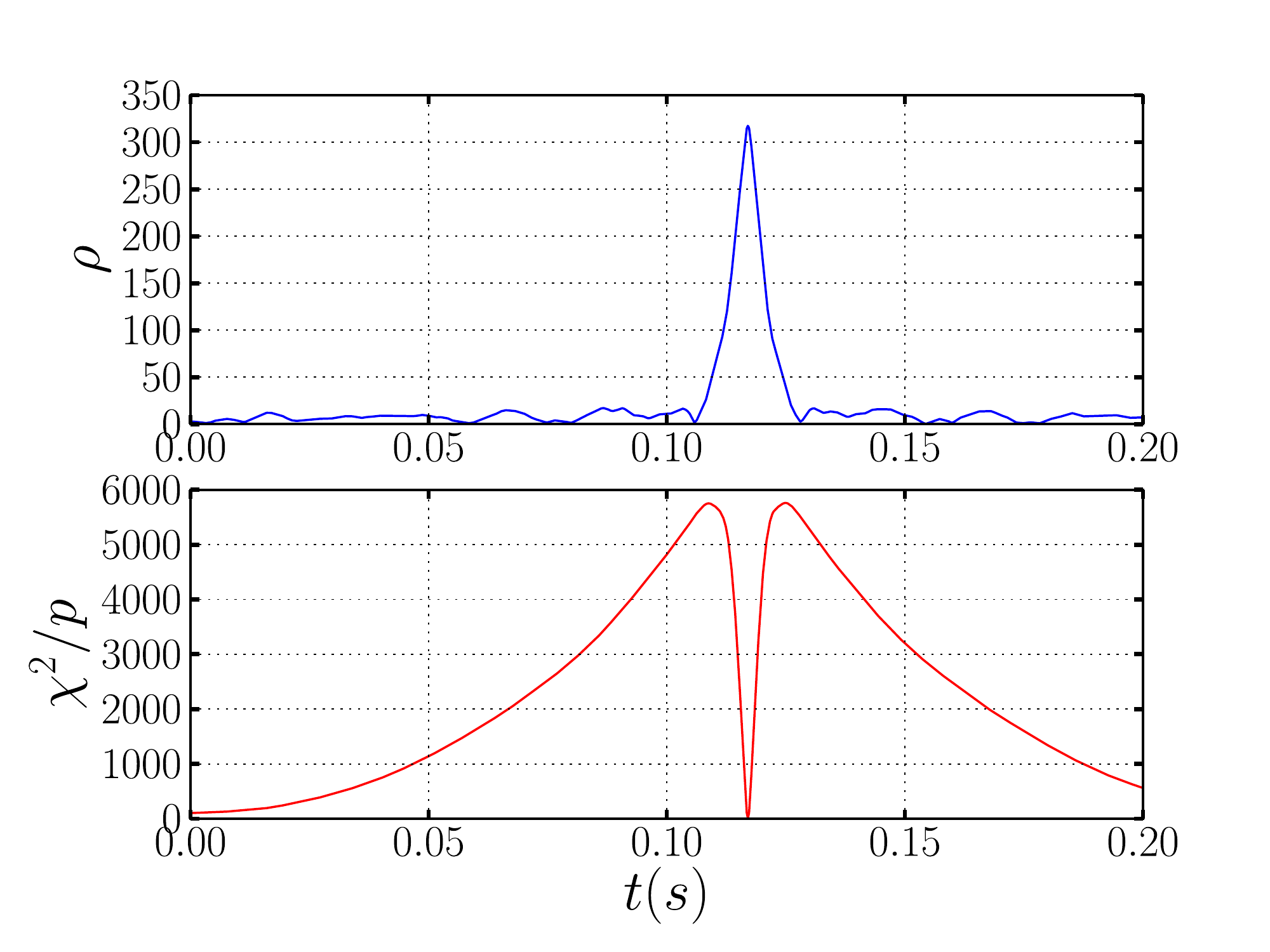}
\caption{\label{fig:ihope_c2_timeseries}
Value of \ac{SNR} and $\chi^2$ as a function of time, for a simulated
\ac{CBC} signal with SNR=300 in a stretch of S5 data from the H1
detector.  The \ac{SNR} shows a characteristic rise and fall around the
signal.  The $\chi^{2}$ value is small at the time of the signal, but
increases steeply to either side as the template waveform is offset from
the signal in the data.}
\end{figure}

We can also test the consistency of the data with a postulated signal by
examining the time series of \acp{SNR} and $\chi^2$s. For a true \ac{GW} signal, this would show a
single sharp peak at the time of the signal, with the width of the falloff
determined by the autocorrelation function of the template
\cite{Hanna:2008,HarryFairhurst:2011}.
Thus, counting the number of time samples around a trigger for which the
\ac{SNR} is above a set threshold provides a useful consistency test
\cite{Shawhan:2004cn}. Examining the behavior of the $\chi^{2}$ time series
provides a more powerful diagnostic \cite{Rodriguez:2007}.
To wit, the $r^{2}$ test sets an upper threshold on the amount of time $\Delta T$ (in a
window $T$ prior to the trigger\footnote{The nonsymmetric window was chosen
because the merger--ringdown phase of \ac{CBC} signals, which is not modeled in
inspiral-only searches, may cause an elevation in the $\chi^2$ time series after
the trigger.}) for which
\begin{equation}\label{eq:rsq_thresh}
  \chi^{2} \ge p \, r^{2},
\end{equation}
where $p$ is the number of subtemplates used to compute the $\chi^2$. We found
empirically that setting \(\unit{T = 6}{\second}\) and $r^2 = 15$ produces a powerful test
\cite{Rodriguez:2007}.
Figure \ref{fig:ihope_c2_timeseries} shows the characteristic shape of the
$\chi^{2}$ time series for \ac{CBC} signals: close to zero when the template is
aligned with the signal, then increasing as the two are offset in time, before
falling off again with larger time offsets.

An effective $\Delta T$ threshold must be a function of SNR; the $\Delta T$
commonly used for \ihope\ searches is
\begin{equation}
\label{eq:deltatthreshold}
\Delta T < \left\{
\begin{array}{ll}
\displaystyle 2 \times 10^{-4} \, \mathrm{s} & \text{for $\rho < 12$}, \\
\displaystyle \rho^{9/8} \times 7.5 \times 10^{-3} \, \mathrm{s}
& \text{for $\rho \geq 12$}.
\end{array} 
\right.
\end{equation}
The threshold for $\rho < 12$ eliminates triggers for which \textit{any} sample
is above the threshold from equation (\ref{eq:rsq_thresh}).

\begin{figure}
\includegraphics[width=\linewidth]{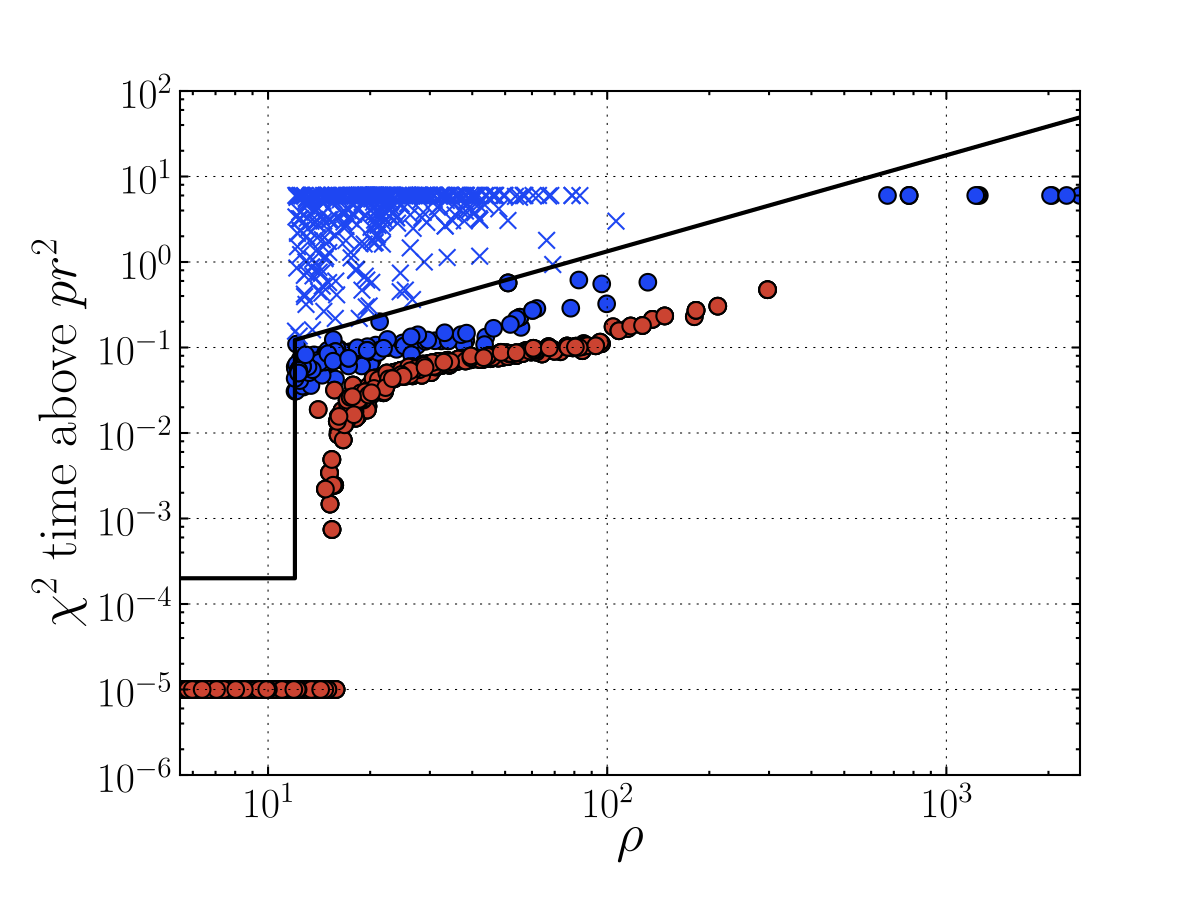}
\caption{\label{fig:ihope_rsq}
The $\chi^2$ time above $p r^2$ as a function of \ac{SNR}, for all
second-stage H1 triggers in a month of representative S5 data. The $r^{2}$
test has already been applied on triggers with $\rho < 12$, and only those
surviving the cut are shown.  The blue
crosses mark all background triggers (with $\rho
> 12$) that fail the cut; blue circles indicate background triggers that pass
it.  Red circles mark simulated-\ac{GW} triggers, none of which are cut.}
\end{figure}

In Fig.\ \ref{fig:ihope_rsq} we show the effect of such an SNR test. For $\rho <
12$, the value of $\Delta T$ is smaller than the sample rate, therefore
triggers are discarded if there are any time samples in the \(\unit{6}{\second}\) prior to
the trigger for which Eq.\ \eqref{eq:rsq_thresh} is satisfied.
(Since the \(\unit{6}{\second}\) window includes the trigger, for some \acp{SNR} this imposes a
more stringent requirement than the $\chi^2$ test \eqref{eq:chisq_thresh},
explaining the notch at $\rho < 12$ and relatively large $\chi^{2}$ values in
Fig.\ \ref{fig:ihope_snr_chi}.) For $\rho \geq 12$, the threshold is
\ac{SNR} dependent. The $r^2$ test is powerful at removing a large number of
high-\ac{SNR} background triggers (the blue crosses), without affecting the
triggers produced by simulated \ac{GW} signals (the red circles).
The cut is chosen to be conservative to allow for 
any imperfect matching between \ac{CBC} signals and template waveforms.

\subsection{Amplitude-consistency tests}
\label{ssec:amplitude_consistency}

\begin{figure}
\includegraphics[width=\linewidth]{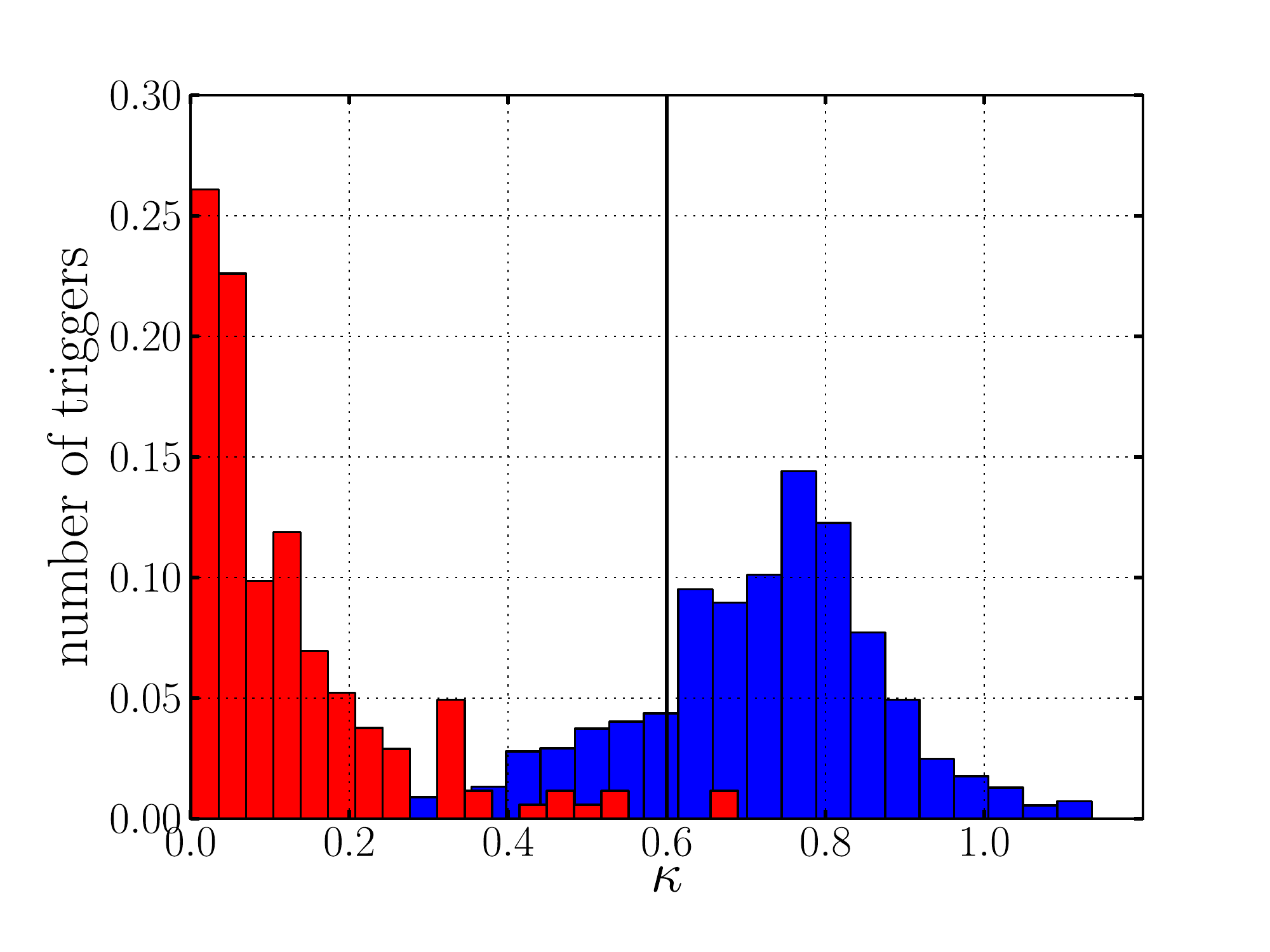}
\caption{\label{fig:ihope_effdistcut}
Distribution of $\kappa$ [Eq.\ \eqref{eq:fracdistcut}], the fractional
difference in the effective distances measured by H1 and H2 for coincident
triggers in those detectors in a month of representative S5 data. Background
triggers (blue) tend to have larger $\kappa$ than simulated-\ac{GW} triggers (red).
}
\end{figure}

The two LIGO Hanford detectors H1 and H2 share the same vacuum tubes,
and therefore expose the same sensitive axes to any incoming \ac{GW}. Thus, the ratio of
the H1 and H2 SNRs for true \ac{GW} signals should equal the ratio of detector
sensitivities. We can formulate a formal test of H1--H2 \emph{amplitude
consistency}\footnote{The detector H2 was not operational during LIGO run S6, so
the H1--H2 amplitude-consistency tests were not applied; they were however used
in searches over data from previous runs.} in terms of a \ac{GW} source's \emph{effective
distance} $D_{\mathrm{eff},A}$---the distance at which an optimally located and
oriented source would give the SNR observed with detector $A$. Namely, we
require that
\begin{equation}
\label{eq:fracdistcut}
\kappa = 2 \frac{\left| D_\mathrm{eff,H1} - D_\mathrm{eff,H2}
\right|}{D_\mathrm{eff,H1} + D_\mathrm{eff,H2}} \leq \kappa^*;
\end{equation}
setting a threshold $\kappa^*$ provides discrimination against noise triggers
while allowing for some measurement uncertainty. In Fig.\
\ref{fig:ihope_effdistcut} we show the distribution of $\kappa$ for simulated-\ac{GW}
triggers and background triggers in a month of representative S5
data.  We found
empirically that setting $\kappa^{*} = 0.6$ produces a powerful test.

An amplitude-consistency test can be defined also for triggers that are seen
in only one of H1 and H2. We do this by removing any triggers from H1 which
are loud enough that we would have expected to observe a trigger in H2 (and
vice-versa).  We proceed by calculating $\sigma_{A}$, the distance at which an
optimally located and oriented source yields an SNR of 1 in detector $A$, and
noting that $D_{\mathrm{eff},A} = \sigma_A / \rho_A$.  Then, by rearranging
(\ref{eq:fracdistcut}), we are led to require that  a trigger that is seen
only in H1 satisfy
\begin{equation}
\rho_\mathrm{H1} < \frac{\sigma_\mathrm{H1}}{\sigma_\mathrm{H2}}
\left( \frac{2 + \kappa^*}{2 - \kappa^*} \right) \rho^*_\mathrm{H2},
\end{equation}
where $\rho^*_\mathrm{H2}$ is the SNR threshold used for H2.  The
effective distance cut removes essentially all H2 triggers for which
there is no H1 coincidence: since H2 typically had around half the distance
sensitivity of H1, a value of $\kappa^* = 0.6$ imposes $\rho_{\mathrm{H2}} <
\rho_{\mathrm{H1}}^*$.

Neither test was used between any other pair of detectors because, in principle,
any ratio of effective distances is possible for a real signal seen in two
nonaligned detectors. However, large values of $\kappa$ are rather unlikely, especially
for the Hanford and Livingston LIGO detectors, which are \emph{almost} aligned.
Therefore amplitude-consistency tests should still be applicable.

\subsection{Data-quality vetoes}
\label{ssec:vetotimes}

Environmental factors can cause periods of elevated detector glitch rate. In the
very worst (but very rare) cases, this makes the data essentially unusable.
More commonly, if these glitchy periods were analyzed together with periods of
relatively clean data, they could produce a large number of high-SNR triggers,
and possibly mask \ac{GW} candidates in clean data. It is therefore necessary to
remove or separate the glitchy periods.

This is accomplished using \emph{data quality} (DQ) flags
\cite{Slutsky:2010ff,Christensen:2010,Aasi:2012wd}. All detectors are equipped with
environmental and instrumental monitors; their output is recorded in the
detector's \emph{auxiliary} channels. Periods of heightened activity in these
channels (e.g., as caused by elevated seismic noise \cite{SeisVeto}) are
automatically marked with DQ flags \cite{glitchmon}. DQ flags can also be added
manually if the detector operators observe poor instrumental behavior.

If a DQ flag is found to be strongly correlated with \ac{CBC} triggers, and if
the flag is \emph{safe} (i.e., not triggered by real \acp{GW}), then it can be
used a DQ \emph{veto}. Veto safety is assessed by comparing the fraction of
hardware \ac{GW} injections that are vetoed with the total fraction of data that
is vetoed.  During the S6 and VSR2-3 runs, a simplified form of \ihope~was run
daily on the preceding 24 hours of data from each detector individually,
specifically looking for non-Gaussian features that could be correlated with
instrumental or environmental effects \cite{Pekowsky:2012,SeisVeto}.  
The results of these daily runs were used to help identify common
glitch mechanisms and to mitigate the effects of non-Gaussian noise
by suggesting data quality vetoes.

 Vetoes are assigned to categories based on the severity of
instrumental problems and on how well the couplings between the \ac{GW} and
auxiliary channels are understood \cite{Slutsky:2010ff,
Christensen:2010,Aasi:2012wd}. Correspondingly, \ac{CBC} searches assign data to
four DQ categories:
\begin{description} 
\item[Category 1] Seriously compromised or missing data. The data are entirely
unusable, to the extent that they would corrupt noise PSD estimates. These
times are excluded from the analysis, as if the detector was not in science
mode (introduced in Sec.\ \ref{ssec:segment_psd}).
\item[Category 2] Instrumental problems with known couplings to the \ac{GW}
channel.  Although the data are compromised, these times can still be used for
PSD estimation. Data flagged as category-2 are analyzed in the pipeline, but any triggers
occurring during these times are discarded. This reduces the fragmentation of
science segments, maximizing the amount of data that can be analyzed.
\item[Category 3] Likely instrumental problems, casting doubt on triggers
found during these times. Data flagged as category-3 are analyzed and triggers are
processed. However, the excess noise in such times may
obscure signals in clean data.  Consequently, the analysis is also performed
excluding time flagged as category-3, allowing weaker signals in clean data to be
extracted. These data are excluded from the estimation of upper limits on
\ac{GW}-event rates.
\item[Good data] Data without any active environmental or instrumental source
of noise transients. These data are analyzed in full.  
\end{description}

Poor quality data are effectively removed from the analysis, reducing the total
amount of analyzed time. For instance, in the third month of the S5 analysis
reported in Ref.\ \cite{Abbott:2009qj}, removing category-1 times left
\(\unit{1.2 \times 10^6}{\second}\) of data when at least two detectors were
operational; removing category-2 and -3 times left \(\unit{1.0 \times
10^6}{\second}\), although the majority of lost time was category-3, and was
therefore analyzed for loud signals.

\subsection{Ranking statistics}
\label{ssec:effective_snr}

\begin{figure}
\includegraphics[width=\linewidth]{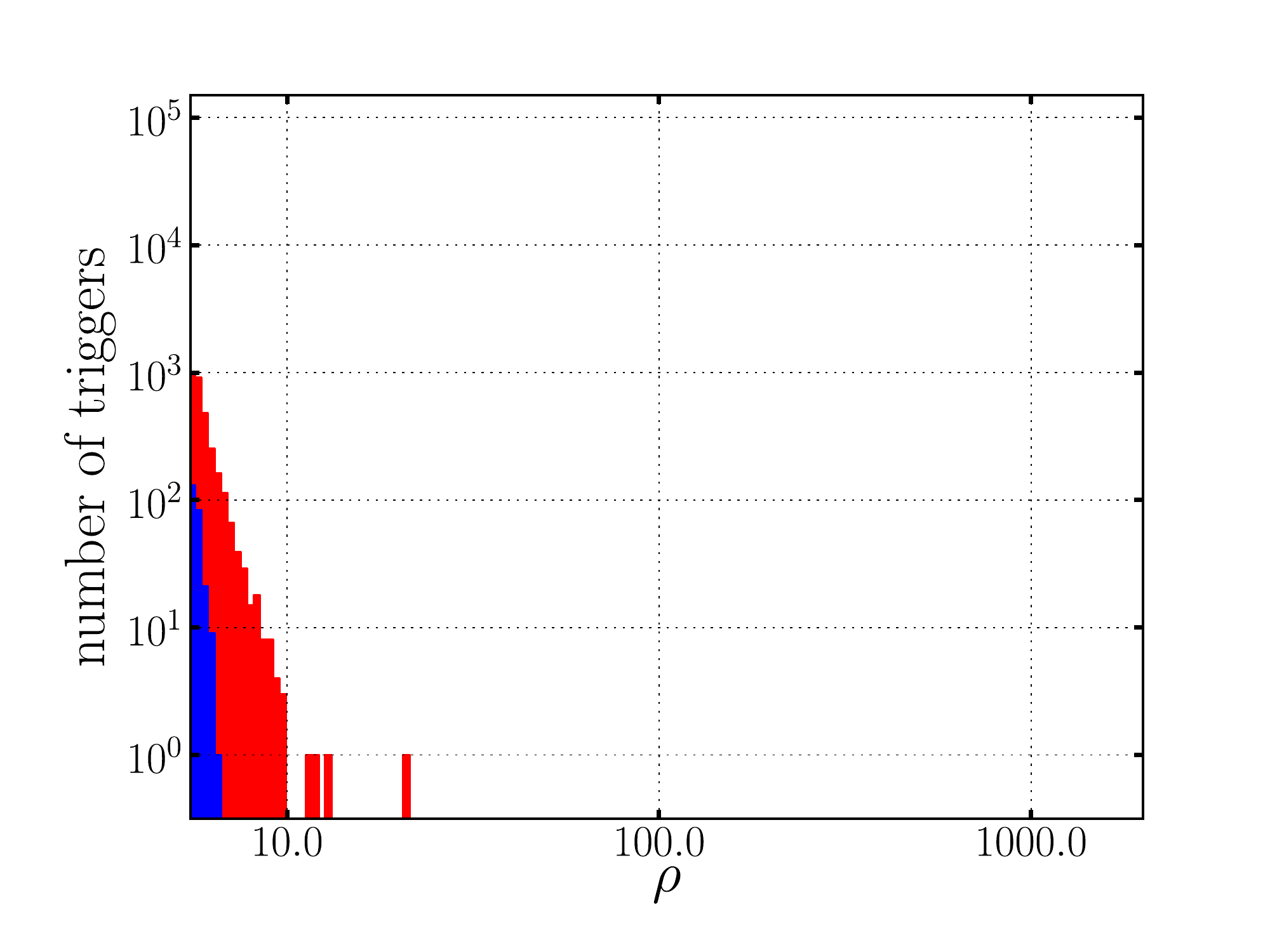}
\caption{\label{fig:finaltrigs}
Distribution of single detector \acp{SNR} for H1 triggers found in coincidence
with L1 triggers (in time shifts) in a month of simulated Gaussian noise
(blue) and representative S5 data (red). These triggers have survived
$\chi^{2}$, $r^{2}$, and H1--H2 amplitude-consistency tests, as well as DQ
vetoes.}
\end{figure}

The application of signal-consistency and amplitude-consistency tests, as well
as data-quality vetoes, is very effective in reducing the non-Gaussian tail of
high-\ac{SNR} triggers. In Fig.\ \ref{fig:finaltrigs} we show the distribution
of H1 triggers that are coincident with triggers in the L1 detector (in time
shifts) and that pass all cuts.  For consistency, identical cuts have been
applied to the simulated, Gaussian data, including vetoing times of poor data
quality in the real data.  The majority of these have minimal impact, although
the data quality vetoes will remove a (random) fraction of the triggers arising
in the simulated data analysis.

Remarkably, in the real data, almost no triggers are left that have
$\mathrm{SNR} > 10$. Nevertheless, a small number of coincident noise triggers
with large \ac{SNR} remain. These triggers have passed all cuts, but they
generally have significantly worse $\chi^2$ values than expected for true
signals, as we showed in Fig.\ \ref{fig:ihope_snr_chi}.

It is therefore useful to rank triggers using a \emph{combination} of \ac{SNR}
and $\chi^2$, by introducing a \textit{re-weighted SNR}.  Over the course of
the LIGO-Virgo analyses, several distinct re-weighted \acp{SNR} have been used.
For the LIGO \ac{S5} run and \ac{VSR1}, we adopted the \emph{effective SNR}
$\rho_\mathrm{eff}$, defined as \cite{Collaboration:2009tt}
\begin{equation}
\rho^2_{\rm eff} =
\frac{\rho^2}{\sqrt{\left(\frac{\chi^2}{n_{\mathrm{dof}}}\right)\left(1 +
\frac{\rho^2}{250}\right)}} \, ,
\end{equation}
where $n_\mathrm{dof} = 2p - 2$ is the number of $\chi^2$
degrees of freedom, and the factor $250$ was tuned empirically to provide
separation between background triggers and simulated \ac{GW} signals. The
normalization of $\rho_\mathrm{eff}$ ensures that a ``quiet'' signal with $\rho
\simeq 8$ and $\chi^{2}$ $\simeq n_\mathrm{dof}$ will have $\rho_\mathrm{eff}
\simeq \rho$.

Figure \ref{fig:ihope_snr_chi} shows contours of constant $\rho_\mathrm{eff}$
in the $\rho$--$\chi^2$ plane.
While $\rho_\mathrm{eff}$ successfully separates
background triggers from simulated-\ac{GW} triggers, it can artificially elevate
the \ac{SNR} of triggers with unusually small $\chi^{2}$. As discussed in
Ref.\ \cite{Collaboration:S5HighMass}, these can sometimes become the most
significant triggers in a search. Thus, a different statistic was adopted for
the LIGO S6 run and \ac{VSR23}. This \emph{new SNR} $\rho_\mathrm{new}$
\cite{Collaboration:S6CBClowmass} was defined as
\begin{equation}
\rho_{\mathrm{new}} = \left\{
\begin{array}{cl}
\rho & \text{for $\chi^2 \le n_{\mathrm{dof}}$}, \\
\rho
\left[\frac{1}{2} \left(1 +
\left(\frac{\chi^2}{n_{\mathrm{dof}}}\right)^{\!3}\right)\right]^{-1/6}
& \text{for $\chi^2 > n_{\mathrm{dof}}$}.
\end{array} 
\right.
\end{equation}
Figure \ref{fig:ihope_snr_chi} also shows contours of constant $\rho_\mathrm{new}$
in the $\rho$--$\chi^2$ plane.
The new SNR was found to provide even better background--signal separation,
especially for low-mass nonspinning inspirals \cite{Collaboration:S6CBClowmass},
and it has the desirable feature that $\rho_\mathrm{new}$ does not take larger
values than $\rho$ when the $\chi^2$ is less than the expected value.
Other ways of defining a 
detection statistic as a function of $\rho$ and $\chi^2$ can be defined
and optimized for analyses covering different regions of parameter space
and different data sets.

For coincident triggers, the re-weighted SNRs measured in the coincident
detectors are added in quadrature to give a \emph{combined}, re-weighted
SNR, which is used to rank the triggers and evaluate their statistical
significance.  Using this ranking statistic, we find that the
distribution of background triggers in real data is remarkably close to
their distribution in simulated Gaussian noise. Thus, our consistency
tests and DQ vetoes have successfully eliminated the vast majority of
high SNR triggers due to non-Gaussian noise from the search. While this
comes at the inevitable cost of missing potential detections at times of
poor data quality, it significantly improves the detection capability of
a search.

\begin{figure}
\includegraphics[width=\linewidth]{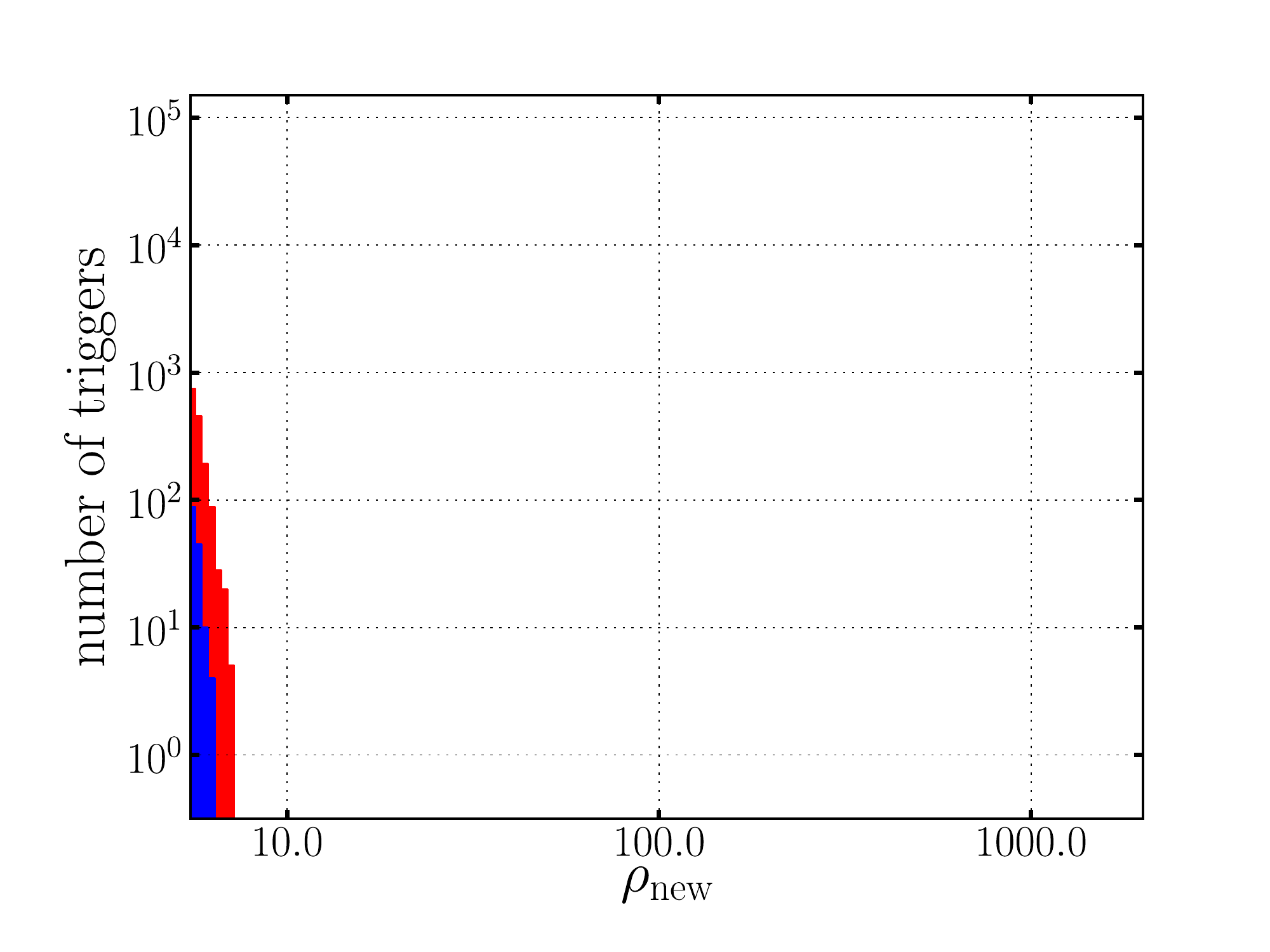}
\caption{\label{fig:new_snr_dist}
Distribution of single detector new SNR, $\rho_\mathrm{new}$, for H1 triggers found
in coincidence with L1 triggers (in time shifts) in a month of simulated
Gaussian noise (blue) and representative S5 data (red).  The tail of
high \ac{SNR} triggers due to non-Gaussian noise has been virtually
eliminated---a remarkable achievement given that the first stage of the
pipeline generated single-detector triggers with $\mathrm{SNR} > 1,000$.
}
\end{figure}

\section{Interpretation of the Results}
\label{sec:interpretation}

At the end of the data processing described above, the \ihope\ pipeline produces a set of coincident triggers
ranked by their combined re-weighted \ac{SNR}; these triggers have passed the various
signal-consistency and data-quality tests outlined above.  While at this stage
the majority of loud background triggers identified in real data have been eliminated or
downweighted, the distribution of triggers is still different from the case of
Gaussian noise, and it depends on the quality of the detector data
and the signal parameter space being searched over.  Therefore
it is not possible to derive an analytical mapping from combined re-weighted \ac{SNR}
to event significance, as characterized by the \ac{FAR}.  Instead, the \ac{FAR} is
evaluated empirically by performing numerous \emph{time-shift} analyses, in
which artificial time shifts are introduced between the data from different
detectors.  (These are discussed in Sec.\ \ref{ssec:background}.) Furthermore,
the rate of triggers as a function of combined re-weighted \ac{SNR} varies over parameter
space; to improve the \ac{FAR} accuracy, we divide triggers into groups with
similar combined re-weighted \ac{SNR} distributions (see Sec.\ \ref{ssec:far}).  The
sensitivity of a search is evaluated by measuring the rate of recovery of a
large number of simulated signals, with parameters drawn from astrophysically
motivated distributions (see Sec.\ \ref{ssec:injections}).  The sensitivity is
then used to estimate the \ac{CBC} event rates or upper limits
as a function of signal parameters (see Sec.~\ref{ssec:ul}).

\subsection{Background event rate from time shifts}
\label{ssec:background}

The rate of coincident triggers as a function of combined re-weighted \ac{SNR}
is estimated by performing numerous time-shift analyses: in each we
artificially introduce different relative time shifts in the data from each
detector \cite{Amaldi:1989}.  The time shifts that are introduced must be large
enough such that each time-shift analysis is statistically independent.

To perform the time-shift analysis in practice, we simply shift the triggers
generated at the first matched-filtering stage of the analysis
(\ref{ssec:inspiral}), and repeat all subsequent stages from multi-detector
coincidence (\ref{ssec:thinca}) onwards.  Shifts are performed on a ring: for
each time-coincidence period (i.e., data segment where a certain set of
detectors is operational), triggers that are shifted past the end are
re-inserted at the beginning.  Since the time-coincidence periods are
determined \emph{before} applying Category-2 and -3 DQ flags, there is some
variation in analyzed time among time-shift analyses. To ensure statistical
independence, time shifts are performed in multiples of \(\unit{5}{\second}\);
this ensures that they are significantly larger than the light travel time
between the detectors, the autocorrelation time of the templates, and the
duration of most non-transient glitches seen in the data.  Therefore, any
coincidences seen in the time shifts cannot be due to a single \ac{GW} source,
and are most likely due to noise-background triggers.  It is possible, however,
for a \ac{GW}-induced trigger in one detector to arise in time-shift
coincidence with noise in another detector.  Indeed, this issue arose in
Ref.~\cite{Collaboration:S6CBClowmass}, where a ``blind injection'' was added
to the data to test the analysis procedure.

The H1 and H2 detectors share
the Hanford beam tubes and are affected by the same environmental
disturbances; furthermore, noise transients in the two detectors have
been observed to be correlated. Thus, time-shift analysis is
ineffective at estimating the coincident background between these co-located detectors,
and it is not used.  
Coincident triggers from H1 and H2 when no other detectors are operational
are excluded from the analysis.
When detectors at additional sites are operational,
we do perform time shifts, keeping H1 and H2 ``in time'' but
shifting both relative to the other detectors.

Our normal practice is to begin by performing 100 time-shift analyses to
provide an estimate of the noise background. If any coincident in-time triggers
are still more significant (i.e., have larger combined re-weighted \ac{SNR})
than all the time-shifted triggers, additional time shifts are performed to
provide an estimate of the \ac{FAR}.  A very significant candidate would have a
very low FAR, and an accurate determination of its FAR requires a large number of
time slides: in Ref.~\cite{Collaboration:S6CBClowmass} over a million were
performed. However, there is a limit to the number of statistically independent
time shifts that are possible to perform, as explored in \cite{Was:2009vh}.
Additionally, as the number of time shifts grows, the computational savings of
our two-stage search are diminished, because a greater fraction of the
templates survive to the second filtering stage where the computationally
costly signal-consistency tests are performed (see Sec.\ \ref{ssec:chisq}).  We
are currently investigating whether it is computationally feasible to run
\ihope\ as a single-stage pipeline and compute $\chi^2$ and $r^2$ for every
trigger.

\subsection{Calculation of false-alarm rates}
\label{ssec:far}

The \ac{FAR} for a coincident trigger is given by the rate at
which background triggers with the same or greater SNR occur due to detector noise.
This rate is computed from the time-shift analyses; for a fixed combined re-weighted
\ac{SNR}, it varies across the template mass space, and it depends on which detectors
were operational and how glitchy they were.  To accurately account for
this, coincident triggers are split into \emph{categories},
and \acp{FAR} are calculated within each, relative to a background of comparable triggers.
The triggers from each category are then re-combined into a single
list and ranked by their \acp{FAR}.
\begin{figure}
\includegraphics[width=\linewidth]{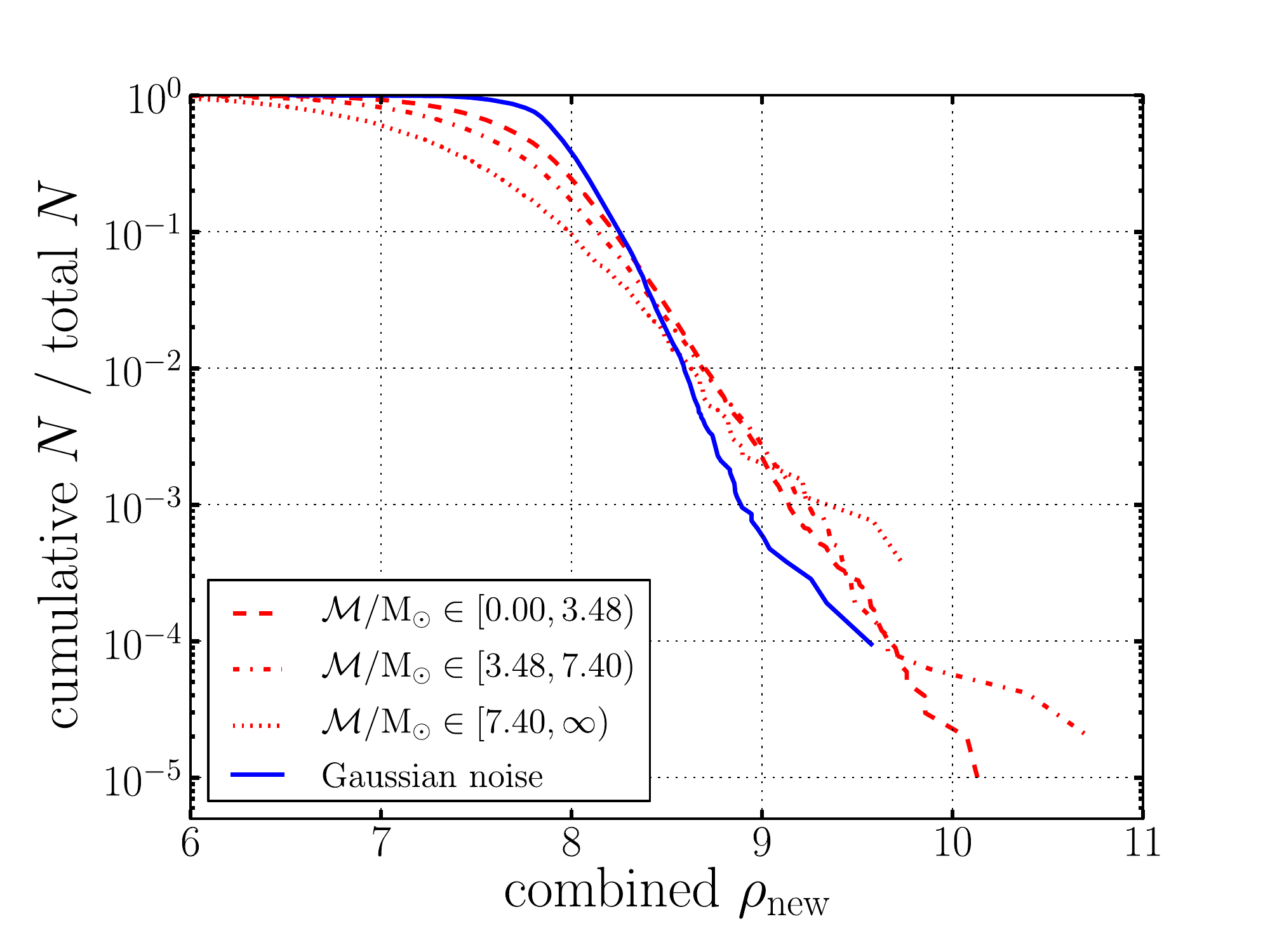}
\caption{\label{fig:mchirp_bins}
Fraction of time-shift coincident triggers between H1 and L1 in a month of
representative S5 data that have
combined new \ac{SNR} greater than or equal to the x-axis value, for three
chirp-mass bins.  The distribution from a month of Gaussian noise is also
shown for comparison.  The tails of the
distributions become more shallow for larger chirp masses $\mathcal{M}$, so
triggers with higher $\mathcal{M}$ are more likely to have higher SNRs.}

\end{figure}

Typically, signal-consistency tests are more powerful for longer-duration
templates than for shorter ones, so the non-Gaussian background is
suppressed better for low-mass templates, while high-mass templates are more
likely to result in triggers with larger combined re-weighted \acp{SNR}.
In recent searches, triggers have been
separated into three bins in \emph{chirp mass} $\mathcal{M}$ \cite{findchirp}:
$\mathcal{M} \le 3.48 \, M_{\odot}$, $3.48
\, M_{\odot} < \mathcal{M} \le 7.4 \, M_{\odot}$, and $\mathcal{M} > 7.4 \, M_{\odot}$.
Figure \ref{fig:mchirp_bins} shows the distribution of
coincident triggers between H1 and L1 as a function of combined
$\rho_\mathrm{new}$ for the triggers in each of these mass bins. As expected,
the high-$\mathcal{M}$ bin has a greater fraction of high-\ac{SNR} triggers.

The combined re-weighted \ac{SNR} is calculated as the quadrature sum of the
\acp{SNR} in the individual detectors.  However, different detectors can have
different rates of non-stationary transients as well as different
sensitivities, so the combined \ac{SNR} is not necessarily the best measure of
the significance of a trigger.  Additionally, background triggers found in
three-detector coincidence will have a different distribution of combined
re-weighted \acp{SNR} than two-detector coincident triggers
\cite{Collaboration:2009tt}.  Therefore, we separate coincident triggers by
their \emph{type}, which is determined by the coincidence itself (e.g., H1H2,
or H1H2L1) and by the availability of data from each detector, known as
``coincident time.'' Thus, the trigger types would include H1L1 coincidences in
H1L1 double-coincident time; H1L1, H1V1, L1V1, and H1L1V1 coincidences in
H1L1V1 triple-coincident time; and so on.  When H1 and H2 are both operational,
we have fewer coincidence types than might be expected as H1H2 triggers are
excluded due to our inability to estimate their background distribution, and
the effective distance cut removes H2L1 or H2V1 coincidences.  The product of
mass bins and trigger types yields all the trigger categories.

For simplicity, we treat times when
different networks of detectors were operational as entirely separate
experiments; this is straightforward to do, as there is no overlap in
time between them.  Furthermore, the data from a long science run is
typically broken down into a number of distinct stretches, often based
upon varying detector sensitivity or glitchiness, and each is handled
independently.

\begin{figure}
\includegraphics[width=\linewidth]{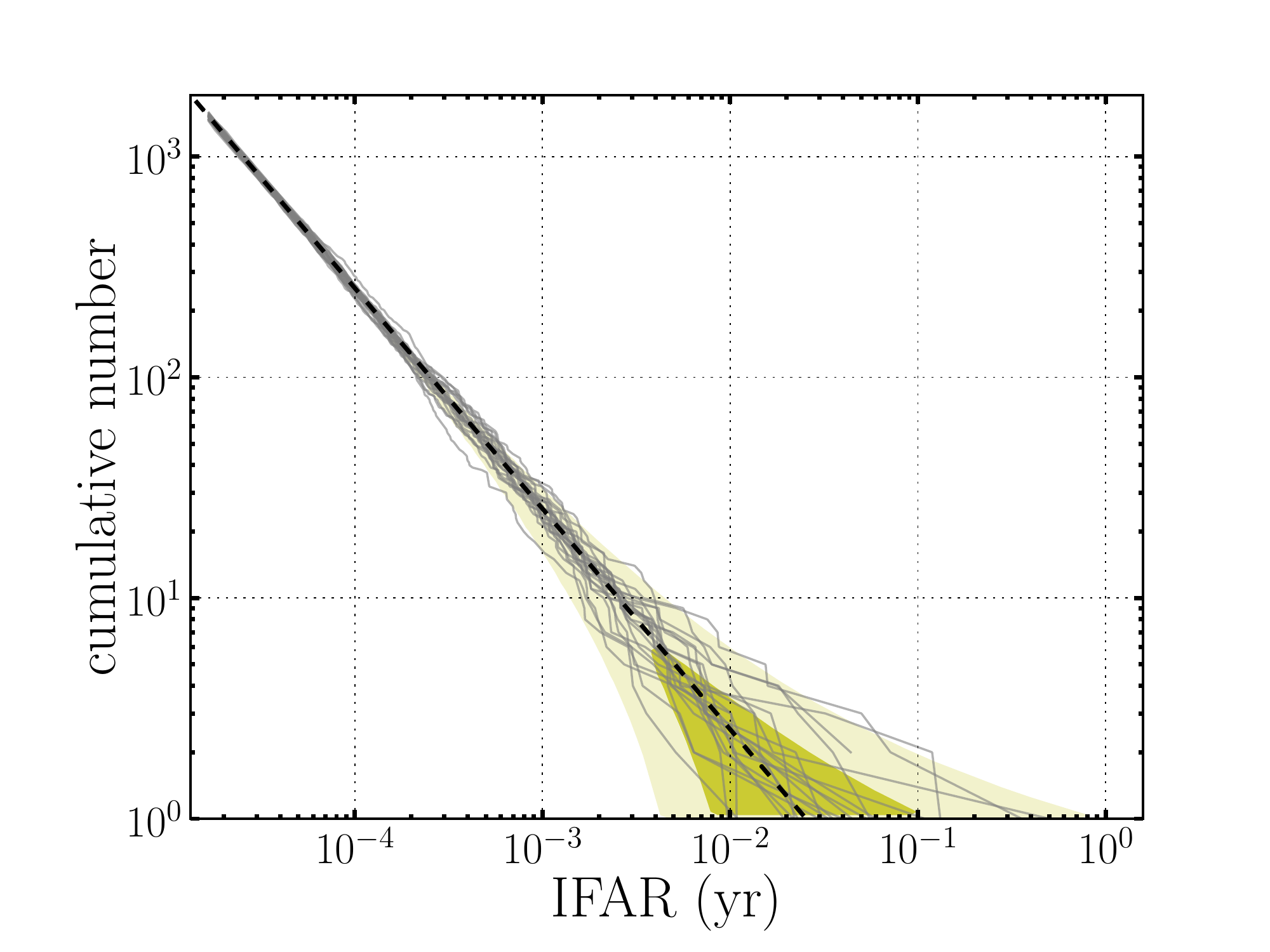}
  \caption[Cumulative histograms of inverse false alarm
rate.]{\label{fig:ihope_ifar}
Cumulative histogram of triggers vs.\ IFAR for all time-shift triggers 
in H1H2L1 triple-coincident time from a representative month of S5 data.  
The black dashed line marks the expected cumulative number,
while the shaded regions
mark its 1- and 2-$\sigma$ variation. The thin grey lines show the cumulative number
for 20 of the time shifts, providing an additional indication
of the expected deviation from the mean.}
\end{figure}

For each category of coincident triggers within an experiment, an additional
clustering stage is applied.  If there is another coincident trigger with a
larger combined re-weighted \ac{SNR} within \(\unit{10}{\second}\) of a given
trigger's end time, the trigger is removed. We then compute the \ac{FAR} as a
function of combined re-weighted SNR as the rate (number over the total
coincident, time-shifted search time) of time-shift coincidences observed with
higher combined re-weighted SNR within each category.  These results must then
be combined to estimate the overall significance of triggers: we calculate a
\emph{combined \ac{FAR}} across categories by ranking all triggers by their
\ac{FAR}, counting the number of more significant time-shift triggers, and
dividing by the total time-shift time.  The resulting combined \ac{FAR} is
essentially the same as the uncombined \ac{FAR}, multiplied by the number of
categories that were combined.  We often quote the inverse \ac{FAR} (IFAR) as
the ranking statistic, so that more significant triggers correspond to larger
values. A loud \ac{GW} may produce triggers in more than one mass bin, and
consequently more than one candidate trigger might be due to a single event.
This is resolved by reporting only the coincident trigger with the largest IFAR
associated with a given event.  Figure \ref{fig:ihope_ifar} shows the expected
mean (the dashed line) and variation (the shaded areas) of the cumulative
number of triggers as a function of IFAR for the analysis of three-detector
H1H2L1 time in a representative month of S5 data. The variations among time
shifts (the thin lines) match the expected distribution.  The duration of the
time-shift analysis is \(\unit{\sim 10^8}{\second}\), but taking into account
the six categories of triggers (three mass bins and two coincidence types),
this yields a minimum \ac{FAR} of \(\unit{\sim 1}{\invyear}\).

Clearly a \ac{FAR} of \(\unit{\sim 1}{\invyear}\) is insufficient to confidently identify \ac{GW}
events.  The challenge of extending background estimation to the level
where a loud trigger can become a detection candidate was met in the
S6--VSR2/3 search \cite{Collaboration:S6CBClowmass, Dent:2012}.
Remarkably, even for \acp{FAR} of one in tens of thousands of years, no
tail of triggers with large combined re-weighted \acp{SNR} was observed.
Evidently, the cuts, tests, and thresholds discussed in Section
\ref{sec:nongauss} are effective at eliminating any evidence of a
non-Gaussian background, at least for low chirp masses.

In calculating the \ac{FAR}, we treat all trigger categories identically, so
we implicitly assign the same weight to each.  However, this is not
appropriate when the detectors have significantly different
sensitivities, since a \ac{GW} is more likely to be observed in the most
sensitive detectors.  In the search of \ac{LIGO} S5 and Virgo VSR1 data
\cite{S5LowMassLV}, this approach was refined by weighting the
categories on the basis of the search sensitivity for each trigger type.
However, if there were an accurate astrophysical model of \ac{CBC} merger rates
for different binary masses, the weighting could easily be extended to
the mass bins.

\subsection{Evaluating search sensitivity}
\label{ssec:injections}

The sensitivity of a search is measured by adding simulated \ac{GW} signals to
the data and verifying their recovery by the pipeline, which also helps tune
the pipeline's performance against expected sources.  The simulated signals can
be added as \emph{hardware injections}
\cite{Brown:2004,Collaboration:S6CBClowmass}, by actuating the end mirrors of
the interferometers to reproduce the response of the interferometer to
\acp{GW}; or as \emph{software injections}, by modifying the data after it has
been read into the pipeline.  Hardware injections provide a better end-to-end
test of the analysis, but only a limited number can be performed, since the
data containing hardware injections cannot be used to search for real \ac{GW}
signals.  Consequently, large-scale injection campaigns are performed in
software.

Software injections are performed into all operational detectors
\emph{coherently} (i.e., with relative time delays, phases and amplitudes
appropriate for the relative location and orientation of the source and
the detectors).  Simulated \ac{GW} sources are generally placed uniformly
over the celestial sphere, with uniformly distributed orientations.  The mass and
spin parameters are generally chosen to uniformly cover the search
parameter space, since they are not well constrained by astrophysical
observations, particularly so for binaries containing black holes \cite{Mandel:2009nx}.
Although sources are expected to be roughly uniform in volume, we do not
follow that distribution for simulations, but instead attempt to place a
greater fraction of injections at distances where they would be
marginally detectable by the pipeline.  The techniques used to reduce the
dimensionality of parameter space, such as analytically maximizing the
detection statistic, cannot be applied to the injections, which must
cover the entire space.  This necessitates large simulation campaigns.

The \ihope\ pipeline is run on the data containing simulated signals using the same
configuration as for the rest of the search.  Injected signals are considered
to be found if there is a coincident trigger within \(\unit{1}{\second}\) of their injection time.
The loudest coincident trigger within the \(\unit{1}{\second}\) window is associated with
the injection, and it may be louder than any trigger in the time-shift
analyses (i.e., it may have a \ac{FAR} of zero).  Using a \(\unit{1}{\second}\) time
window to associate triggers and injections and no requirement on mass
consistency may lead to some of these being found spuriously, in
coincidence with background triggers. However, this effect has negligible
consequences on the estimated search sensitivity near the combined re-weighted
\ac{SNR} of the most significant trigger.
\begin{figure}
\includegraphics[width=\linewidth]{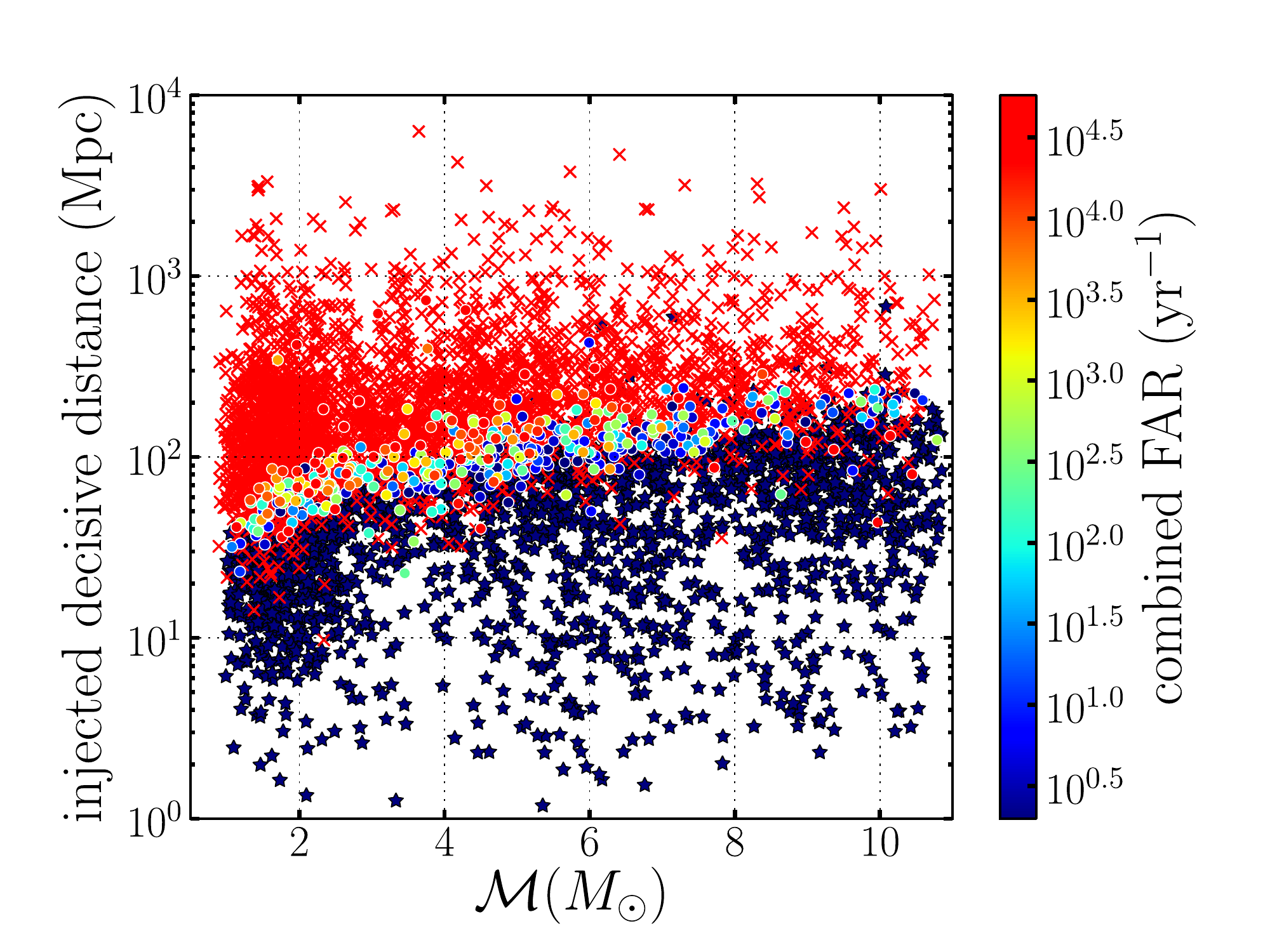}
    \caption{\label{fig:found_missed}Found and missed injections in one
month of S5 data 
    plotted at their chirp mass $\mathcal{M}$ and \emph{decisive distance} (see main text for definition).
    Red crosses are missed injections; colored circles are injections found
    with non-zero combined \ac{FAR}, which can be read off the colormap on the right;
    black stars are injections found with \ac{FAR} = 0
    (i.e., associated with triggers louder than any in the background from 100 time shifts).
	Nearby injections that are missed or found with high \acp{FAR} are followed up
    to check for problems in the pipeline, and to improve data quality.
    }
\end{figure}

Figure \ref{fig:found_missed} shows the results of a large
number of software injections performed in one month of S5 data. 
For each injection, we indicate whether the signal was missed (red crosses) or found (circles, and stars for \ac{FAR} = 0).
The recovery of simulated signals can be compared with the theoretically expected
sensitivity of the search, taking into account variations over parameter space:
the expected \ac{SNR} of a signal is proportional to $\mathcal{M}^{5/6}$ (for low-mass binaries),
inversely proportional to effective distance (see Sec.\ \ref{ssec:amplitude_consistency}),
and a function of the detectors' noise PSD.
An insightful way to display injections, used in Fig.\ \ref{fig:found_missed},
is to show their chirp mass $\mathcal{M}$ and \emph{decisive distance}---the second largest
effective distance for the detectors that were operating at the time of
the injection (in a coincidence search, it is the second most
sensitive detector that limits the overall sensitivity).
Indeed, our empirical results are in good agreement with the
stated sensitivity of the detectors \cite{PSD:S5,Collaboration:2012wu}.
A small number of signals are missed at low distances: these
are typically found to lie close to loud non-Gaussian glitches in the
detector data. 

\subsection{Bounding the binary coalescence rate}
\label{ssec:ul}

The results of a search can be used to estimate (if positive detections
are reported) or bound the rate of binary coalescences. An upper limit
on the merger rate is calculated by evaluating the sensitivity of the
search at the loudest observed trigger \cite{loudestGWDAW03, Fairhurst:2007qj,
Biswas:2007ni, keppel:thesis}. Heuristically, the 90\% rate upper limit corresponds to a few
(order 2--3) signals occurring over the search time within a small enough
distance to generate a trigger with IFAR larger than the loudest observed trigger.

More specifically, we assume that \ac{CBC}
events occur randomly and independently, and that the event rate
is proportional to the star-formation rate, which is
itself assumed proportional to blue-light galaxy luminosity \cite{Phinney:1991ei}.
For searches sensitive out to tens or hundreds of megaparsecs, it is
reasonable to approximate the blue-light luminosity as uniform in volume,
and quote rates per unit volume and time \cite{ratesdoc}.  
We follow \cite{Biswas:2007ni, Collaboration:2009tt}
and infer the probability density for the merger rate $R$, given
that in an observation time $T$ no other trigger was seen with IFAR larger
than its loudest-event value, $\alpha_m$:
\begin{equation}\label{eq:posterior}
p(R | \alpha_m,T ) \propto p(R) \, e^{-RV(\alpha_m)T} \left(
1 + \Lambda(\alpha_m) R\,T\,V(\alpha_m) \right);
\end{equation}
here $p(R)$ is the prior probability density for $R$,
usually taken as the result of previous searches or as a uniform distribution
for the first search of a kind;
$V(\alpha)$ is the volume of space in which the search could have seen
a signal with $\mathrm{IFAR} \geq \alpha$;
and the quantity $\Lambda$ is the relative probability that the loudest trigger
was due to a \acp{GW} rather than noise,
\begin{equation}
\Lambda = \frac{\lvert
V^{\prime}(\alpha_m)\rvert}{V(\alpha_m)}\frac{P_B(\alpha_m)}{P^{\prime}
_B(\alpha_m)}, \quad \text{with} \quad
P_B(\alpha) = e^{-T/\alpha},
\end{equation}
with the prime denoting differentiation with respect to $\alpha$.
For a chosen confidence level $\gamma$ (typically 0.9 = 90\%),
the upper limit $R_\ast$ on the rate is then given by
\begin{equation}
\gamma = \int^{R_\ast}_0 p(R | \alpha_m,T) \, dR.
\end{equation}
\begin{figure}
\includegraphics[width=\linewidth]{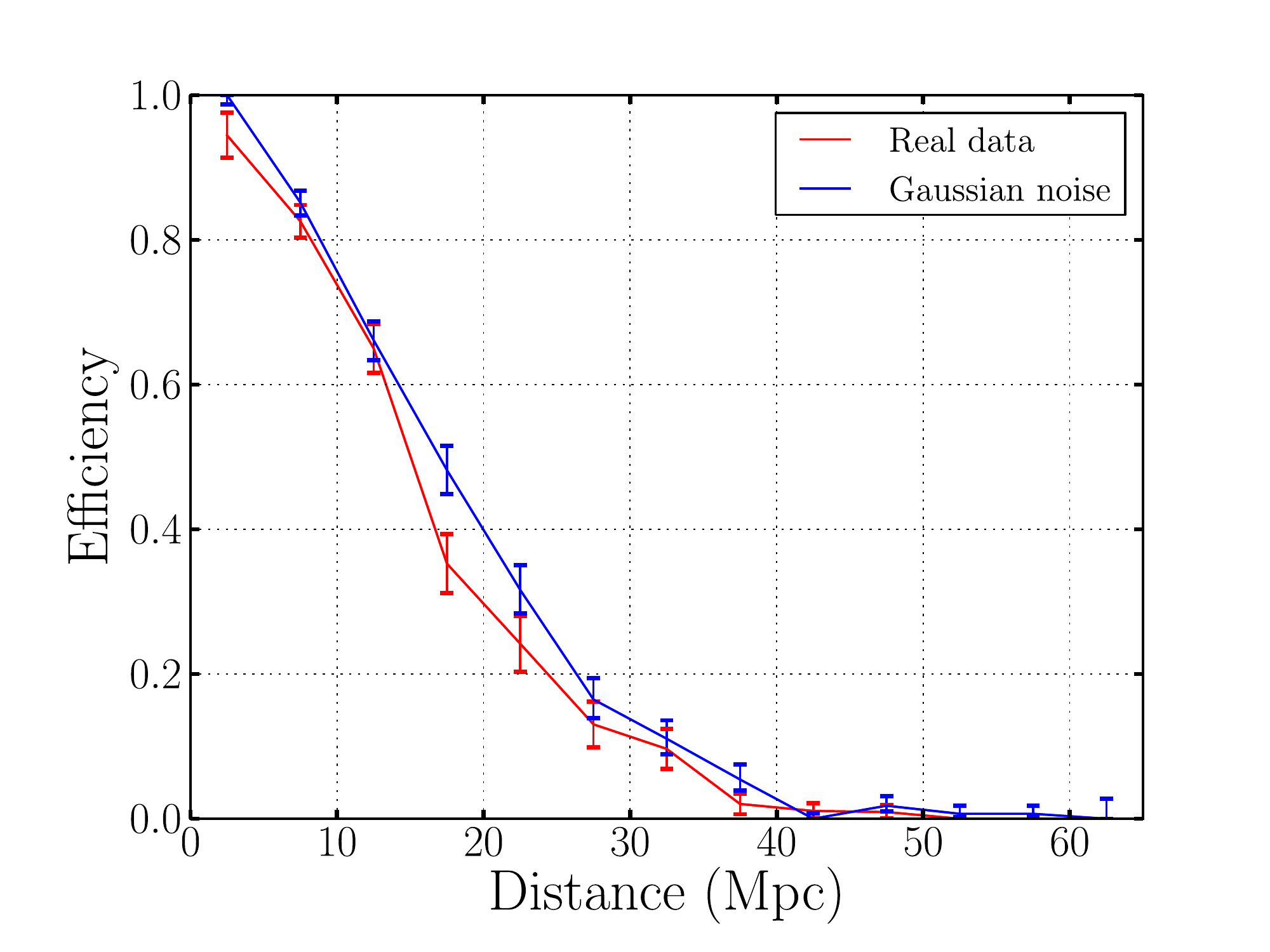}
\caption[Software injection recovery for BNS
injections.]{\label{fig:efficiency}
Search efficiency for \ac{BNS} injections in
a month of representative S5 data (blue) and in Gaussian noise (red),
for a false-alarm rate equal to the \ac{FAR} of the loudest foreground
trigger in each analysis.
}
\end{figure}

It is clear from Eq.\ \eqref{eq:posterior} that the decay of
$p(R|\alpha_m,T)$ and the resulting $R_\ast$ depend critically on the
\emph{sensitive volume} $V(\alpha_m)$.  In previous sections we have
shown how \ihope\ is highly effective at filtering out triggers due to
non-Gaussian noise, thus improving sensitivity, and in the context of
computing upper limits, we can quantify the residual effects of
non-Gaussian features on $V(\alpha_m)$.  In Fig.\ \ref{fig:efficiency}
we show the search \emph{efficiency} for \ac{BNS} signals, i.e. the
fraction of \ac{BNS} injections found with IFAR above a fiducial value,
here set to the IFAR of the loudest in-time noise trigger as a function of
distance, for one month of S5 data and for a month of Gaussian noise
with the same PSDs.\footnote{For Gaussian noise, we do not actually run
injections through the pipeline, but compute the expected \ac{SNR},
given the sensitivity of the detectors at that time, and compare with
the largest SNR among Gaussian-noise in-time triggers.} Despite the
significant non-Gaussianity of real data, the distance at which
efficiency is 50\% is reduced by $\sim 10$\% and the sensitive search
volume by $\sim 30$\%, compared to Gaussian-noise expectations.

\section{Discussion and future developments}
\label{sec:discussion}

In this paper we have given a detailed description of the \ihope\ software pipeline,
developed to search for \acp{GW} from \ac{CBC} events in LIGO and Virgo data, and we
have provided several examples of its performance on a sample stretch of data from the LIGO S5 run.
The pipeline is based on a matched-filtering engine augmented by a substantial number of
additional modules that implement coincidence, signal-consistency tests, data-quality cuts,
tunable ranking statistics, background estimation by time shifts, and sensitivity evaluation by injections.
Indeed, with the \ihope\ pipeline we can run analyses that go all the way from detector
strain data to event significance and upper limits on \ac{CBC} rates.

The pipeline was developed over a number of years, from the early
versions used in LIGO's S2 BNS search to its mature incarnation used in
the analysis of S6 and VSR3 data. One of the major successes of the \ihope\ pipeline
was the mitigation of spurious triggers from non-Gaussian noise
transients, to such an extent that the overall volume sensitivity is
reduced by less than 20\%  compared to what would be possible if noise
was Gaussian.  Nevertheless, there are still significant improvements
that can and must be made to \ac{CBC} searches if we are to meet the
challenges posed by analyzing the data of advanced detectors. In the
following paragraphs, we briefly discuss some of these improvements and
challenges.

\paragraph*{Coherent analysis.}
As discussed above, the \ihope\ pipeline comes close to the sensitivity that
would be achieved if noise was Gaussian, with the same PSD.  Therefore, while
some improvement could be obtained by implementing more sophisticated
signal-consistency tests and data-quality cuts, it will not be significant.  If
three or more detectors are active, sensitivity \emph{would} be improved in a
\emph{coherent} \cite{FinnChernoff:1993, PaiDhurandharBose2001,
HarryFairhurst:2011} (rather than coincident) analysis that filters the data
from all operating detectors simultaneously, requiring consistency between the
times of arrival and relative amplitudes of \ac{GW} signals, as observed in
each data stream. Such a search is challenging to implement because the data
from the detectors must be combined differently for each sky position,
significantly increasing computational cost.

Coherent searches \emph{have} already been run for unmodeled burst-like
transients \cite{S5VSR1Burst}, and for \ac{CBC} signals in coincidence with
gamma-ray-burst observations \cite{Briggs:2012ce}, but a full all-sky, all-time 
pipeline like  \ihope\ would require
significantly more computation. A promising compromise may be a hierarchical
search consisting of a first coincidence stage followed by the coherent
analysis of candidates, although the estimation of background trigger rates would
prove challenging as time shifts in a coherent analysis cannot be performed using
only the recorded single detector triggers but require the full \ac{SNR} time series.

\paragraph*{Background estimation.}
The first positive \ac{GW} detection requires that we assign a very low
false-alarm probability to a candidate trigger
\cite{Collaboration:S6CBClowmass}. In the \ihope\ pipeline, this would
necessitate a large number of time shifts, thus negating the computational
savings of splitting matched filtering between two stages, or a different
method of background estimation \cite{Dent:2012,Cannon:2012}.  Whichever the
solution, it will need to be automated to identify signal candidates rapidly
for possible astronomical follow up.

\paragraph*{Event-rate estimation.}
After the first detections, we will begin to quote event-rate estimates rather
than upper limits. The loudest-event method can be used for this
\cite{Biswas:2007ni}, provided that the data are broken up so that much less
than one gravitational wave signal is expected in each analyzed stretch. There are however other
approaches \cite{Messenger:2012} that should be considered for implementation.

\paragraph*{Template length.}
The sensitive band of advanced detectors will extend to lower frequencies
(\(\unit{\sim 10}{\hertz}\)) than their first-generation counterparts, greatly increasing the
length and number of templates required in a matched-filtering search.
Increasing computational resources may not be sufficient, so we are
investigating alternative approaches to filtering \cite{Marion:2004,
Cannon:2010, Cannon:2011tb, Cannon:2011xk, Cannon:2011vi} and possibly the use
of graphical processing units (GPUs).

\paragraph*{Latency.}
The latency of \ac{CBC} searches (i.e., the ``wall-clock'' time necessary for search
results to become available) has decreased over the course of successive
science runs, but further progress is needed to perform prompt follow-up
observations of \ac{GW} candidate with conventional (electromagnetic) telescopes
\cite{Virgo:2011aa, Metzger:2011bv}. The target should be posting candidate
triggers within minutes to hours of data taking, which was in fact achieved in
the S6--VSR3 analysis with the MBTA pipeline \cite{Marion:2004}.

\paragraph*{Template accuracy.}
While the templates currently used in \ihope\ are very accurate approximations
to \ac{BNS} signals, they could still be improved for the purpose of \ac{NSBH}
and \ac{BBH} searches \cite{BuonannoIyerOchsnerYiSathya2009}. It is
straightforward to extend \ihope\ to include the effects of spin on the
progress of inspiral (i.e., its \emph{phasing}), but it is harder to include
the orbital precession caused by spins and the resulting waveform modulations.
The first extension would already improve sensitivity to \ac{BBH} signals
\cite{Ajith:2009bn, Santamaria:2010}, but precessional effects are expected to
be more significant for \ac{NSBH} systems \cite{Pan:2003qt, Ajith:2011hq}.

\paragraph*{Parameter estimation.}
Last, while \ihope\ effectively searches the entire template parameter space to
identify candidate triggers, at the end of the pipeline the only information
available about these are the estimated binary masses, arrival time, and
effective distance. Dedicated follow-up analyses can provide much more detailed
and reliable estimates of all parameters \cite{Sluys:2008a, Sluys:2008b,
Veitch:2010, Feroz:2009}, but \ihope\ itself could be modified to provide rough
first-cut estimates.

\acknowledgments 

The authors would like to thank their colleagues in the LIGO Scientific
Collaboration and Virgo Collaboration, and particularly the other members of
the Compact Binary Coalescence Search Group.

The authors gratefully acknowledge the support of the United States National
Science Foundation, the Science and Technology Facilities Council of the United
Kingdom, the Royal Society, the Max Planck Society, the National Aeronautics
and Space Administration, Industry Canada and the Province of Ontario through
the Ministry of Research \& Innovation.  LIGO was constructed by the California
Institute of Technology and Massachusetts Institute of Technology with funding
from the National Science Foundation and operates under cooperative agreement
PHY-0757058. 

\bibliographystyle{apsrev4-1}
\bibliography{../bibtex/iulpapers.bib}

\begin{thebibliography}{93}%
\makeatletter
\providecommand \@ifxundefined [1]{%
 \@ifx{#1\undefined}
}%
\providecommand \@ifnum [1]{%
 \ifnum #1\expandafter \@firstoftwo
 \else \expandafter \@secondoftwo
 \fi
}%
\providecommand \@ifx [1]{%
 \ifx #1\expandafter \@firstoftwo
 \else \expandafter \@secondoftwo
 \fi
}%
\providecommand \natexlab [1]{#1}%
\providecommand \enquote  [1]{``#1''}%
\providecommand \bibnamefont  [1]{#1}%
\providecommand \bibfnamefont [1]{#1}%
\providecommand \citenamefont [1]{#1}%
\providecommand \href@noop [0]{\@secondoftwo}%
\providecommand \href [0]{\begingroup \@sanitize@url \@href}%
\providecommand \@href[1]{\@@startlink{#1}\@@href}%
\providecommand \@@href[1]{\endgroup#1\@@endlink}%
\providecommand \@sanitize@url [0]{\catcode `\\12\catcode `\$12\catcode
  `\&12\catcode `\#12\catcode `\^12\catcode `\_12\catcode `\%12\relax}%
\providecommand \@@startlink[1]{}%
\providecommand \@@endlink[0]{}%
\providecommand \url  [0]{\begingroup\@sanitize@url \@url }%
\providecommand \@url [1]{\endgroup\@href {#1}{\urlprefix }}%
\providecommand \urlprefix  [0]{URL }%
\providecommand \Eprint [0]{\href }%
\providecommand \doibase [0]{http://dx.doi.org/}%
\providecommand \selectlanguage [0]{\@gobble}%
\providecommand \bibinfo  [0]{\@secondoftwo}%
\providecommand \bibfield  [0]{\@secondoftwo}%
\providecommand \translation [1]{[#1]}%
\providecommand \BibitemOpen [0]{}%
\providecommand \bibitemStop [0]{}%
\providecommand \bibitemNoStop [0]{.\EOS\space}%
\providecommand \EOS [0]{\spacefactor3000\relax}%
\providecommand \BibitemShut  [1]{\csname bibitem#1\endcsname}%
\let\auto@bib@innerbib\@empty
\bibitem [{\citenamefont {Abbott}\ \emph
  {et~al.}(2009{\natexlab{a}})\citenamefont {Abbott} \emph
  {et~al.}}]{Abbott:2007kv}%
  \BibitemOpen
  \bibfield  {author} {\bibinfo {author} {\bibfnamefont {B.}~\bibnamefont
  {Abbott}} \emph {et~al.} (\bibinfo {collaboration} {LIGO Scientific
  Collaboration}),\ }\href {\doibase 10.1088/0034-4885/72/7/076901} {\bibfield
  {journal} {\bibinfo  {journal} {Rep.\ Prog.\ Phys.}\ }\textbf {\bibinfo
  {volume} {72}},\ \bibinfo {pages} {076901} (\bibinfo {year}
  {2009}{\natexlab{a}})},\ \Eprint {http://arxiv.org/abs/0711.3041}
  {arXiv:0711.3041 [gr-qc]} \BibitemShut {NoStop}%
\bibitem [{\citenamefont {Accadia}\ \emph {et~al.}(2012)\citenamefont {Accadia}
  \emph {et~al.}}]{Accadia:2012zz}%
  \BibitemOpen
  \bibfield  {author} {\bibinfo {author} {\bibfnamefont {T.}~\bibnamefont
  {Accadia}} \emph {et~al.} (\bibinfo {collaboration} {VIRGO Collaboration}),\
  }\href@noop {} {\bibfield  {journal} {\bibinfo  {journal} {JINST}\ }\textbf
  {\bibinfo {volume} {7}},\ \bibinfo {pages} {P03012} (\bibinfo {year}
  {2012})}\BibitemShut {NoStop}%
\bibitem [{\citenamefont {Grote}(2008)}]{Grote:2008}%
  \BibitemOpen
  \bibfield  {author} {\bibinfo {author} {\bibfnamefont {H.}~\bibnamefont
  {Grote}} (\bibinfo {collaboration} {LIGO Scientific Collaboration}),\
  }\href@noop {} {\bibfield  {journal} {\bibinfo  {journal} {Class. Quant.
  Grav.}\ }\textbf {\bibinfo {volume} {25}},\ \bibinfo {pages} {114043}
  (\bibinfo {year} {2008})}\BibitemShut {NoStop}%
\bibitem [{\citenamefont {Abbott}\ \emph {et~al.}(2004)\citenamefont {Abbott}
  \emph {et~al.}}]{Abbott:2003pj}%
  \BibitemOpen
  \bibfield  {author} {\bibinfo {author} {\bibfnamefont {B.}~\bibnamefont
  {Abbott}} \emph {et~al.} (\bibinfo {collaboration} {LIGO Scientific
  Collaboration}),\ }\href {\doibase 10.1103/PhysRevD.69.122001} {\bibfield
  {journal} {\bibinfo  {journal} {Phys.Rev.}\ }\textbf {\bibinfo {volume}
  {D69}},\ \bibinfo {pages} {122001} (\bibinfo {year} {2004})},\ \Eprint
  {http://arxiv.org/abs/gr-qc/0308069} {arXiv:gr-qc/0308069} \BibitemShut
  {NoStop}%
\bibitem [{\citenamefont {Abbott}\ \emph
  {et~al.}(2005{\natexlab{a}})\citenamefont {Abbott} \emph
  {et~al.}}]{LIGOS2iul}%
  \BibitemOpen
  \bibfield  {author} {\bibinfo {author} {\bibfnamefont {B.}~\bibnamefont
  {Abbott}} \emph {et~al.} (\bibinfo {collaboration} {LIGO Scientific
  Collaboration}),\ }\href {\doibase 10.1103/PhysRevD.72.082001} {\bibfield
  {journal} {\bibinfo  {journal} {Phys.Rev.}\ }\textbf {\bibinfo {volume}
  {D72}},\ \bibinfo {pages} {082001} (\bibinfo {year} {2005}{\natexlab{a}})},\
  \Eprint {http://arxiv.org/abs/gr-qc/0505041} {arXiv:gr-qc/0505041}
  \BibitemShut {NoStop}%
\bibitem [{\citenamefont {Abbott}\ \emph
  {et~al.}(2006{\natexlab{a}})\citenamefont {Abbott} \emph
  {et~al.}}]{LIGOS2bbh}%
  \BibitemOpen
  \bibfield  {author} {\bibinfo {author} {\bibfnamefont {B.}~\bibnamefont
  {Abbott}} \emph {et~al.} (\bibinfo {collaboration} {LIGO Scientific
  Collaboration}),\ }\href {\doibase 10.1103/PhysRevD.73.062001} {\bibfield
  {journal} {\bibinfo  {journal} {Phys.Rev.}\ }\textbf {\bibinfo {volume}
  {D73}},\ \bibinfo {pages} {062001} (\bibinfo {year} {2006}{\natexlab{a}})},\
  \Eprint {http://arxiv.org/abs/gr-qc/0509129} {arXiv:gr-qc/0509129}
  \BibitemShut {NoStop}%
\bibitem [{\citenamefont {Abbott}\ \emph
  {et~al.}(2005{\natexlab{b}})\citenamefont {Abbott} \emph
  {et~al.}}]{LIGOS2macho}%
  \BibitemOpen
  \bibfield  {author} {\bibinfo {author} {\bibfnamefont {B.}~\bibnamefont
  {Abbott}} \emph {et~al.} (\bibinfo {collaboration} {{LIGO} Scientific
  Collaboration}),\ }\href@noop {} {\bibfield  {journal} {\bibinfo  {journal}
  {Phys.~Rev.~D}\ }\textbf {\bibinfo {volume} {72}},\ \bibinfo {pages} {082002}
  (\bibinfo {year} {2005}{\natexlab{b}})},\ \Eprint
  {http://arxiv.org/abs/arXiv:gr-qc/0505042} {arXiv:gr-qc/0505042} \BibitemShut
  {NoStop}%
\bibitem [{\citenamefont {Abbott}\ \emph
  {et~al.}(2006{\natexlab{b}})\citenamefont {Abbott} \emph
  {et~al.}}]{ligotama}%
  \BibitemOpen
  \bibfield  {author} {\bibinfo {author} {\bibfnamefont {B.}~\bibnamefont
  {Abbott}} \emph {et~al.} (\bibinfo {collaboration} {LIGO Scientific
  Collaboration, TAMA Collaboration}),\ }\href {\doibase
  10.1103/PhysRevD.73.102002} {\bibfield  {journal} {\bibinfo  {journal}
  {Phys.Rev.}\ }\textbf {\bibinfo {volume} {D73}},\ \bibinfo {pages} {102002}
  (\bibinfo {year} {2006}{\natexlab{b}})},\ \Eprint
  {http://arxiv.org/abs/gr-qc/0512078} {arXiv:gr-qc/0512078} \BibitemShut
  {NoStop}%
\bibitem [{\citenamefont {Abbott}\ \emph
  {et~al.}(2008{\natexlab{a}})\citenamefont {Abbott} \emph
  {et~al.}}]{S3_BCVSpin}%
  \BibitemOpen
  \bibfield  {author} {\bibinfo {author} {\bibfnamefont {B.}~\bibnamefont
  {Abbott}} \emph {et~al.} (\bibinfo {collaboration} {LIGO Scientific
  Collaboration}),\ }\href {\doibase 10.1103/PhysRevD.78.042002} {\bibfield
  {journal} {\bibinfo  {journal} {Phys.Rev.}\ }\textbf {\bibinfo {volume}
  {D78}},\ \bibinfo {pages} {042002} (\bibinfo {year} {2008}{\natexlab{a}})},\
  \Eprint {http://arxiv.org/abs/0712.2050} {arXiv:0712.2050 [gr-qc]}
  \BibitemShut {NoStop}%
\bibitem [{\citenamefont {Abbott}\ \emph
  {et~al.}(2008{\natexlab{b}})\citenamefont {Abbott} \emph
  {et~al.}}]{LIGOS3S4all}%
  \BibitemOpen
  \bibfield  {author} {\bibinfo {author} {\bibfnamefont {B.}~\bibnamefont
  {Abbott}} \emph {et~al.} (\bibinfo {collaboration} {LIGO Scientific
  Collaboration}),\ }\href {\doibase 10.1103/PhysRevD.77.062002} {\bibfield
  {journal} {\bibinfo  {journal} {Phys.Rev.}\ }\textbf {\bibinfo {volume}
  {D77}},\ \bibinfo {pages} {062002} (\bibinfo {year} {2008}{\natexlab{b}})},\
  \Eprint {http://arxiv.org/abs/0704.3368} {arXiv:0704.3368 [gr-qc]}
  \BibitemShut {NoStop}%
\bibitem [{\citenamefont {Abbott}\ \emph
  {et~al.}(2009{\natexlab{b}})\citenamefont {Abbott} \emph
  {et~al.}}]{Collaboration:2009tt}%
  \BibitemOpen
  \bibfield  {author} {\bibinfo {author} {\bibfnamefont {B.}~\bibnamefont
  {Abbott}} \emph {et~al.} (\bibinfo {collaboration} {LIGO Scientific
  Collaboration}),\ }\href {\doibase 10.1103/PhysRevD.79.122001} {\bibfield
  {journal} {\bibinfo  {journal} {Phys.~Rev.~D}\ }\textbf {\bibinfo {volume}
  {79}},\ \bibinfo {pages} {122001} (\bibinfo {year} {2009}{\natexlab{b}})},\
  \Eprint {http://arxiv.org/abs/0901.0302} {arXiv:0901.0302 [gr-qc]}
  \BibitemShut {NoStop}%
\bibitem [{\citenamefont {Abbott}\ \emph
  {et~al.}(2009{\natexlab{c}})\citenamefont {Abbott} \emph
  {et~al.}}]{Abbott:2009qj}%
  \BibitemOpen
  \bibfield  {author} {\bibinfo {author} {\bibfnamefont {B.}~\bibnamefont
  {Abbott}} \emph {et~al.} (\bibinfo {collaboration} {LIGO Scientific
  Collaboration}),\ }\href {\doibase 10.1103/PhysRevD.80.047101} {\bibfield
  {journal} {\bibinfo  {journal} {Phys.Rev.}\ }\textbf {\bibinfo {volume}
  {D80}},\ \bibinfo {pages} {047101} (\bibinfo {year} {2009}{\natexlab{c}})},\
  \Eprint {http://arxiv.org/abs/0905.3710} {arXiv:0905.3710 [gr-qc]}
  \BibitemShut {NoStop}%
\bibitem [{\citenamefont {Abadie}\ \emph
  {et~al.}(2010{\natexlab{a}})\citenamefont {Abadie} \emph
  {et~al.}}]{S5LowMassLV}%
  \BibitemOpen
  \bibfield  {author} {\bibinfo {author} {\bibfnamefont {J.}~\bibnamefont
  {Abadie}} \emph {et~al.} (\bibinfo {collaboration} {LIGO Scientific
  Collaboration and Virgo Collaboration}),\ }\href {\doibase
  10.1103/PhysRevD.82.102001} {\bibfield  {journal} {\bibinfo  {journal}
  {\prd}\ }\textbf {\bibinfo {volume} {82}},\ \bibinfo {pages} {102001}
  (\bibinfo {year} {2010}{\natexlab{a}})},\ \Eprint
  {http://arxiv.org/abs/1005.4655} {arXiv:1005.4655 [gr-qc]} \BibitemShut
  {NoStop}%
\bibitem [{\citenamefont {Abadie}\ \emph
  {et~al.}(2012{\natexlab{a}})\citenamefont {Abadie} \emph
  {et~al.}}]{Collaboration:S6CBClowmass}%
  \BibitemOpen
  \bibfield  {author} {\bibinfo {author} {\bibfnamefont {J.}~\bibnamefont
  {Abadie}} \emph {et~al.} (\bibinfo {collaboration} {LIGO Scientific
  Collaboration and Virgo Collaboration}),\ }\href {\doibase
  10.1103/PhysRevD.85.082002} {\bibfield  {journal} {\bibinfo  {journal}
  {Phys.\ Rev.\ D}\ }\textbf {\bibinfo {volume} {85}},\ \bibinfo {pages}
  {082002} (\bibinfo {year} {2012}{\natexlab{a}})},\ \Eprint
  {http://arxiv.org/abs/arXiv:1111.7314} {arXiv:1111.7314} \BibitemShut
  {NoStop}%
\bibitem [{\citenamefont {Brown}(2005)}]{brown-2005-22}%
  \BibitemOpen
  \bibfield  {author} {\bibinfo {author} {\bibfnamefont {D.~A.}\ \bibnamefont
  {Brown}} (\bibinfo {collaboration} {LIGO Collaboration}),\ }\href {\doibase
  10.1088/0264-9381/22/18/S24} {\bibfield  {journal} {\bibinfo  {journal}
  {Class.Quant.Grav.}\ }\textbf {\bibinfo {volume} {22}},\ \bibinfo {pages}
  {S1097} (\bibinfo {year} {2005})},\ \Eprint
  {http://arxiv.org/abs/gr-qc/0505102} {arXiv:gr-qc/0505102 [gr-qc]}
  \BibitemShut {NoStop}%
\bibitem [{\citenamefont {Blanchet}(2002)}]{Blanchet:2002av}%
  \BibitemOpen
  \bibfield  {author} {\bibinfo {author} {\bibfnamefont {L.}~\bibnamefont
  {Blanchet}},\ }\href@noop {} {\bibfield  {journal} {\bibinfo  {journal}
  {Living Rev. Rel.}\ }\textbf {\bibinfo {volume} {5}},\ \bibinfo {pages} {3}
  (\bibinfo {year} {2002})},\ \Eprint
  {http://arxiv.org/abs/arXiv:gr-qc/0202016} {arXiv:gr-qc/0202016} \BibitemShut
  {NoStop}%
\bibitem [{\citenamefont {Centrella}\ \emph {et~al.}(2010)\citenamefont
  {Centrella}, \citenamefont {Baker}, \citenamefont {Kelly},\ and\
  \citenamefont {van Meter}}]{Centrella:2010}%
  \BibitemOpen
  \bibfield  {author} {\bibinfo {author} {\bibfnamefont {J.~M.}\ \bibnamefont
  {Centrella}}, \bibinfo {author} {\bibfnamefont {J.~G.}\ \bibnamefont
  {Baker}}, \bibinfo {author} {\bibfnamefont {B.~J.}\ \bibnamefont {Kelly}}, \
  and\ \bibinfo {author} {\bibfnamefont {J.~R.}\ \bibnamefont {van Meter}},\
  }\href@noop {} {\bibfield  {journal} {\bibinfo  {journal} {Rev. Mod. Phys.}\
  }\textbf {\bibinfo {volume} {82}},\ \bibinfo {pages} {3069} (\bibinfo {year}
  {2010})},\ \Eprint {http://arxiv.org/abs/1010.5260} {arXiv:1010.5260 [gr-qc]}
  \BibitemShut {NoStop}%
\bibitem [{\citenamefont {Hannam}(2009)}]{Hannam:2009rd}%
  \BibitemOpen
  \bibfield  {author} {\bibinfo {author} {\bibfnamefont {M.}~\bibnamefont
  {Hannam}},\ }\href@noop {} {\bibfield  {journal} {\bibinfo  {journal} {Class.
  Quant. Grav.}\ }\textbf {\bibinfo {volume} {26}},\ \bibinfo {pages} {114001}
  (\bibinfo {year} {2009})},\ \Eprint {http://arxiv.org/abs/arXiv:0901.2931}
  {arXiv:0901.2931} \BibitemShut {NoStop}%
\bibitem [{\citenamefont {Sperhake}\ \emph {et~al.}(2011)\citenamefont
  {Sperhake}, \citenamefont {Berti},\ and\ \citenamefont
  {Cardoso}}]{2011arXiv1107.2819S}%
  \BibitemOpen
  \bibfield  {author} {\bibinfo {author} {\bibfnamefont {U.}~\bibnamefont
  {Sperhake}}, \bibinfo {author} {\bibfnamefont {E.}~\bibnamefont {Berti}}, \
  and\ \bibinfo {author} {\bibfnamefont {V.}~\bibnamefont {Cardoso}},\
  }\href@noop {} {\  (\bibinfo {year} {2011})},\ \Eprint
  {http://arxiv.org/abs/1107.2819} {arXiv:1107.2819 [gr-qc]} \BibitemShut
  {NoStop}%
\bibitem [{\citenamefont {{Berti}}\ \emph {et~al.}(2007)\citenamefont
  {{Berti}}, \citenamefont {{Cardoso}}, \citenamefont {{Gonzalez}},
  \citenamefont {{Sperhake}}, \citenamefont {{Hannam}}, \citenamefont
  {{Husa}},\ and\ \citenamefont {{Br{\"u}gmann}}}]{Berti:2007gv}%
  \BibitemOpen
  \bibfield  {author} {\bibinfo {author} {\bibfnamefont {E.}~\bibnamefont
  {{Berti}}}, \bibinfo {author} {\bibfnamefont {V.}~\bibnamefont {{Cardoso}}},
  \bibinfo {author} {\bibfnamefont {J.~A.}\ \bibnamefont {{Gonzalez}}},
  \bibinfo {author} {\bibfnamefont {U.}~\bibnamefont {{Sperhake}}}, \bibinfo
  {author} {\bibfnamefont {M.}~\bibnamefont {{Hannam}}}, \bibinfo {author}
  {\bibfnamefont {S.}~\bibnamefont {{Husa}}}, \ and\ \bibinfo {author}
  {\bibfnamefont {B.}~\bibnamefont {{Br{\"u}gmann}}},\ }\href {\doibase
  10.1103/PhysRevD.76.064034} {\bibfield  {journal} {\bibinfo  {journal}
  {\prd}\ }\textbf {\bibinfo {volume} {76}},\ \bibinfo {pages} {064034}
  (\bibinfo {year} {2007})},\ \Eprint
  {http://arxiv.org/abs/arXiv:gr-qc/0703053} {arXiv:gr-qc/0703053} \BibitemShut
  {NoStop}%
\bibitem [{\citenamefont {Wainstein}\ and\ \citenamefont
  {Zubakov}(1962)}]{wainstein:1962}%
  \BibitemOpen
  \bibfield  {author} {\bibinfo {author} {\bibfnamefont {L.~A.}\ \bibnamefont
  {Wainstein}}\ and\ \bibinfo {author} {\bibfnamefont {V.~D.}\ \bibnamefont
  {Zubakov}},\ }\href@noop {} {\emph {\bibinfo {title} {Extraction of signals
  from noise}}}\ (\bibinfo  {publisher} {Prentice-Hall},\ \bibinfo {address}
  {Englewood Cliffs, NJ},\ \bibinfo {year} {1962})\BibitemShut {NoStop}%
\bibitem [{\citenamefont {Cokelaer}\ and\ \citenamefont
  {Pathak}(2009)}]{Cokelaer:2009hj}%
  \BibitemOpen
  \bibfield  {author} {\bibinfo {author} {\bibfnamefont {T.}~\bibnamefont
  {Cokelaer}}\ and\ \bibinfo {author} {\bibfnamefont {D.}~\bibnamefont
  {Pathak}},\ }\href@noop {} {\bibfield  {journal} {\bibinfo  {journal} {Class.
  Quant. Grav.}\ }\textbf {\bibinfo {volume} {26}},\ \bibinfo {pages} {045013}
  (\bibinfo {year} {2009})},\ \Eprint {http://arxiv.org/abs/0903.4791}
  {arXiv:0903.4791 [gr-qc]} \BibitemShut {NoStop}%
\bibitem [{\citenamefont {Brown}\ and\ \citenamefont
  {Zimmerman}(2010)}]{Brown:2009ng}%
  \BibitemOpen
  \bibfield  {author} {\bibinfo {author} {\bibfnamefont {D.~A.}\ \bibnamefont
  {Brown}}\ and\ \bibinfo {author} {\bibfnamefont {P.~J.}\ \bibnamefont
  {Zimmerman}},\ }\href {\doibase 10.1103/PhysRevD.81.024007} {\bibfield
  {journal} {\bibinfo  {journal} {Phys.Rev.}\ }\textbf {\bibinfo {volume}
  {D81}},\ \bibinfo {pages} {024007} (\bibinfo {year} {2010})},\ \Eprint
  {http://arxiv.org/abs/0909.0066} {arXiv:0909.0066 [gr-qc]} \BibitemShut
  {NoStop}%
\bibitem [{\citenamefont {Van Den~Broeck}\ \emph {et~al.}(2009)\citenamefont
  {Van Den~Broeck} \emph {et~al.}}]{VanDenBroeck:2009gd}%
  \BibitemOpen
  \bibfield  {author} {\bibinfo {author} {\bibfnamefont {C.}~\bibnamefont {Van
  Den~Broeck}} \emph {et~al.},\ }\href {\doibase 10.1103/PhysRevD.80.024009}
  {\bibfield  {journal} {\bibinfo  {journal} {Phys.Rev.}\ }\textbf {\bibinfo
  {volume} {D80}},\ \bibinfo {pages} {024009} (\bibinfo {year} {2009})},\
  \Eprint {http://arxiv.org/abs/0904.1715} {arXiv:0904.1715 [gr-qc]}
  \BibitemShut {NoStop}%
\bibitem [{\citenamefont {Allen}\ \emph {et~al.}(2012)\citenamefont {Allen},
  \citenamefont {Anderson}, \citenamefont {Brady}, \citenamefont {Brown},\ and\
  \citenamefont {Creighton}}]{Allen:2005fk}%
  \BibitemOpen
  \bibfield  {author} {\bibinfo {author} {\bibfnamefont {B.}~\bibnamefont
  {Allen}}, \bibinfo {author} {\bibfnamefont {W.~G.}\ \bibnamefont {Anderson}},
  \bibinfo {author} {\bibfnamefont {P.~R.}\ \bibnamefont {Brady}}, \bibinfo
  {author} {\bibfnamefont {D.~A.}\ \bibnamefont {Brown}}, \ and\ \bibinfo
  {author} {\bibfnamefont {J.~D.~E.}\ \bibnamefont {Creighton}},\ }\href@noop
  {} {\  (\bibinfo {year} {2012})},\ \Eprint
  {http://arxiv.org/abs/gr-qc/0509116} {arXiv:gr-qc/0509116} \BibitemShut
  {NoStop}%
\bibitem [{\citenamefont {Apostolatos}\ \emph {et~al.}(1994)\citenamefont
  {Apostolatos} \emph {et~al.}}]{PhysRevD.49.6274}%
  \BibitemOpen
  \bibfield  {author} {\bibinfo {author} {\bibfnamefont {T.~A.}\ \bibnamefont
  {Apostolatos}} \emph {et~al.},\ }\href {\doibase 10.1103/PhysRevD.49.6274}
  {\bibfield  {journal} {\bibinfo  {journal} {Phys. Rev. D}\ }\textbf {\bibinfo
  {volume} {49}},\ \bibinfo {pages} {6274} (\bibinfo {year}
  {1994})}\BibitemShut {NoStop}%
\bibitem [{\citenamefont {Apostolatos}(1995)}]{Apostolatos:1995}%
  \BibitemOpen
  \bibfield  {author} {\bibinfo {author} {\bibfnamefont {T.~A.}\ \bibnamefont
  {Apostolatos}},\ }\href@noop {} {\bibfield  {journal} {\bibinfo  {journal}
  {Phys.~Rev.~D}\ }\textbf {\bibinfo {volume} {52}},\ \bibinfo {pages} {605}
  (\bibinfo {year} {1995})}\BibitemShut {NoStop}%
\bibitem [{\citenamefont {Buonanno}\ \emph {et~al.}(2003)\citenamefont
  {Buonanno}, \citenamefont {Chen},\ and\ \citenamefont
  {Vallisneri}}]{BuonannoChenVallisneri:2003b}%
  \BibitemOpen
  \bibfield  {author} {\bibinfo {author} {\bibfnamefont {A.}~\bibnamefont
  {Buonanno}}, \bibinfo {author} {\bibfnamefont {Y.}~\bibnamefont {Chen}}, \
  and\ \bibinfo {author} {\bibfnamefont {M.}~\bibnamefont {Vallisneri}},\
  }\href@noop {} {\bibfield  {journal} {\bibinfo  {journal} {Phys.~Rev.~D}\
  }\textbf {\bibinfo {volume} {67}},\ \bibinfo {pages} {104025} (\bibinfo
  {year} {2003})},\ \bibinfo {note} {erratum-ibid. 74 (2006)
  029904(E)}\BibitemShut {NoStop}%
\bibitem [{\citenamefont {Pan}\ \emph {et~al.}(2004)\citenamefont {Pan},
  \citenamefont {Buonanno}, \citenamefont {Chen},\ and\ \citenamefont
  {Vallisneri}}]{Pan:2003qt}%
  \BibitemOpen
  \bibfield  {author} {\bibinfo {author} {\bibfnamefont {Y.}~\bibnamefont
  {Pan}}, \bibinfo {author} {\bibfnamefont {A.}~\bibnamefont {Buonanno}},
  \bibinfo {author} {\bibfnamefont {Y.-b.}\ \bibnamefont {Chen}}, \ and\
  \bibinfo {author} {\bibfnamefont {M.}~\bibnamefont {Vallisneri}},\
  }\href@noop {} {\bibfield  {journal} {\bibinfo  {journal} {\prd}\ }\textbf
  {\bibinfo {volume} {69}},\ \bibinfo {pages} {104017} (\bibinfo {year}
  {2004})},\ \bibinfo {note} {erratum-ibid. 74 (2006) 029905(E)},\ \Eprint
  {http://arxiv.org/abs/gr-qc/0310034} {gr-qc/0310034} \BibitemShut {NoStop}%
\bibitem [{\citenamefont {Allen}(2005)}]{Allen:2004}%
  \BibitemOpen
  \bibfield  {author} {\bibinfo {author} {\bibfnamefont {B.}~\bibnamefont
  {Allen}},\ }\href@noop {} {\bibfield  {journal} {\bibinfo  {journal}
  {Phys.~Rev.~D}\ }\textbf {\bibinfo {volume} {71}},\ \bibinfo {pages} {062001}
  (\bibinfo {year} {2005})}\BibitemShut {NoStop}%
\bibitem [{\citenamefont {Brady}\ and\ \citenamefont
  {Fairhurst}(2008)}]{Fairhurst:2007qj}%
  \BibitemOpen
  \bibfield  {author} {\bibinfo {author} {\bibfnamefont {P.~R.}\ \bibnamefont
  {Brady}}\ and\ \bibinfo {author} {\bibfnamefont {S.}~\bibnamefont
  {Fairhurst}},\ }\href {\doibase 10.1088/0264-9381/25/10/105002} {\bibfield
  {journal} {\bibinfo  {journal} {Class. Quant. Grav.}\ }\textbf {\bibinfo
  {volume} {25}},\ \bibinfo {pages} {105002} (\bibinfo {year} {2008})},\
  \Eprint {http://arxiv.org/abs/arXiv:0707.2410} {arXiv:0707.2410} \BibitemShut
  {NoStop}%
\bibitem [{\citenamefont {Brown}(2004)}]{findchirp}%
  \BibitemOpen
  \bibfield  {author} {\bibinfo {author} {\bibfnamefont {D.~A.}\ \bibnamefont
  {Brown}},\ }\emph {\bibinfo {title} {Search for gravitational radiation from
  black hole {MACHOs} in the {G}alactic halo}},\ \href@noop {} {Ph.D. thesis},\
  \bibinfo  {school} {University of Wisconsin--Milwaukee} (\bibinfo {year}
  {2004})\BibitemShut {NoStop}%
\bibitem [{\citenamefont {Cokelaer}(2007)}]{hexabank}%
  \BibitemOpen
  \bibfield  {author} {\bibinfo {author} {\bibfnamefont {T.}~\bibnamefont
  {Cokelaer}},\ }\href@noop {} {\bibfield  {journal} {\bibinfo  {journal}
  {Phys.~Rev.~D}\ }\textbf {\bibinfo {volume} {76}},\ \bibinfo {pages} {102004}
  (\bibinfo {year} {2007})},\ \Eprint {http://arxiv.org/abs/arXiv:0706.4437}
  {arXiv:0706.4437} \BibitemShut {NoStop}%
\bibitem [{\citenamefont {Babak}\ \emph {et~al.}(2006)\citenamefont {Babak},
  \citenamefont {Balasubramanian}, \citenamefont {Churches}, \citenamefont
  {Cokelaer},\ and\ \citenamefont {Sathyaprakash}}]{BBCCS:2006}%
  \BibitemOpen
  \bibfield  {author} {\bibinfo {author} {\bibfnamefont {S.}~\bibnamefont
  {Babak}}, \bibinfo {author} {\bibfnamefont {R.}~\bibnamefont
  {Balasubramanian}}, \bibinfo {author} {\bibfnamefont {D.}~\bibnamefont
  {Churches}}, \bibinfo {author} {\bibfnamefont {T.}~\bibnamefont {Cokelaer}},
  \ and\ \bibinfo {author} {\bibfnamefont {B.~S.}\ \bibnamefont
  {Sathyaprakash}},\ }\href@noop {} {\bibfield  {journal} {\bibinfo  {journal}
  {Class.\ Quant.\ Grav.}\ }\textbf {\bibinfo {volume} {23}},\ \bibinfo {pages}
  {5477} (\bibinfo {year} {2006})},\ \Eprint
  {http://arxiv.org/abs/gr-qc/0604037} {gr-qc/0604037} \BibitemShut {NoStop}%
\bibitem [{\citenamefont {Owen}\ and\ \citenamefont
  {Sathyaprakash}(1999)}]{Owen:1998dk}%
  \BibitemOpen
  \bibfield  {author} {\bibinfo {author} {\bibfnamefont {B.~J.}\ \bibnamefont
  {Owen}}\ and\ \bibinfo {author} {\bibfnamefont {B.~S.}\ \bibnamefont
  {Sathyaprakash}},\ }\href@noop {} {\bibfield  {journal} {\bibinfo  {journal}
  {Phys.~Rev.~D}\ }\textbf {\bibinfo {volume} {60}},\ \bibinfo {pages} {022002}
  (\bibinfo {year} {1999})}\BibitemShut {NoStop}%
\bibitem [{\citenamefont {Owen}(1996)}]{Owen:1995tm}%
  \BibitemOpen
  \bibfield  {author} {\bibinfo {author} {\bibfnamefont {B.~J.}\ \bibnamefont
  {Owen}},\ }\href@noop {} {\bibfield  {journal} {\bibinfo  {journal}
  {Phys.~Rev.~D}\ }\textbf {\bibinfo {volume} {53}},\ \bibinfo {pages} {6749}
  (\bibinfo {year} {1996})}\BibitemShut {NoStop}%
\bibitem [{\citenamefont {Balasubramanian}\ \emph {et~al.}(1996)\citenamefont
  {Balasubramanian}, \citenamefont {Sathyaprakash},\ and\ \citenamefont
  {Dhurandhar}}]{Balasubramanian:1995bm}%
  \BibitemOpen
  \bibfield  {author} {\bibinfo {author} {\bibfnamefont {R.}~\bibnamefont
  {Balasubramanian}}, \bibinfo {author} {\bibfnamefont {B.~S.}\ \bibnamefont
  {Sathyaprakash}}, \ and\ \bibinfo {author} {\bibfnamefont {S.~V.}\
  \bibnamefont {Dhurandhar}},\ }\href {\doibase 10.1103/PhysRevD.53.3033}
  {\bibfield  {journal} {\bibinfo  {journal} {Phys.~Rev.~D}\ }\textbf {\bibinfo
  {volume} {53}},\ \bibinfo {pages} {3033} (\bibinfo {year} {1996})},\ \Eprint
  {http://arxiv.org/abs/gr-qc/9508011} {arXiv:gr-qc/9508011} \BibitemShut
  {NoStop}%
\bibitem [{\citenamefont {Dhurandhar}\ and\ \citenamefont
  {Sathyaprakash}(1994)}]{Dhurandhar:1992mw}%
  \BibitemOpen
  \bibfield  {author} {\bibinfo {author} {\bibfnamefont {S.~V.}\ \bibnamefont
  {Dhurandhar}}\ and\ \bibinfo {author} {\bibfnamefont {B.~S.}\ \bibnamefont
  {Sathyaprakash}},\ }\href@noop {} {\bibfield  {journal} {\bibinfo  {journal}
  {Phys. Rev.}\ }\textbf {\bibinfo {volume} {D49}},\ \bibinfo {pages} {1707}
  (\bibinfo {year} {1994})}\BibitemShut {NoStop}%
\bibitem [{\citenamefont {Sathyaprakash}\ and\ \citenamefont
  {Dhurandhar}(1991)}]{SathyaDhurandhar:1991}%
  \BibitemOpen
  \bibfield  {author} {\bibinfo {author} {\bibfnamefont {B.~S.}\ \bibnamefont
  {Sathyaprakash}}\ and\ \bibinfo {author} {\bibfnamefont {S.~V.}\ \bibnamefont
  {Dhurandhar}},\ }\href@noop {} {\bibfield  {journal} {\bibinfo  {journal}
  {Phys. Rev. D}\ }\textbf {\bibinfo {volume} {44}},\ \bibinfo {pages} {3819}
  (\bibinfo {year} {1991})}\BibitemShut {NoStop}%
\bibitem [{\citenamefont {Sathyaprakash}(1994)}]{Sathyaprakash:1994nj}%
  \BibitemOpen
  \bibfield  {author} {\bibinfo {author} {\bibfnamefont {B.}~\bibnamefont
  {Sathyaprakash}},\ }\href {\doibase 10.1103/PhysRevD.50.R7111} {\bibfield
  {journal} {\bibinfo  {journal} {Phys.Rev.}\ }\textbf {\bibinfo {volume}
  {D50}},\ \bibinfo {pages} {7111} (\bibinfo {year} {1994})},\ \Eprint
  {http://arxiv.org/abs/gr-qc/9411043} {arXiv:gr-qc/9411043 [gr-qc]}
  \BibitemShut {NoStop}%
\bibitem [{\citenamefont {Droz}\ \emph {et~al.}(1999)\citenamefont {Droz},
  \citenamefont {Knapp}, \citenamefont {Poisson},\ and\ \citenamefont
  {Owen}}]{Droz:1999qx}%
  \BibitemOpen
  \bibfield  {author} {\bibinfo {author} {\bibfnamefont {S.}~\bibnamefont
  {Droz}}, \bibinfo {author} {\bibfnamefont {D.~J.}\ \bibnamefont {Knapp}},
  \bibinfo {author} {\bibfnamefont {E.}~\bibnamefont {Poisson}}, \ and\
  \bibinfo {author} {\bibfnamefont {B.~J.}\ \bibnamefont {Owen}},\ }\href@noop
  {} {\bibfield  {journal} {\bibinfo  {journal} {Phys.~Rev.~D}\ }\textbf
  {\bibinfo {volume} {59}},\ \bibinfo {pages} {124016} (\bibinfo {year}
  {1999})}\BibitemShut {NoStop}%
\bibitem [{FFT()}]{FFTW}%
  \BibitemOpen
  \href@noop {} {\enquote {\bibinfo {title} {{FFTW - Fastest Fourier Transform
  in the West}},}\ }\bibinfo {howpublished}
  {\url{http://www.fftw.org/}}\BibitemShut {NoStop}%
\bibitem [{\citenamefont {Sengupta}\ \emph {et~al.}(2006)\citenamefont
  {Sengupta}, \citenamefont {Gupchup},\ and\ \citenamefont
  {Robinson}}]{SenguptaTrigScan:2008}%
  \BibitemOpen
  \bibfield  {author} {\bibinfo {author} {\bibfnamefont {A.~S.}\ \bibnamefont
  {Sengupta}}, \bibinfo {author} {\bibfnamefont {J.~A.}\ \bibnamefont
  {Gupchup}}, \ and\ \bibinfo {author} {\bibfnamefont {C.~A.~K.}\ \bibnamefont
  {Robinson}},\ }\href
  {http://www.ego-gw.it/ILIAS-GW/documents/WG2CARDIFF06/Anand_talk.pdf} {\
  (\bibinfo {year} {2006})}\BibitemShut {NoStop}%
\bibitem [{\citenamefont {Capano}(2012)}]{Capano:2012}%
  \BibitemOpen
  \bibfield  {author} {\bibinfo {author} {\bibfnamefont {C.}~\bibnamefont
  {Capano}},\ }\emph {\bibinfo {title} {Searching for Gravitational Waves from
  Compact Binary Coalescence using LIGO and Virgo Data}},\ \href
  {https://gwic.ligo.org/thesisprize/2011/capano_thesis.pdf} {Ph.D. thesis},\
  \bibinfo  {school} {Syracuse Univerisity} (\bibinfo {year}
  {2012})\BibitemShut {NoStop}%
\bibitem [{\citenamefont {Robinson}\ \emph {et~al.}(2008)\citenamefont
  {Robinson}, \citenamefont {Sathyaprakash},\ and\ \citenamefont
  {Sengupta}}]{Robinson:2008}%
  \BibitemOpen
  \bibfield  {author} {\bibinfo {author} {\bibfnamefont {C.~A.~K.}\
  \bibnamefont {Robinson}}, \bibinfo {author} {\bibfnamefont {B.~S.}\
  \bibnamefont {Sathyaprakash}}, \ and\ \bibinfo {author} {\bibfnamefont
  {A.~S.}\ \bibnamefont {Sengupta}},\ }\href {\doibase
  10.1103/PhysRevD.78.062002} {\bibfield  {journal} {\bibinfo  {journal}
  {Phys.~Rev.~D}\ }\textbf {\bibinfo {volume} {78}},\ \bibinfo {eid} {062002}
  (\bibinfo {year} {2008})}\BibitemShut {NoStop}%
\bibitem [{\citenamefont {Schutz}(1989)}]{Schutz:1989cu}%
  \BibitemOpen
  \bibfield  {author} {\bibinfo {author} {\bibfnamefont {B.~F.}\ \bibnamefont
  {Schutz}},\ }\href@noop {} {\bibfield  {journal} {\bibinfo  {journal} {NATO
  Adv.Study Inst.Ser.C.Math.Phys.Sci.}\ }\textbf {\bibinfo {volume} {253}},\
  \bibinfo {pages} {1} (\bibinfo {year} {1989})}\BibitemShut {NoStop}%
\bibitem [{\citenamefont {Cutler}\ \emph {et~al.}(1993)\citenamefont {Cutler}
  \emph {et~al.}}]{Cutler:1992tc}%
  \BibitemOpen
  \bibfield  {author} {\bibinfo {author} {\bibfnamefont {C.}~\bibnamefont
  {Cutler}} \emph {et~al.},\ }\href@noop {} {\bibfield  {journal} {\bibinfo
  {journal} {Phys. Rev. Lett.}\ }\textbf {\bibinfo {volume} {70}},\ \bibinfo
  {pages} {2984} (\bibinfo {year} {1993})}\BibitemShut {NoStop}%
\bibitem [{\citenamefont {Babak}\ \emph {et~al.}(2005)\citenamefont {Babak},
  \citenamefont {Grote}, \citenamefont {Hewitson}, \citenamefont {L{\"u}ck},\
  and\ \citenamefont {Strain}}]{Babak:2005iz}%
  \BibitemOpen
  \bibfield  {author} {\bibinfo {author} {\bibfnamefont {S.}~\bibnamefont
  {Babak}}, \bibinfo {author} {\bibfnamefont {H.}~\bibnamefont {Grote}},
  \bibinfo {author} {\bibfnamefont {M.}~\bibnamefont {Hewitson}}, \bibinfo
  {author} {\bibfnamefont {H.}~\bibnamefont {L{\"u}ck}}, \ and\ \bibinfo
  {author} {\bibfnamefont {K.}~\bibnamefont {Strain}},\ }\href@noop {}
  {\bibfield  {journal} {\bibinfo  {journal} {Physical Review D}\ }\textbf
  {\bibinfo {volume} {72}} (\bibinfo {year} {2005})}\BibitemShut {NoStop}%
\bibitem [{\citenamefont {Buonanno}\ \emph {et~al.}(2009)\citenamefont
  {Buonanno}, \citenamefont {Iyer}, \citenamefont {Ochsner}, \citenamefont
  {Pan},\ and\ \citenamefont
  {Sathyaprakash}}]{BuonannoIyerOchsnerYiSathya2009}%
  \BibitemOpen
  \bibfield  {author} {\bibinfo {author} {\bibfnamefont {A.}~\bibnamefont
  {Buonanno}}, \bibinfo {author} {\bibfnamefont {B.~R.}\ \bibnamefont {Iyer}},
  \bibinfo {author} {\bibfnamefont {E.}~\bibnamefont {Ochsner}}, \bibinfo
  {author} {\bibfnamefont {Y.}~\bibnamefont {Pan}}, \ and\ \bibinfo {author}
  {\bibfnamefont {B.~S.}\ \bibnamefont {Sathyaprakash}},\ }\href {\doibase
  10.1103/PhysRevD.80.084043} {\bibfield  {journal} {\bibinfo  {journal} {Phys.
  Rev. D}\ }\textbf {\bibinfo {volume} {80}},\ \bibinfo {pages} {084043}
  (\bibinfo {year} {2009})}\BibitemShut {NoStop}%
\bibitem [{\citenamefont {Abadie}\ \emph
  {et~al.}(2010{\natexlab{b}})\citenamefont {Abadie} \emph
  {et~al.}}]{S5calibration}%
  \BibitemOpen
  \bibfield  {author} {\bibinfo {author} {\bibfnamefont {J.}~\bibnamefont
  {Abadie}} \emph {et~al.} (\bibinfo {collaboration} {LIGO Scientific
  Collaboration}),\ }\href@noop {} {\bibfield  {journal} {\bibinfo  {journal}
  {Nuclear Instruments and Methods in Physics Research}\ }\textbf {\bibinfo
  {volume} {A624}},\ \bibinfo {pages} {223} (\bibinfo {year}
  {2010}{\natexlab{b}})},\ \Eprint {http://arxiv.org/abs/arxiv:1007.3973}
  {arxiv:1007.3973} \BibitemShut {NoStop}%
\bibitem [{\citenamefont {Hanna}(2008)}]{Hanna:2008}%
  \BibitemOpen
  \bibfield  {author} {\bibinfo {author} {\bibfnamefont {C.}~\bibnamefont
  {Hanna}},\ }\emph {\bibinfo {title} {Searching for gravitational waves from
  binary systems in non-stationary data}},\ \href
  {http://etd.lsu.edu/docs/available/etd-03272008-092832/} {Ph.D. thesis},\
  \bibinfo  {school} {Louisiana State University} (\bibinfo {year}
  {2008})\BibitemShut {NoStop}%
\bibitem [{\citenamefont {Harry}\ and\ \citenamefont
  {Fairhurst}(2011)}]{HarryFairhurst:2011}%
  \BibitemOpen
  \bibfield  {author} {\bibinfo {author} {\bibfnamefont {I.~W.}\ \bibnamefont
  {Harry}}\ and\ \bibinfo {author} {\bibfnamefont {S.}~\bibnamefont
  {Fairhurst}},\ }\href {\doibase 10.1103/PhysRevD.83.084002} {\bibfield
  {journal} {\bibinfo  {journal} {Phys. Rev. D}\ }\textbf {\bibinfo {volume}
  {83}},\ \bibinfo {pages} {084002} (\bibinfo {year} {2011})}\BibitemShut
  {NoStop}%
\bibitem [{\citenamefont {Shawhan}\ and\ \citenamefont
  {Ochsner}(2004)}]{Shawhan:2004cn}%
  \BibitemOpen
  \bibfield  {author} {\bibinfo {author} {\bibfnamefont {P.}~\bibnamefont
  {Shawhan}}\ and\ \bibinfo {author} {\bibfnamefont {E.}~\bibnamefont
  {Ochsner}},\ }\href@noop {} {\bibfield  {journal} {\bibinfo  {journal}
  {Class. Quant. Grav}\ }\textbf {\bibinfo {volume} {21}},\ \bibinfo {pages}
  {S1757} (\bibinfo {year} {2004})}\BibitemShut {NoStop}%
\bibitem [{\citenamefont {Rodr\'iguez}(2007)}]{Rodriguez:2007}%
  \BibitemOpen
  \bibfield  {author} {\bibinfo {author} {\bibfnamefont {A.}~\bibnamefont
  {Rodr\'iguez}},\ }\emph {\bibinfo {title} {Reducing false alarms in searches
  for gravitational waves from coalescing binary systems}},\ \href@noop {}
  {Master's thesis},\ \bibinfo  {school} {Louisiana State University} (\bibinfo
  {year} {2007}),\ \Eprint {http://arxiv.org/abs/arXiv:0802.1376}
  {arXiv:0802.1376} \BibitemShut {NoStop}%
\bibitem [{\citenamefont {Slutsky}\ \emph {et~al.}(2010)\citenamefont {Slutsky}
  \emph {et~al.}}]{Slutsky:2010ff}%
  \BibitemOpen
  \bibfield  {author} {\bibinfo {author} {\bibfnamefont {J.}~\bibnamefont
  {Slutsky}} \emph {et~al.},\ }\href {\doibase 10.1088/0264-9381/27/16/165023}
  {\bibfield  {journal} {\bibinfo  {journal} {Class.\ Quant.\ Grav.}\ }\textbf
  {\bibinfo {volume} {27}},\ \bibinfo {pages} {165023} (\bibinfo {year}
  {2010})},\ \Eprint {http://arxiv.org/abs/1004.0998} {arXiv:1004.0998 [gr-qc]}
  \BibitemShut {NoStop}%
\bibitem [{\citenamefont {{Christensen (for the LIGO Scientific Collaboration
  and the Virgo Collaboration)}}(2010)}]{Christensen:2010}%
  \BibitemOpen
  \bibfield  {author} {\bibinfo {author} {\bibfnamefont {N.}~\bibnamefont
  {{Christensen (for the LIGO Scientific Collaboration and the Virgo
  Collaboration)}}},\ }\href@noop {} {\bibfield  {journal} {\bibinfo  {journal}
  {Class.\ Quant.\ Grav.}\ }\textbf {\bibinfo {volume} {27}},\ \bibinfo {pages}
  {194010} (\bibinfo {year} {2010})}\BibitemShut {NoStop}%
\bibitem [{\citenamefont {Aasi}\ \emph {et~al.}(2012)\citenamefont {Aasi} \emph
  {et~al.}}]{Aasi:2012wd}%
  \BibitemOpen
  \bibfield  {author} {\bibinfo {author} {\bibfnamefont {J.}~\bibnamefont
  {Aasi}} \emph {et~al.} (\bibinfo {collaboration} {VIRGO Collaboration}),\
  }\href@noop {} {\  (\bibinfo {year} {2012})},\ \Eprint
  {http://arxiv.org/abs/1203.5613} {arXiv:1203.5613 [gr-qc]} \BibitemShut
  {NoStop}%
\bibitem [{\citenamefont {MacLeod}\ \emph {et~al.}(2012)\citenamefont {MacLeod}
  \emph {et~al.}}]{SeisVeto}%
  \BibitemOpen
  \bibfield  {author} {\bibinfo {author} {\bibfnamefont {D.}~\bibnamefont
  {MacLeod}} \emph {et~al.},\ }\href@noop {} {\bibfield  {journal} {\bibinfo
  {journal} {Class. Quant. Grav.}\ }\textbf {\bibinfo {volume} {29}},\ \bibinfo
  {pages} {055006} (\bibinfo {year} {2012})},\ \Eprint
  {http://arxiv.org/abs/1108.0312} {arXiv:1108.0312 [gr-qc]} \BibitemShut
  {NoStop}%
\bibitem [{\citenamefont {Ito}()}]{glitchmon}%
  \BibitemOpen
  \bibfield  {author} {\bibinfo {author} {\bibfnamefont {M.}~\bibnamefont
  {Ito}},\ }\href@noop {} {\enquote {\bibinfo {title} {glitchmon: A {DMT}
  monitor to look for transient signals in selected channels},}\ }\bibinfo
  {note} {Developed using the {LIGO} Data Monitoring Tool (DMT)
  library}\BibitemShut {NoStop}%
\bibitem [{\citenamefont {Pekowsky}(2012)}]{Pekowsky:2012}%
  \BibitemOpen
  \bibfield  {author} {\bibinfo {author} {\bibfnamefont {L.}~\bibnamefont
  {Pekowsky}},\ }\emph {\bibinfo {title} {Characterization of Enhanced
  Interferometric Gravitational-wave Detectors and Studies of Numeric
  Simulations for Compact-binary Coalescences}},\ \href
  {https://gwic.ligo.org/thesisprize/2011/pekowsky_thesis.pdf} {Ph.D. thesis},\
  \bibinfo  {school} {Syracuse Univerisity} (\bibinfo {year}
  {2012})\BibitemShut {NoStop}%
\bibitem [{\citenamefont {Abadie}\ \emph {et~al.}(2011)\citenamefont {Abadie}
  \emph {et~al.}}]{Collaboration:S5HighMass}%
  \BibitemOpen
  \bibfield  {author} {\bibinfo {author} {\bibfnamefont {J.}~\bibnamefont
  {Abadie}} \emph {et~al.} (\bibinfo {collaboration} {LIGO Scientific
  Collaboration and Virgo Collaboration}),\ }\href {\doibase
  10.1103/PhysRevD.83.122005} {\bibfield  {journal} {\bibinfo  {journal} {Phys.
  Rev. D}\ }\textbf {\bibinfo {volume} {83}},\ \bibinfo {pages} {122005}
  (\bibinfo {year} {2011})},\ \Eprint {http://arxiv.org/abs/arXiv:1102.3781}
  {arXiv:1102.3781} \BibitemShut {NoStop}%
\bibitem [{\citenamefont {{Amaldi}}\ \emph {et~al.}(1989)\citenamefont
  {{Amaldi}}, \citenamefont {{Aguiar}}, \citenamefont {{Bassan}}, \citenamefont
  {{Bonifazi}}, \citenamefont {{Carelli}}, \citenamefont {{Castellano}},
  \citenamefont {{Cavallari}}, \citenamefont {{Coccia}}, \citenamefont
  {{Cosmelli}}, \citenamefont {{Fairbank}}, \citenamefont {{Frasca}},
  \citenamefont {{Foglietti}}, \citenamefont {{Habel}}, \citenamefont
  {{Hamilton}}, \citenamefont {{Henderson}}, \citenamefont {{Johnson}},
  \citenamefont {{Lane}}, \citenamefont {{Mann}}, \citenamefont {{McAshan}},
  \citenamefont {{Michelson}}, \citenamefont {{Modena}}, \citenamefont
  {{Pallottino}}, \citenamefont {{Pizzela}}, \citenamefont {{Price}},
  \citenamefont {{Rapagnani}}, \citenamefont {{Ricci}}, \citenamefont
  {{Solomonson}}, \citenamefont {{Stevenson}}, \citenamefont {{Taber}},\ and\
  \citenamefont {{Xu}}}]{Amaldi:1989}%
  \BibitemOpen
  \bibfield  {author} {\bibinfo {author} {\bibfnamefont {E.}~\bibnamefont
  {{Amaldi}}}, \bibinfo {author} {\bibfnamefont {O.}~\bibnamefont {{Aguiar}}},
  \bibinfo {author} {\bibfnamefont {M.}~\bibnamefont {{Bassan}}}, \bibinfo
  {author} {\bibfnamefont {P.}~\bibnamefont {{Bonifazi}}}, \bibinfo {author}
  {\bibfnamefont {P.}~\bibnamefont {{Carelli}}}, \bibinfo {author}
  {\bibfnamefont {M.~G.}\ \bibnamefont {{Castellano}}}, \bibinfo {author}
  {\bibfnamefont {G.}~\bibnamefont {{Cavallari}}}, \bibinfo {author}
  {\bibfnamefont {E.}~\bibnamefont {{Coccia}}}, \bibinfo {author}
  {\bibfnamefont {C.}~\bibnamefont {{Cosmelli}}}, \bibinfo {author}
  {\bibfnamefont {W.~M.}\ \bibnamefont {{Fairbank}}}, \bibinfo {author}
  {\bibfnamefont {S.}~\bibnamefont {{Frasca}}}, \bibinfo {author}
  {\bibfnamefont {V.}~\bibnamefont {{Foglietti}}}, \bibinfo {author}
  {\bibfnamefont {R.}~\bibnamefont {{Habel}}}, \bibinfo {author} {\bibfnamefont
  {W.~O.}\ \bibnamefont {{Hamilton}}}, \bibinfo {author} {\bibfnamefont
  {J.}~\bibnamefont {{Henderson}}}, \bibinfo {author} {\bibfnamefont
  {W.}~\bibnamefont {{Johnson}}}, \bibinfo {author} {\bibfnamefont {K.~R.}\
  \bibnamefont {{Lane}}}, \bibinfo {author} {\bibfnamefont {A.~G.}\
  \bibnamefont {{Mann}}}, \bibinfo {author} {\bibfnamefont {M.~S.}\
  \bibnamefont {{McAshan}}}, \bibinfo {author} {\bibfnamefont {P.~F.}\
  \bibnamefont {{Michelson}}}, \bibinfo {author} {\bibfnamefont
  {I.}~\bibnamefont {{Modena}}}, \bibinfo {author} {\bibfnamefont {G.~V.}\
  \bibnamefont {{Pallottino}}}, \bibinfo {author} {\bibfnamefont
  {G.}~\bibnamefont {{Pizzela}}}, \bibinfo {author} {\bibfnamefont {J.~C.}\
  \bibnamefont {{Price}}}, \bibinfo {author} {\bibfnamefont {R.}~\bibnamefont
  {{Rapagnani}}}, \bibinfo {author} {\bibfnamefont {F.}~\bibnamefont
  {{Ricci}}}, \bibinfo {author} {\bibfnamefont {N.}~\bibnamefont
  {{Solomonson}}}, \bibinfo {author} {\bibfnamefont {T.~R.}\ \bibnamefont
  {{Stevenson}}}, \bibinfo {author} {\bibfnamefont {R.~C.}\ \bibnamefont
  {{Taber}}}, \ and\ \bibinfo {author} {\bibfnamefont {B.~X.}\ \bibnamefont
  {{Xu}}},\ }\href@noop {} {\bibfield  {journal} {\bibinfo  {journal} {Astron.
  Astrophys.}\ }\textbf {\bibinfo {volume} {216}},\ \bibinfo {pages} {325}
  (\bibinfo {year} {1989})}\BibitemShut {NoStop}%
\bibitem [{\citenamefont {Was}\ \emph {et~al.}(2010)\citenamefont {Was} \emph
  {et~al.}}]{Was:2009vh}%
  \BibitemOpen
  \bibfield  {author} {\bibinfo {author} {\bibfnamefont {M.}~\bibnamefont
  {Was}} \emph {et~al.},\ }\href {\doibase 10.1088/0264-9381/27/1/015005}
  {\bibfield  {journal} {\bibinfo  {journal} {Class. Quant. Grav.}\ }\textbf
  {\bibinfo {volume} {27}},\ \bibinfo {pages} {015005} (\bibinfo {year}
  {2010})},\ \Eprint {http://arxiv.org/abs/0906.2120} {arXiv:0906.2120 [gr-qc]}
  \BibitemShut {NoStop}%
\bibitem [{\citenamefont {Dent}\ \emph {et~al.}()\citenamefont {Dent} \emph
  {et~al.}}]{Dent:2012}%
  \BibitemOpen
  \bibfield  {author} {\bibinfo {author} {\bibfnamefont {T.}~\bibnamefont
  {Dent}} \emph {et~al.},\ }\href@noop {} {\bibinfo  {journal} {In
  Preparation}\ }\BibitemShut {NoStop}%
\bibitem [{\citenamefont {Brown (for~the {LIGO}
  Scientific~Collaboration)}(2004)}]{Brown:2004}%
  \BibitemOpen
\bibfield  {journal} {  }\bibfield  {author} {\bibinfo {author} {\bibfnamefont
  {D.~A.}\ \bibnamefont {Brown (for~the {LIGO} Scientific~Collaboration)}},\
  }\href@noop {} {\bibfield  {journal} {\bibinfo  {journal} {Class. Quant.
  Grav.}\ }\textbf {\bibinfo {volume} {21}},\ \bibinfo {pages} {S797} (\bibinfo
  {year} {2004})}\BibitemShut {NoStop}%
\bibitem [{\citenamefont {Mandel}\ and\ \citenamefont
  {O'Shaughnessy}(2010)}]{Mandel:2009nx}%
  \BibitemOpen
  \bibfield  {author} {\bibinfo {author} {\bibfnamefont {I.}~\bibnamefont
  {Mandel}}\ and\ \bibinfo {author} {\bibfnamefont {R.}~\bibnamefont
  {O'Shaughnessy}},\ }\href {\doibase 10.1088/0264-9381/27/11/114007}
  {\bibfield  {journal} {\bibinfo  {journal} {Class. Quant. Grav.}\ }\textbf
  {\bibinfo {volume} {27}},\ \bibinfo {pages} {114007} (\bibinfo {year}
  {2010})},\ \Eprint {http://arxiv.org/abs/0912.1074} {arXiv:0912.1074
  [astro-ph.HE]} \BibitemShut {NoStop}%
\bibitem [{\citenamefont {Abadie}\ \emph
  {et~al.}(2010{\natexlab{c}})\citenamefont {Abadie} \emph {et~al.}}]{PSD:S5}%
  \BibitemOpen
  \bibfield  {author} {\bibinfo {author} {\bibfnamefont {J.}~\bibnamefont
  {Abadie}} \emph {et~al.} (\bibinfo {collaboration} {LIGO Scientic
  Collaboration and the Virgo Collaboration}),\ }\href@noop {} {\  (\bibinfo
  {year} {2010}{\natexlab{c}})},\ \Eprint {http://arxiv.org/abs/1003.2481}
  {arXiv:1003.2481 [gr-qc]} \BibitemShut {NoStop}%
\bibitem [{\citenamefont {Abadie}\ \emph
  {et~al.}(2012{\natexlab{b}})\citenamefont {Abadie} \emph
  {et~al.}}]{Collaboration:2012wu}%
  \BibitemOpen
  \bibfield  {author} {\bibinfo {author} {\bibfnamefont {J.}~\bibnamefont
  {Abadie}} \emph {et~al.} (\bibinfo {collaboration} {LIGO Scientific
  Collaboration and Virgo Collaboration}),\ }\href@noop {} {\  (\bibinfo {year}
  {2012}{\natexlab{b}})},\ \Eprint {http://arxiv.org/abs/1203.2674}
  {arXiv:1203.2674 [gr-qc]} \BibitemShut {NoStop}%
\bibitem [{\citenamefont {Brady}\ \emph {et~al.}(2004)\citenamefont {Brady},
  \citenamefont {Creighton},\ and\ \citenamefont {Wiseman}}]{loudestGWDAW03}%
  \BibitemOpen
  \bibfield  {author} {\bibinfo {author} {\bibfnamefont {P.~R.}\ \bibnamefont
  {Brady}}, \bibinfo {author} {\bibfnamefont {J.~D.~E.}\ \bibnamefont
  {Creighton}}, \ and\ \bibinfo {author} {\bibfnamefont {A.~G.}\ \bibnamefont
  {Wiseman}},\ }\href@noop {} {\bibfield  {journal} {\bibinfo  {journal}
  {Class. Quant. Grav.}\ }\textbf {\bibinfo {volume} {21}},\ \bibinfo {pages}
  {S1775} (\bibinfo {year} {2004})}\BibitemShut {NoStop}%
\bibitem [{\citenamefont {Biswas}\ \emph {et~al.}(2009)\citenamefont {Biswas},
  \citenamefont {Brady}, \citenamefont {Creighton},\ and\ \citenamefont
  {Fairhurst}}]{Biswas:2007ni}%
  \BibitemOpen
  \bibfield  {author} {\bibinfo {author} {\bibfnamefont {R.}~\bibnamefont
  {Biswas}}, \bibinfo {author} {\bibfnamefont {P.~R.}\ \bibnamefont {Brady}},
  \bibinfo {author} {\bibfnamefont {J.~D.~E.}\ \bibnamefont {Creighton}}, \
  and\ \bibinfo {author} {\bibfnamefont {S.}~\bibnamefont {Fairhurst}},\ }\href
  {\doibase 10.1088/0264-9381/26/17/175009} {\bibfield  {journal} {\bibinfo
  {journal} {Class. Quant. Grav.}\ }\textbf {\bibinfo {volume} {26}},\ \bibinfo
  {pages} {175009} (\bibinfo {year} {2009})},\ \Eprint
  {http://arxiv.org/abs/0710.0465} {arXiv:0710.0465 [gr-qc]} \BibitemShut
  {NoStop}%
\bibitem [{\citenamefont {Keppel}(2009)}]{keppel:thesis}%
  \BibitemOpen
  \bibfield  {author} {\bibinfo {author} {\bibfnamefont {D.}~\bibnamefont
  {Keppel}},\ }\emph {\bibinfo {title} {Signatures and dynamics of compact
  binary coalescences and a search in LIGO's S5 data}},\ \href
  {http://resolver.caltech.edu/CaltechETD:etd-05202009-115750} {Ph.D. thesis},\
  \bibinfo  {school} {Caltech}, \bibinfo {address} {Pasadena, CA} (\bibinfo
  {year} {2009})\BibitemShut {NoStop}%
\bibitem [{\citenamefont {Phinney}(1991)}]{Phinney:1991ei}%
  \BibitemOpen
  \bibfield  {author} {\bibinfo {author} {\bibfnamefont {E.~S.}\ \bibnamefont
  {Phinney}},\ }\href@noop {} {\bibfield  {journal} {\bibinfo  {journal}
  {Astrophys. J.}\ }\textbf {\bibinfo {volume} {380}},\ \bibinfo {pages} {L17}
  (\bibinfo {year} {1991})}\BibitemShut {NoStop}%
\bibitem [{\citenamefont {Abadie}\ \emph
  {et~al.}(2010{\natexlab{d}})\citenamefont {Abadie} \emph
  {et~al.}}]{ratesdoc}%
  \BibitemOpen
  \bibfield  {author} {\bibinfo {author} {\bibfnamefont {J.}~\bibnamefont
  {Abadie}} \emph {et~al.} (\bibinfo {collaboration} {LIGO Scientific
  Collaboration and Virgo Collaboration}),\ }\href@noop {} {\bibfield
  {journal} {\bibinfo  {journal} {Class. Quant. Grav.}\ }\textbf {\bibinfo
  {volume} {27}},\ \bibinfo {pages} {173001} (\bibinfo {year}
  {2010}{\natexlab{d}})}\BibitemShut {NoStop}%
\bibitem [{\citenamefont {{Finn}}\ and\ \citenamefont
  {{Chernoff}}(1993)}]{FinnChernoff:1993}%
  \BibitemOpen
  \bibfield  {author} {\bibinfo {author} {\bibfnamefont {L.~S.}\ \bibnamefont
  {{Finn}}}\ and\ \bibinfo {author} {\bibfnamefont {D.~F.}\ \bibnamefont
  {{Chernoff}}},\ }\href {\doibase 10.1103/PhysRevD.47.2198} {\bibfield
  {journal} {\bibinfo  {journal} {\prd}\ }\textbf {\bibinfo {volume} {47}},\
  \bibinfo {pages} {2198} (\bibinfo {year} {1993})},\ \Eprint
  {http://arxiv.org/abs/arXiv:gr-qc/9301003} {arXiv:gr-qc/9301003} \BibitemShut
  {NoStop}%
\bibitem [{\citenamefont {Pai}\ \emph {et~al.}(2001)\citenamefont {Pai},
  \citenamefont {Dhurandhar},\ and\ \citenamefont
  {Bose}}]{PaiDhurandharBose2001}%
  \BibitemOpen
  \bibfield  {author} {\bibinfo {author} {\bibfnamefont {A.}~\bibnamefont
  {Pai}}, \bibinfo {author} {\bibfnamefont {S.}~\bibnamefont {Dhurandhar}}, \
  and\ \bibinfo {author} {\bibfnamefont {S.}~\bibnamefont {Bose}},\ }\href
  {\doibase 10.1103/PhysRevD.64.042004} {\bibfield  {journal} {\bibinfo
  {journal} {Phys.~Rev.~D}\ }\textbf {\bibinfo {volume} {64}},\ \bibinfo
  {pages} {042004} (\bibinfo {year} {2001})},\ \Eprint
  {http://arxiv.org/abs/gr-qc/0009078} {arXiv:gr-qc/0009078} \BibitemShut
  {NoStop}%
\bibitem [{\citenamefont {Abadie}\ \emph
  {et~al.}(2010{\natexlab{e}})\citenamefont {Abadie} \emph
  {et~al.}}]{S5VSR1Burst}%
  \BibitemOpen
  \bibfield  {author} {\bibinfo {author} {\bibfnamefont {J.}~\bibnamefont
  {Abadie}} \emph {et~al.} (\bibinfo {collaboration} {LIGO Scientific
  Collaboration and Virgo Collaboration}),\ }\href@noop {} {\bibfield
  {journal} {\bibinfo  {journal} {Phys. Rev. D}\ }\textbf {\bibinfo {volume}
  {81}},\ \bibinfo {pages} {102001} (\bibinfo {year}
  {2010}{\natexlab{e}})}\BibitemShut {NoStop}%
\bibitem [{\citenamefont {Briggs}\ \emph {et~al.}(2012)\citenamefont {Briggs}
  \emph {et~al.}}]{Briggs:2012ce}%
  \BibitemOpen
  \bibfield  {author} {\bibinfo {author} {\bibfnamefont {M.}~\bibnamefont
  {Briggs}} \emph {et~al.} (\bibinfo {collaboration} {LIGO Scientific
  Collaboration and Virgo Collaboration}),\ }\href@noop {} {\  (\bibinfo {year}
  {2012})},\ \Eprint {http://arxiv.org/abs/1205.2216} {arXiv:1205.2216
  [astro-ph.HE]} \BibitemShut {NoStop}%
\bibitem [{\citenamefont {Cannon}\ \emph {et~al.}()\citenamefont {Cannon},
  \citenamefont {Hanna},\ and\ \citenamefont {Keppel}}]{Cannon:2012}%
  \BibitemOpen
  \bibfield  {author} {\bibinfo {author} {\bibfnamefont {K.}~\bibnamefont
  {Cannon}}, \bibinfo {author} {\bibfnamefont {C.}~\bibnamefont {Hanna}}, \
  and\ \bibinfo {author} {\bibfnamefont {D.}~\bibnamefont {Keppel}},\
  }\href@noop {} {\bibinfo  {journal} {In Preparation}\ }\BibitemShut {NoStop}%
\bibitem [{\citenamefont {Messenger}\ and\ \citenamefont
  {Veitch}()}]{Messenger:2012}%
  \BibitemOpen
\bibfield  {journal} {  }\bibfield  {author} {\bibinfo {author} {\bibfnamefont
  {C.}~\bibnamefont {Messenger}}\ and\ \bibinfo {author} {\bibfnamefont
  {J.}~\bibnamefont {Veitch}},\ }\href@noop {} {\bibinfo  {journal} {In
  Preparation}\ }\BibitemShut {NoStop}%
\bibitem [{\citenamefont {Marion}\ \emph {et~al.}(2004)\citenamefont {Marion}
  \emph {et~al.}}]{Marion:2004}%
  \BibitemOpen
\bibfield  {journal} {  }\bibfield  {author} {\bibinfo {author} {\bibfnamefont
  {F.}~\bibnamefont {Marion}} \emph {et~al.} (\bibinfo {collaboration} {Virgo
  Collaboration}),\ }in\ \href@noop {} {\emph {\bibinfo {booktitle}
  {Proceedings of the Rencontres de Moriond 2003}}}\ (\bibinfo {year}
  {2004})\BibitemShut {NoStop}%
\bibitem [{\citenamefont {{Cannon}}\ \emph {et~al.}(2010)\citenamefont
  {{Cannon}}, \citenamefont {{Chapman}}, \citenamefont {{Hanna}}, \citenamefont
  {{Keppel}}, \citenamefont {{Searle}},\ and\ \citenamefont
  {{Weinstein}}}]{Cannon:2010}%
  \BibitemOpen
  \bibfield  {author} {\bibinfo {author} {\bibfnamefont {K.}~\bibnamefont
  {{Cannon}}}, \bibinfo {author} {\bibfnamefont {A.}~\bibnamefont {{Chapman}}},
  \bibinfo {author} {\bibfnamefont {C.}~\bibnamefont {{Hanna}}}, \bibinfo
  {author} {\bibfnamefont {D.}~\bibnamefont {{Keppel}}}, \bibinfo {author}
  {\bibfnamefont {A.~C.}\ \bibnamefont {{Searle}}}, \ and\ \bibinfo {author}
  {\bibfnamefont {A.~J.}\ \bibnamefont {{Weinstein}}},\ }\href {\doibase
  10.1103/PhysRevD.82.044025} {\bibfield  {journal} {\bibinfo  {journal}
  {\prd}\ }\textbf {\bibinfo {volume} {82}},\ \bibinfo {eid} {044025} (\bibinfo
  {year} {2010})},\ \Eprint {http://arxiv.org/abs/1005.0012} {arXiv:1005.0012
  [gr-qc]} \BibitemShut {NoStop}%
\bibitem [{\citenamefont {Cannon}\ \emph
  {et~al.}(2011{\natexlab{a}})\citenamefont {Cannon}, \citenamefont {Hanna},
  \citenamefont {Keppel},\ and\ \citenamefont {Searle}}]{Cannon:2011tb}%
  \BibitemOpen
  \bibfield  {author} {\bibinfo {author} {\bibfnamefont {K.}~\bibnamefont
  {Cannon}}, \bibinfo {author} {\bibfnamefont {C.}~\bibnamefont {Hanna}},
  \bibinfo {author} {\bibfnamefont {D.}~\bibnamefont {Keppel}}, \ and\ \bibinfo
  {author} {\bibfnamefont {A.~C.}\ \bibnamefont {Searle}},\ }\href {\doibase
  10.1103/PhysRevD.83.084053} {\bibfield  {journal} {\bibinfo  {journal}
  {Phys.Rev.}\ }\textbf {\bibinfo {volume} {D83}},\ \bibinfo {pages} {084053}
  (\bibinfo {year} {2011}{\natexlab{a}})},\ \Eprint
  {http://arxiv.org/abs/1101.0584} {arXiv:1101.0584 [physics.data-an]}
  \BibitemShut {NoStop}%
\bibitem [{\citenamefont {Cannon}\ \emph
  {et~al.}(2011{\natexlab{b}})\citenamefont {Cannon}, \citenamefont {Hanna},\
  and\ \citenamefont {Keppel}}]{Cannon:2011xk}%
  \BibitemOpen
  \bibfield  {author} {\bibinfo {author} {\bibfnamefont {K.}~\bibnamefont
  {Cannon}}, \bibinfo {author} {\bibfnamefont {C.}~\bibnamefont {Hanna}}, \
  and\ \bibinfo {author} {\bibfnamefont {D.}~\bibnamefont {Keppel}},\ }\href
  {\doibase 10.1103/PhysRevD.84.084003} {\bibfield  {journal} {\bibinfo
  {journal} {Phys.Rev.}\ }\textbf {\bibinfo {volume} {D84}},\ \bibinfo {pages}
  {084003} (\bibinfo {year} {2011}{\natexlab{b}})},\ \Eprint
  {http://arxiv.org/abs/1101.4939} {arXiv:1101.4939 [gr-qc]} \BibitemShut
  {NoStop}%
\bibitem [{\citenamefont {Cannon}\ \emph {et~al.}(2012)\citenamefont {Cannon},
  \citenamefont {Cariou}, \citenamefont {Chapman}, \citenamefont
  {Crispin-Ortuzar}, \citenamefont {Fotopoulos} \emph
  {et~al.}}]{Cannon:2011vi}%
  \BibitemOpen
  \bibfield  {author} {\bibinfo {author} {\bibfnamefont {K.}~\bibnamefont
  {Cannon}}, \bibinfo {author} {\bibfnamefont {R.}~\bibnamefont {Cariou}},
  \bibinfo {author} {\bibfnamefont {A.}~\bibnamefont {Chapman}}, \bibinfo
  {author} {\bibfnamefont {M.}~\bibnamefont {Crispin-Ortuzar}}, \bibinfo
  {author} {\bibfnamefont {N.}~\bibnamefont {Fotopoulos}},  \emph {et~al.},\
  }\href@noop {} {\bibfield  {journal} {\bibinfo  {journal} {Astrophys.J.}\
  }\textbf {\bibinfo {volume} {748}},\ \bibinfo {pages} {136} (\bibinfo {year}
  {2012})},\ \Eprint {http://arxiv.org/abs/1107.2665} {arXiv:1107.2665
  [astro-ph.IM]} \BibitemShut {NoStop}%
\bibitem [{\citenamefont {{Abadie}}\ \emph {et~al.}(2012)\citenamefont
  {{Abadie}}, \citenamefont {{Abbott}}, \citenamefont {{Abbott}}, \citenamefont
  {{Abbott}}, \citenamefont {{Abernathy}}, \citenamefont {{Accadia}},
  \citenamefont {{Acernese}}, \citenamefont {{Adams}}, \citenamefont
  {{Adhikari}}, \citenamefont {{Affeldt}},\ and\ \citenamefont
  {et~al.}}]{Virgo:2011aa}%
  \BibitemOpen
  \bibfield  {author} {\bibinfo {author} {\bibfnamefont {J.}~\bibnamefont
  {{Abadie}}}, \bibinfo {author} {\bibfnamefont {B.~P.}\ \bibnamefont
  {{Abbott}}}, \bibinfo {author} {\bibfnamefont {R.}~\bibnamefont {{Abbott}}},
  \bibinfo {author} {\bibfnamefont {T.~D.}\ \bibnamefont {{Abbott}}}, \bibinfo
  {author} {\bibfnamefont {M.}~\bibnamefont {{Abernathy}}}, \bibinfo {author}
  {\bibfnamefont {T.}~\bibnamefont {{Accadia}}}, \bibinfo {author}
  {\bibfnamefont {F.}~\bibnamefont {{Acernese}}}, \bibinfo {author}
  {\bibfnamefont {C.}~\bibnamefont {{Adams}}}, \bibinfo {author} {\bibfnamefont
  {R.}~\bibnamefont {{Adhikari}}}, \bibinfo {author} {\bibfnamefont
  {C.}~\bibnamefont {{Affeldt}}}, \ and\ \bibinfo {author} {\bibnamefont
  {et~al.}},\ }\href {\doibase 10.1051/0004-6361/201218860} {\bibfield
  {journal} {\bibinfo  {journal} {Astron Astrophys}\ }\textbf {\bibinfo
  {volume} {541}},\ \bibinfo {eid} {A155} (\bibinfo {year} {2012})}\BibitemShut
  {NoStop}%
\bibitem [{\citenamefont {Metzger}\ and\ \citenamefont
  {Berger}(2012)}]{Metzger:2011bv}%
  \BibitemOpen
  \bibfield  {author} {\bibinfo {author} {\bibfnamefont {B.}~\bibnamefont
  {Metzger}}\ and\ \bibinfo {author} {\bibfnamefont {E.}~\bibnamefont
  {Berger}},\ }\href {\doibase 10.1088/0004-637X/746/1/48} {\bibfield
  {journal} {\bibinfo  {journal} {Astrophys.J.}\ }\textbf {\bibinfo {volume}
  {746}},\ \bibinfo {pages} {48} (\bibinfo {year} {2012})},\ \Eprint
  {http://arxiv.org/abs/1108.6056} {arXiv:1108.6056 [astro-ph.HE]} \BibitemShut
  {NoStop}%
\bibitem [{\citenamefont {Ajith}\ \emph {et~al.}(2011)\citenamefont {Ajith}
  \emph {et~al.}}]{Ajith:2009bn}%
  \BibitemOpen
  \bibfield  {author} {\bibinfo {author} {\bibfnamefont {P.}~\bibnamefont
  {Ajith}} \emph {et~al.},\ }\href {\doibase 10.1103/PhysRevLett.106.241101}
  {\bibfield  {journal} {\bibinfo  {journal} {Phys. Rev. Lett.}\ }\textbf
  {\bibinfo {volume} {106}},\ \bibinfo {pages} {241101} (\bibinfo {year}
  {2011})},\ \Eprint {http://arxiv.org/abs/0909.2867} {arXiv:0909.2867 [gr-qc]}
  \BibitemShut {NoStop}%
\bibitem [{\citenamefont {Santamaria}\ \emph {et~al.}(2010)\citenamefont
  {Santamaria} \emph {et~al.}}]{Santamaria:2010}%
  \BibitemOpen
  \bibfield  {author} {\bibinfo {author} {\bibfnamefont {L.}~\bibnamefont
  {Santamaria}} \emph {et~al.},\ }\href {\doibase 10.1103/PhysRevD.82.064016}
  {\bibfield  {journal} {\bibinfo  {journal} {Phys. Rev. D}\ }\textbf {\bibinfo
  {volume} {82}},\ \bibinfo {pages} {064016} (\bibinfo {year} {2010})},\
  \Eprint {http://arxiv.org/abs/1005.3306} {arXiv:1005.3306 [gr-qc]}
  \BibitemShut {NoStop}%
\bibitem [{\citenamefont {Ajith}(2011)}]{Ajith:2011hq}%
  \BibitemOpen
  \bibfield  {author} {\bibinfo {author} {\bibfnamefont {P.}~\bibnamefont
  {Ajith}},\ }\href@noop {} {\bibfield  {journal} {\bibinfo  {journal}
  {Physical Review D}\ }\textbf {\bibinfo {volume} {84}} (\bibinfo {year}
  {2011})}\BibitemShut {NoStop}%
\bibitem [{\citenamefont {{van der Sluys}}\ \emph
  {et~al.}(2008{\natexlab{a}})\citenamefont {{van der Sluys}} \emph
  {et~al.}}]{Sluys:2008a}%
  \BibitemOpen
  \bibfield  {author} {\bibinfo {author} {\bibfnamefont {M.}~\bibnamefont {{van
  der Sluys}}} \emph {et~al.},\ }\href {\doibase
  10.1088/0264-9381/25/18/184011} {\bibfield  {journal} {\bibinfo  {journal}
  {Class. Quant. Grav}\ }\textbf {\bibinfo {volume} {25}},\ \bibinfo {pages}
  {184011} (\bibinfo {year} {2008}{\natexlab{a}})},\ \Eprint
  {http://arxiv.org/abs/0805.1689} {arXiv:0805.1689 [gr-qc]} \BibitemShut
  {NoStop}%
\bibitem [{\citenamefont {{van der Sluys}}\ \emph
  {et~al.}(2008{\natexlab{b}})\citenamefont {{van der Sluys}} \emph
  {et~al.}}]{Sluys:2008b}%
  \BibitemOpen
  \bibfield  {author} {\bibinfo {author} {\bibfnamefont {M.~V.}\ \bibnamefont
  {{van der Sluys}}} \emph {et~al.},\ }\href {\doibase 10.1086/595279}
  {\bibfield  {journal} {\bibinfo  {journal} {The Astrophysical Journal
  Letters}\ }\textbf {\bibinfo {volume} {688}},\ \bibinfo {pages} {L61}
  (\bibinfo {year} {2008}{\natexlab{b}})},\ \Eprint
  {http://arxiv.org/abs/0710.1897} {arXiv:0710.1897} \BibitemShut {NoStop}%
\bibitem [{\citenamefont {{Veitch}}\ and\ \citenamefont
  {{Vecchio}}(2010)}]{Veitch:2010}%
  \BibitemOpen
  \bibfield  {author} {\bibinfo {author} {\bibfnamefont {J.}~\bibnamefont
  {{Veitch}}}\ and\ \bibinfo {author} {\bibfnamefont {A.}~\bibnamefont
  {{Vecchio}}},\ }\href@noop {} {\bibfield  {journal} {\bibinfo  {journal}
  {\prd}\ }\textbf {\bibinfo {volume} {81}},\ \bibinfo {pages} {062003}
  (\bibinfo {year} {2010})},\ \Eprint {http://arxiv.org/abs/0911.3820}
  {arXiv:0911.3820 [astro-ph.CO]} \BibitemShut {NoStop}%
\bibitem [{\citenamefont {{Feroz}}\ \emph {et~al.}(2009)\citenamefont
  {{Feroz}}, \citenamefont {{Hobson}},\ and\ \citenamefont
  {{Bridges}}}]{Feroz:2009}%
  \BibitemOpen
  \bibfield  {author} {\bibinfo {author} {\bibfnamefont {F.}~\bibnamefont
  {{Feroz}}}, \bibinfo {author} {\bibfnamefont {M.~P.}\ \bibnamefont
  {{Hobson}}}, \ and\ \bibinfo {author} {\bibfnamefont {M.}~\bibnamefont
  {{Bridges}}},\ }\href {\doibase 10.1111/j.1365-2966.2009.14548.x} {\bibfield
  {journal} {\bibinfo  {journal} {Monthly Notices of the Royal Astronomical
  Society}\ }\textbf {\bibinfo {volume} {398}},\ \bibinfo {pages} {1601}
  (\bibinfo {year} {2009})},\ \Eprint {http://arxiv.org/abs/0809.3437}
  {arXiv:0809.3437} \BibitemShut {NoStop}%
\end{thebibliography}%

\end{document}